\documentclass[11pt]{article}
\usepackage[margin=1in]{geometry}
\usepackage{amsmath,amssymb,amsthm,mathtools,bm}
\usepackage{graphicx,booktabs,array,enumitem,microtype,needspace,placeins}
\usepackage{caption}
\DeclareCaptionType{algorithm}[Algorithm][List of Algorithms]
\usepackage[round,authoryear]{natbib}
\usepackage[colorlinks=true,citecolor=blue,linkcolor=blue,urlcolor=blue]{hyperref}
\usepackage[capitalise,nameinlink,noabbrev]{cleveref}
\crefname{algorithm}{Algorithm}{Algorithms}
\Crefname{algorithm}{Algorithm}{Algorithms}
\crefname{definition}{Definition}{Definitions}
\crefname{appendix}{Appendix}{Appendices}
\Crefname{appendix}{Appendix}{Appendices}
\crefname{example}{Example}{Examples}
\crefname{remark}{Remark}{Remarks}
\Crefname{definition}{Definition}{Definitions}
\Crefname{example}{Example}{Examples}
\Crefname{remark}{Remark}{Remarks}
\crefname{proposition}{Proposition}{Propositions}
\Crefname{proposition}{Proposition}{Propositions}
\hypersetup{pdftitle={The stable false discovery rate},
pdfauthor={Aaditya Ramdas},pdfsubject={Multiple testing with fractional allocations},
pdfkeywords={FDR, stable multiple testing, e-values, PRDS, anchoring, e-closure}}
\setlist{itemsep=2pt,topsep=4pt}
\numberwithin{equation}{section}
\newtheorem{theorem}{Theorem}[section]
\newtheorem{lemma}[theorem]{Lemma}
\newtheorem{proposition}[theorem]{Proposition}
\newtheorem{corollary}[theorem]{Corollary}
\theoremstyle{definition}
\newtheorem{definition}[theorem]{Definition}
\newtheorem{example}[theorem]{Example}
\theoremstyle{remark}
\newtheorem{remark}[theorem]{Remark}
\newcommand{\E}{\mathbb E}
\newcommand{\Prob}{\mathbb P}
\newcommand{\R}{\mathbb R}
\newcommand{\cP}{\mathcal P}
\newcommand{\cA}{\mathcal A}
\newcommand{\cB}{\mathcal B}

\newcommand{\one}{\mathbf 1}
\newcommand{\sFDP}{\operatorname{sFDP}}
\newcommand{\sFDR}{\operatorname{sFDR}}
\newcommand{\FDP}{\operatorname{FDP}}
\newcommand{\FDR}{\operatorname{FDR}}
\newcommand{\pos}[1]{\left(#1\right)_{+}}
\newcommand{\ms}{[m]}
\newcommand{\dd}{\,\mathrm d}

\title{The stable false discovery rate}
\author{Aaditya Ramdas\\Stanford University\\\texttt{aramdas@stanford.edu}}
\date{November 2026}

\begin{document}
\maketitle

\begin{abstract}
When a discovery depends on analysis choices, a list of selected
hypotheses that controls the false discovery rate (FDR) gives little indication of how easily the finding could
change: the selected set may be very sensitive to a variety of analyst choices, like the target FDR level, which machine learning algorithms are used in the pipeline, the choice of test statistics, and so on.
We propose that scientists should measure and control a new metric called the \emph{stable FDR} (sFDR), and give methods to control it, including simple wrappers around existing FDR analysis pipelines.
sFDR is a natural continuous extension of the FDR: we assign each hypothesis an allocation between zero and one, and the sFDR equals the expected fraction of null allocation to total allocation. If one runs multiple analyses of the same data by varying different reasonable choices, allocations can be determined by simply dividing each run's selection indicators by its discovery count, then averaging across analyses and
rescaling. 
The resulting sFDR equals
the average of the ordinary FDRs of the runs, even though the runs are
dependent: thus sFDR control holds under the same assumptions that the original FDR control held. (In contrast, naively reporting just fractions of times each variable was selected need not control the sFDR.) It is interpretable: hypotheses that were always selected are allocated 1 and those never selected are allocated 0. In leukemia expression and functional neuroimaging analyses,
findings with larger allocations were more often supported in separate
samples. We also develop procedures that can assign stable allocations directly from
p-values, e-values, or knockoff statistics, resulting in stable-BH, stable e-BH and stable knockoff procedures. A representation by
e-values describes which allocation rules control the error rate
and gives ways to optimize allocations for different follow-up goals.
\end{abstract}

\section{Introduction}

Scientific conclusions can change when the data stay the same but the
analysis changes. In a study of soccer referees' decisions, 29 teams
examined the same data and question: 20 found a statistically significant
association between players' skin tone and red cards, whereas nine did
not \citep{Silberzahn18}. In neuroimaging, 70 teams tested the same nine
hypotheses on one data set and reached substantially different conclusions
\citep{BN20}. A social science study involving 161 researchers in 73 teams
also found wide variation in estimates and conclusions about immigration
and support for social policies \citep{Breznau22}. These studies document
how strongly conclusions can depend on choices made during analysis.

Such choices arise in ordinary scientific work. An expression analysis
may change its treatment of measurements near the background level.
A brain imaging analysis may use a different group statistic.
A method that learns from one part of the sample and tests on another
may change the split or the fraction reserved for testing. Even the
nominal error level can change which findings cross a reporting threshold.
Randomized procedures introduce another source of variation. For
example, redrawing the knockoffs or changing a random data split can
change which hypotheses are selected on the same observations
\citep{CFJL18,RB24}.

The predictability, computability, and stability (PCS) framework of
\citet{YK20} treats changes to data, preprocessing, and models as
perturbations whose effects on reported results can be assessed and
documented. In multiple testing, an analysis reports a set of
discoveries. A perturbation can change both the membership and the
size of that set. Discovery counts matter because a false discovery
in a short list changes its false discovery proportion more than the
same false discovery in a long list.

We use allocations in $[0,1]$ to report how discoveries persist across
a user-specified (but not data-dependent) collection of analyses. The \emph{stable false discovery
rate} (sFDR) is the expectation of the ratio of null allocation
to total allocation.
For binary allocations, it equals the ordinary FDR \citep{BH95}.
A scientist can use larger allocations to prioritize findings for
review or follow-up.

\Cref{sec:wrapper} gives a simple wrapper for controlling sFDR by combining the results of re-analyses of a dataset using FDR controlling methods with different analyst choices. 
It divides each run's discovery indicators by its discovery
count, averages, and restores a scale on which always-selected
hypotheses receive one. The output has stable FDR equal to the mean
FDR of the runs, despite reusing the same data, and thus the sFDR is controlled under the same assumptions that the FDR is controlled.
\Cref{sec:applications} varies scientifically relevant choices in
Benjamini--Hochberg (BH) analyses of leukemia expression and functional
neuroimaging. Findings assigned larger allocations are more often
supported in separate assessment samples. 
We also exhibit a BH example
where raw selection frequencies exceed the target and study when raw
frequencies give similar reports.
We also develop stable analogs of the BH, e-BH and knockoff procedures, 
give universal representations using stable e-closure and compound
e-values, and study optimization and admissibility.

\paragraph{Related work.}
Repeated subsampling underlies stability selection
\citep{MB10,SS13}. Other methods average evidence to form a new binary
selection \citep{IWR24,RB24}. Here the graded report itself carries an
error guarantee. Continuous and fuzzy testing have several precedents
\citep{GM05,KL09,Habiger15,Koning24}. SMARA develops strategic resource
allocation and a general continuous closure principle
\citep{SMARA26}. Koning's E-measure work studies continuous evidence
over logically structured hypotheses \citep{Koning26}. Our closure
results specialize continuous closure to the reference-unit loss.
The e-value constructions use e-BH, compound e-values, e-closure, and
anchoring \citep{WR22,IWR24,Closure25,Anchor26}.
\Cref{sec:related-details} relates the elementary wrapper to
binary compound e-value aggregation.

\paragraph{Paper outline.}
\Cref{sec:metric,sec:wrapper} define stable FDR, explain why pooling
requires normalization, and give the wrapper. \Cref{sec:applications}
presents the two applications, and \Cref{sec:known-truth} studies when
raw frequencies succeed or fail. \Cref{sec:stablebh,sec:pbh,sec:alpha,sec:ebh,sec:fuzzy-knockoffs}
develop allocation methods for p-values, e-values, and knockoffs.
\Cref{sec:closure,sec:optimization} give representations and ways to
optimize allocations. \Cref{sec:discussion} discusses interpretation
and follow-up. The Appendix Outline after the references locates
the proofs and empirical protocols. \Cref{sec:anchor}
gives improvements by anchoring, and \Cref{sec:other-metrics} extends
the framework to stable familywise error and false coverage.

\section{Allocations and stable FDR}\label{sec:metric}

Let $\cP$ be the statistical model and let $H_i\subseteq\cP$,
$i\in[m]=\{1,\ldots,m\}$, be its hypotheses. Under $P\in\cP$,
$N(P)=\{i:P\in H_i\}$ is the true-null set and $m_0=|N(P)|$.
Write $X$ for the observations, including any auxiliary randomization
held fixed across the analyses under consideration. All procedures
are measurable in their inputs and any further randomization. Capital Latin letters denote random
variables, and lowercase letters denote realizations or deterministic
inputs and outputs. In particular, $A=a(\bm P)$ is the random output
of an allocation map applied to a p-value vector.
For $a=(a_1,\ldots,a_m)\in[0,1]^m$, define
\[
 s(a)=\sum_{i=1}^m a_i,\qquad \mathcal D(a)=1\vee s(a),\qquad
 L_S(a)=\frac{\sum_{i\in S}a_i}{\mathcal D(a)},\quad S\subseteq\ms.
\]

\begin{definition}[Stable FDP and FDR]\label[definition]{def:fuzzy}
For a deterministic allocation $a$ and a random allocation $A$, the stable
false discovery proportion and rate are
\begin{equation}\label{eq:fuzzy}
 \sFDP_N(a)=L_N(a),\qquad
 \sFDR_P(A)=\E_P[L_{N(P)}(A)],\quad A\in[0,1]^m.
\end{equation}
A procedure controls sFDR at level $\alpha$ if
$\sFDR_P(A)\leq\alpha$ for every $P\in\cP$.
\end{definition}

The ordinary false discovery proportion of a rejection set $r$ is
$\FDP_N(r)=|r\cap N|/(1\vee|r|)$, which equals $\sFDP_N(a)$ for
$a_i=\one\{i\in r\}$. The ordinary FDR of a random rejection set $R$ is
$\FDR_P(R)=\E_P[\FDP_{N(P)}(R)]$. For fractional allocations with
$s(a)\geq1$, it is the fraction of allocation assigned to null hypotheses.
For $s(a)<1$, it measures null allocation in units of one full allocation.
Thus the reference unit matters. If the application has a different minimum
meaningful unit $\tau$, the analogous denominator is $\tau\vee s(a)$;
throughout we use units in which $\tau=1$.

\subsection{Combining results across analyses}

Suppose several analyses report allocations for the same hypotheses.
An analysis that gives out ten units and one that gives out one unit
can have the same false discovery proportion. Averaging their raw
allocations gives the first analysis ten times the weight in the
combined proportion. Dividing each allocation vector by its own
denominator $\mathcal D$ gives equal weight to the false discovery
proportions of the individual analyses.

\begin{proposition}[Exact pooling of normalized allocations]\label[proposition]{prop:normalized-pool}
Let $A^{(1)},\ldots,A^{(B)}$ be allocation rules on the same hypotheses.
Normalize each vector by its stable FDP denominator and average the
normalized vectors:
\begin{equation}\label{eq:normalized-pool}
 \bar A_i=\frac1B\sum_{b=1}^B \frac{A_i^{(b)}}{1\vee\sum_j A_j^{(b)}}.
\end{equation}
For every subset $S$ and every realization,
$L_S(\bar A)=B^{-1}\sum_b L_S(A^{(b)})$. Consequently, its sFDR
is the mean of the component risks under every $P$.
The conditional expectation of a randomized rule's normalized allocation
given the observed data likewise has sFDR equal to that of the
randomized rule.
\end{proposition}
\begin{proof}
Each normalized vector is nonnegative and has sum at most one, so the
same holds for its average or conditional expectation. Its
stable FDP therefore equals the sum of its coordinates in $S$. Linearity
gives the pointwise identity, and the tower property gives the
conditional-expectation version.
\end{proof}

\subsection{Reporting raw frequencies of selection does not control the stable FDR}

\begin{figure}[htbp]
 \centering\includegraphics[width=.83\textwidth]{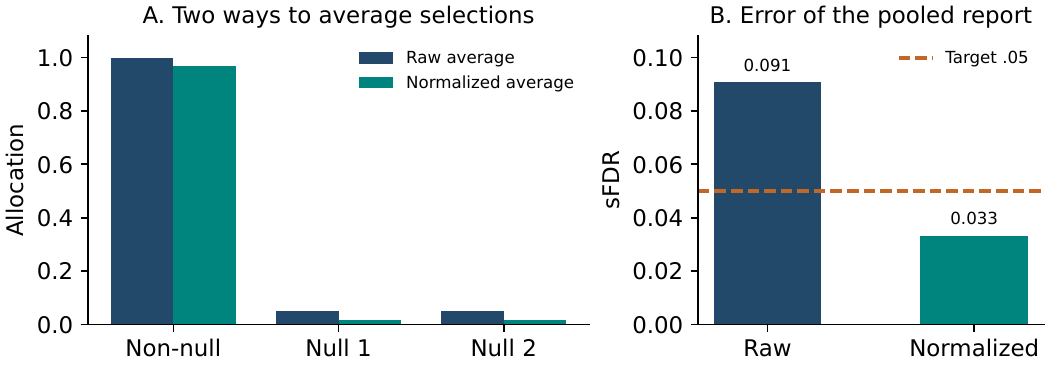}
 \caption{The raw and normalized averages in \Cref{ex:randomization}. The dashed line marks the target $.05$.}
 \label{fig:pooling-loss}
\end{figure}

\Needspace{8\baselineskip}
\begin{example}[Why raw averaging can fail]\label[example]{ex:randomization}
Let $N=\{2,3\}$ and let the observed label $J$ be uniform on
$\{1,\ldots,20\}$. Rule $b$ returns
$A^{(b)}=(1,\one\{J=b\},\one\{J=b\})$. Each has stable FDR
$2/(3\cdot20)=1/30$. Their arithmetic mean is always
$(1,.05,.05)$, whose stable FDP is $1/11>.05$. In each batch,
the one rule that selects both nulls also has the largest total
allocation. The normalized mean $(29/30,1/60,1/60)$ has stable FDP
$1/30$, equal to the mean of the individual rules' FDRs.
\Cref{fig:pooling-loss} shows the two mean reports.
\end{example}

The next section applies \Cref{prop:normalized-pool} to binary
reports and restores a scale suitable for follow-up.

\section{A wrapper for stable multiple testing}\label{sec:wrapper}

Consider $B$ prespecified runs that may differ in their preprocessing,
test statistic, nominal level, or randomization. Each run tests the
same hypotheses. \Cref{alg:wrapper} describes the wrapper when all runs
use level $\alpha$. Its identity transfers the average of the individual
FDR guarantees to the stable allocation.

\begin{algorithm}[!htb]
\centering
\begin{minipage}{\textwidth}
\centering
\fbox{\begin{minipage}{.94\linewidth}
\small
\textbf{Elementary stable FDR wrapper}\par\medskip
\textbf{Inputs.} Observed data $x$; $m$ hypotheses; an FDR-controlling
procedure; level $\alpha\in(0,1)$; number of runs $B\geq1$;
a perturbation scheme that preserves each run's FDR guarantee.
\begin{enumerate}[leftmargin=1.8em,itemsep=4pt,topsep=5pt]
\item \textbf{Rerun.} Redraw the randomization or vary the specified
data, statistic, or tuning choice. Record rejection sets
$r_1,\ldots,r_B$ from $B$ runs, each at level $\alpha$.
\item \textbf{Normalize and average.} Set
\[
 d_i=\frac1B\sum_{b=1}^B
      \frac{\one\{i\in r_b\}}{1\vee|r_b|},
 \qquad s=\sum_i d_i,\qquad v=\max_i d_i.
\]
\item \textbf{Lift.} Return $a\in[0,1]^m$ with coordinates
\[
 a_i=
 \begin{cases}
 d_i,&s<1,\\
 d_i/v,&s=1.
 \end{cases}
\]
Here $s$ is the fraction of nonempty runs, so $s\leq1$;
the second branch has $v>0$. If all runs are empty, $a=0$.
\end{enumerate}
\textbf{Guarantee.} For $P\in\cP$, let $N=N(P)$, let $R_b$ be
the random rejection sets, and let $A$ be the random output. Then
\begin{equation}\label{eq:wrapper-guarantee}
 \E_P\!\left[\frac{\sum_{i\in N}A_i}{1\vee\sum_j A_j}\right]
 =\frac1B\sum_{b=1}^B
   \E_P\!\left[\frac{|R_b\cap N|}{1\vee|R_b|}\right]\leq\alpha .
\end{equation}
Indeed, $a_i/(1\vee\sum_j a_j)=d_i$ in both branches.
Every always-selected hypothesis receives one, and every
never-selected hypothesis receives zero.
\end{minipage}}
\captionsetup{hypcap=false}
\caption{The elementary wrapper at the original level
$\alpha$. The proof is in \Cref{sec:pooling-details}.}
\label{alg:wrapper}
\end{minipage}
\end{algorithm}

The same construction permits prespecified levels $\lambda_1,\ldots,
\lambda_B$ whose arithmetic mean is at most $\alpha$. The identity in
\Cref{eq:wrapper-guarantee} then gives a bound of
$B^{-1}\sum_b\lambda_b$. A symmetric grid around $\alpha$ is one example. The identity holds under arbitrary dependence between runs.

\subsection{Reading the report}

The allocation scale is part of the report. For a follow-up effort budget measured in full allocation units,
$a_i=.6$ assigns six tenths of a unit to hypothesis $i$. Effect estimates
and their uncertainty can be reported alongside the allocation.
When $s(a)\geq1$ and effort is spent proportionally to $a$, stable FDP
is exactly the fraction of that effort assigned to nulls. When
$s(a)<1$, it measures null effort relative to one full unit. Under
the global null it is $\min\{s(a),1\}$; the reference-unit floor permits
small responses to weak evidence.

The lift makes persistent findings interpretable on this scale.
When every rerun is nonempty, write
\[
 d_i=\frac1B\sum_{b=1}^B
       \frac{\one\{i\in r_b\}}{|r_b|},\qquad
 a_i=\frac{d_i}{\max_j d_j}.
\]
These are inverse-discovery-count-weighted selection frequencies,
scaled so their largest coordinate is one. A hypothesis selected in
every run attains that maximum and receives one; a never-selected
hypothesis receives zero. If all runs select the same number of
hypotheses and at least one is always selected, the output equals
the ordinary selection frequencies. Different discovery counts are
where the normalization matters.

\begin{figure}[tb]
 \centering\includegraphics[width=.94\textwidth]{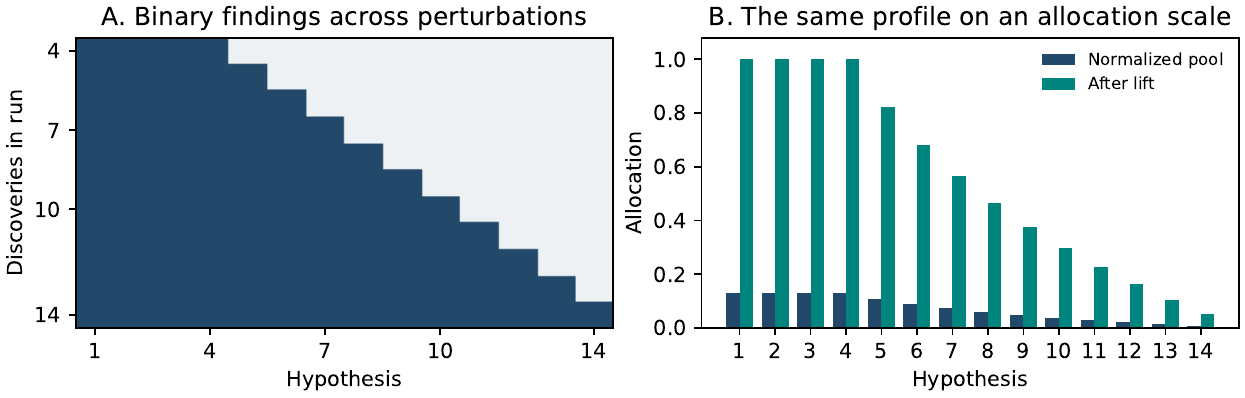}
 \caption{Lifting makes the report usable on a full-unit scale.
 Eleven reruns make between 4 and 14 discoveries. The first four
 hypotheses are selected in every run. Normalization alone assigns
 them about $.13$ each; \Cref{alg:wrapper} lifts them to one and retains
 the relative profile and the normalized error exactly.}
 \label{fig:lift-overview}
\end{figure}

When a persistent core exists, each allocation is the fraction of
inverse-discovery-count weight contributed by runs selecting that feature.
Without a persistent core, the strongest weighted selection frequency
still receives one, so its scale describes relative persistence.
We report raw frequencies alongside allocations in \Cref{fig:leukemia}.
Redrawing knockoffs probes algorithmic randomness;
varying the level probes a decision boundary; removing observations
probes dependence on the available sample. The scheme, its levels,
the number of reruns, and their discovery counts belong in the report.
The allocation guarantee concerns the vector that is reported.
A subsequent binary selection or a different resource allocation
requires its own error analysis.

There is a scale change in the elementary wrapper when the last
empty rerun becomes nonempty. If some reruns are empty, its total
normalized mass is below one and its output is that normalized vector.
\Cref{sec:perturb} gives a continuous lift that also retains
zero allocation for absent findings.

\subsection{What may be changed between runs}

Every run addresses the same hypotheses. Independent random row deletion
from i.i.d.\ observations preserves the retained sample's distribution.
A fixed, label-blind preprocessing transformation preserves a two-group
distributional null. Changing a statistic is allowed when its p-value
remains valid for that common null. A randomized method can have a
guarantee over its randomization without having one at every fixed seed.
The wrapper uses the FDR guarantee averaged over the run's
randomization and allows overlap and dependence between runs. A predetermined finite grid needs no distribution
on the grid beyond the equal averaging in \Cref{alg:wrapper}.

Choices based on the observed outcomes are in general not allowed and may require their own calibration.
An uncalibrated bootstrap need not preserve null p-values, and choosing
the smallest of several valid p-values is generally invalid. Likewise,
repeating an analysis until its report looks satisfactory is still not allowed by our results. The sensitivity scheme and the reported family must be
specified before examining held-out outcomes.

\FloatBarrier
\section{Sensitivity analyses in leukemia and neuroimaging}\label{sec:applications}

\subsection{BH assumptions in the applications}\label{sec:bh-assumptions}

For $m$ p-values, BH sorts $p_{(1)}\leq\cdots\leq p_{(m)}$, sets
$k=\max(\{0\}\cup\{j:p_{(j)}\leq\alpha j/m\})$, and rejects
$\{i:p_i\leq\alpha k/m\}$, with an empty set when $k=0$.
Write $\bm P=(P_1,\ldots,P_m)$ for the random p-value vector.
The null p-values are \emph{superuniform}, meaning
$\Prob_P(P_i\leq u)\leq u$ for every $i\in N(P)$ and $u\in[0,1]$.
Positive regression dependence on the true-null subset (PRDS) is used
in the following weak form: for every null index $i$ and
every measurable coordinatewise increasing set $B\subseteq[0,1]^m$,
\begin{equation}\label{eq:prds}
 u\longmapsto\Prob_P(\bm P\in B\mid P_i\leq u)
 \quad\text{is nondecreasing wherever defined.}
\end{equation}
The usual PRDS condition with conditioning on $P_i=u$ implies
\Cref{eq:prds}. Independent p-values satisfy this condition, as do one-sided
Gaussian p-values with nonnegative correlations in the usual PRDS setting
\citep{BY01}. We write ``nonincreasing'' and ``nondecreasing'' in the
coordinatewise, non-strict sense.

Under valid marginal p-values and PRDS, BH controls FDR at
$\alpha m_0/m$ \citep{BH95,BY01}.

For both applications, stable FDR control follows if the constituent BH
analyses satisfy their FDR assumptions. The observed allocation profiles
and held-out support quantify sensitivity and reproducibility.
\Cref{sec:known-truth} supplies a separate experiment with known nulls
and a Gaussian model in which those assumptions hold.

\subsection{Leukemia: background correction, statistic, and patient influence}
\label{sec:case-study}

The Golub leukemia data contain 7,129 expression probes measured in
38 discovery patients: 27 with acute lymphoblastic leukemia (ALL) and
11 with acute myeloid leukemia (AML) \citep{Golub99}. The hypotheses
assert equality of the two expression distributions, one per probe.
All probes remain in the testing family. The original separate cohort
of 34 patients, 20 ALL and 14 AML, is used only for follow-up assessment.

We vary three choices with direct scientific interpretations.
First, log expression is $\log_2(\max\{x,c\})$, with background floor
$c=20$ or $100$. A floor of $100$ appears in earlier analyses of
these data \citep{DFS02}; the lower floor examines how conclusions
depend on weak expression measurements. Second, the test statistic
is either the absolute difference in mean log expression or a two-sided
rank sum. The former uses effect magnitudes; the latter reduces the
influence of extreme measurements. Third, each run retains $80\%$,
$90\%$, or $100\%$ of each diagnostic group, rounded down. This probes
patient influence without changing the diagnostic comparison.

For each retention fraction we draw ten patient subsets and apply both
floors and both tests, giving 120 equally weighted BH runs, all at
$\alpha=.10$. The mean statistic uses a plus-one permutation p-value
from 1,999 random label permutations. The rank-sum test uses its exact
null distribution after independent random tie breaking. Under each
probe's distributional null, both are marginally valid by label
exchangeability. \Cref{sec:protocols} gives implementation details.
The grid is an exploratory sensitivity analysis fixed before the
test-cohort assessment.

\begin{figure}[tb]
\centering\includegraphics[width=\textwidth]{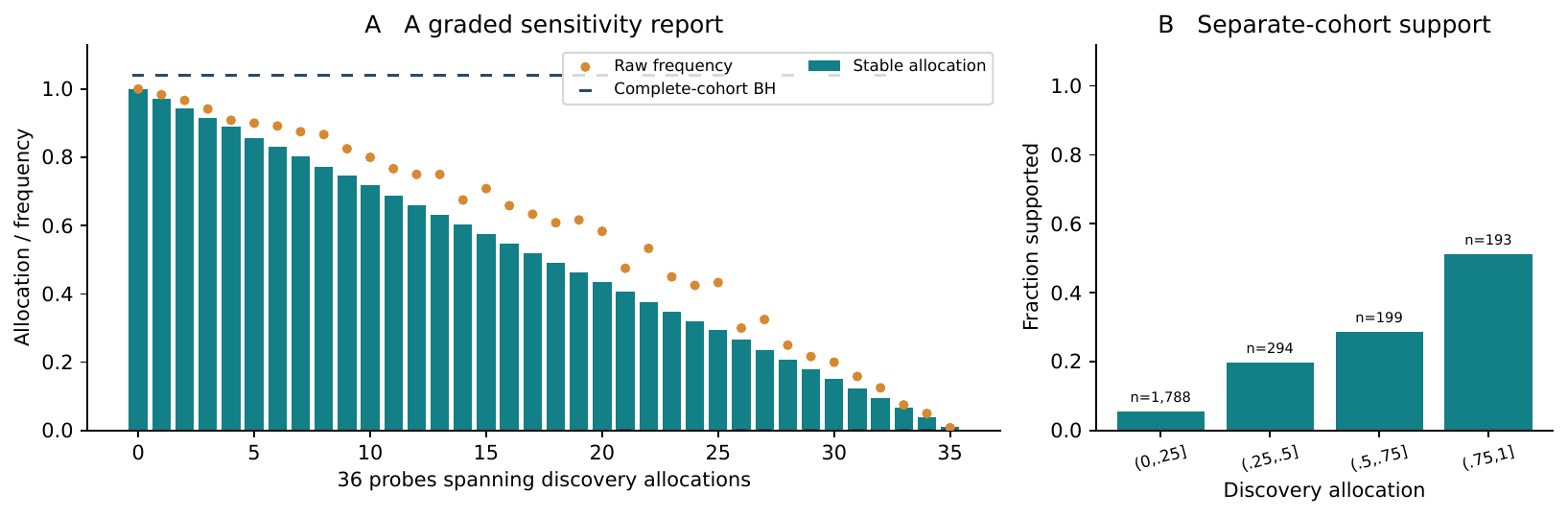}
\caption{Leukemia sensitivity under ordinary BH at $.10$. Left:
36 probes chosen to span the discovery allocation range, with raw
frequencies and the complete-discovery-cohort BH analysis for comparison.
Right: directional support in the separate test cohort, by discovery
allocation bin. Support means test-cohort BH rejection with the same
log-mean effect direction. The bin counts are numbers of probes.}
\label{fig:leukemia}
\end{figure}

The number of discoveries ranges from 178 to 951. All runs are nonempty,
66 probes are always selected, and 4,655 are never selected. The wrapper
assigns these probes one and zero, respectively, with 2,408 fractional
allocations between them. Total allocation is $478.9$ full units.
Before lifting, each persistent probe has normalized allocation only
$.00209$; the lift restores one without changing its normalized error.
\Cref{fig:leukemia} shows probes spanning the discovery allocation range;
their choice uses no test-cohort information.

The reference complete-cohort analysis uses floor $100$, the mean-log
permutation statistic, and BH at $.10$. Applied separately to the test
cohort, that same specification defines support, together with agreement
of effect direction. The permutation reference analyses use 9,999 draws.
The fraction supported rises from $5.5\%$ for allocations in $(0,.25]$
to $51.3\%$ in $(.75,1]$. \Cref{tab:applications} reports both the
support fraction and the amount of supported allocation. The stable
report spends less effort than the reference BH set and has a higher
supported share; its absolute supported allocation is also smaller.
Support is a reproducibility outcome, and unconfirmed findings cannot
be classified as known nulls from this small cohort.

\subsection{Neuroimaging: group statistic and participant influence}
\label{sec:imaging}

The functional localizer data provide subject-level contrast maps for
left versus right button presses \citep{Pinel07,Brainomics17}.
We use 94 available participant maps on their original 3-mm grid.
A fixed external MNI152 brain mask defines 67,969 voxel hypotheses,
and a seeded random split assigns 47 participants to discovery and 47
to held-out assessment. The contrast, spatial grid, mask, and direction
of the comparison are fixed throughout. No voxel enters or leaves the
family because it happens to be significant.

Each hypothesis concerns a nonpositive group location for the contrast
among participants with usable measurements at that voxel. We compare
the usual one-sided one-sample Student t-test with a one-sided signed-rank
test, which uses ranks of absolute participant contrasts rather than
their magnitudes. The comparison examines sensitivity to extreme
participant effects. A common sufficient marginal model is independent
Gaussian usable contrasts at each voxel with its own mean and variance.
The signed-rank test is also valid under the corresponding symmetric
location model. Independence or PRDS across voxel p-values remains a
separate BH assumption, as explained in \Cref{sec:bh-assumptions}.

We retain 37, 42, or all 47 discovery participants, draw 20 subsets at
each size, and apply both statistics to each subset. All 120 runs use
ordinary BH at $.10$. Signed ranks use the exact fair-sign distribution
and independent tie breaking. Missing measurements are omitted at their
voxel; the inferential model is for the usable-observation population.
The family remains the full fixed mask, including voxels assigned
$p=1$ because their data are insufficient.

\begin{figure}[p]
\centering\includegraphics[width=.94\textwidth]{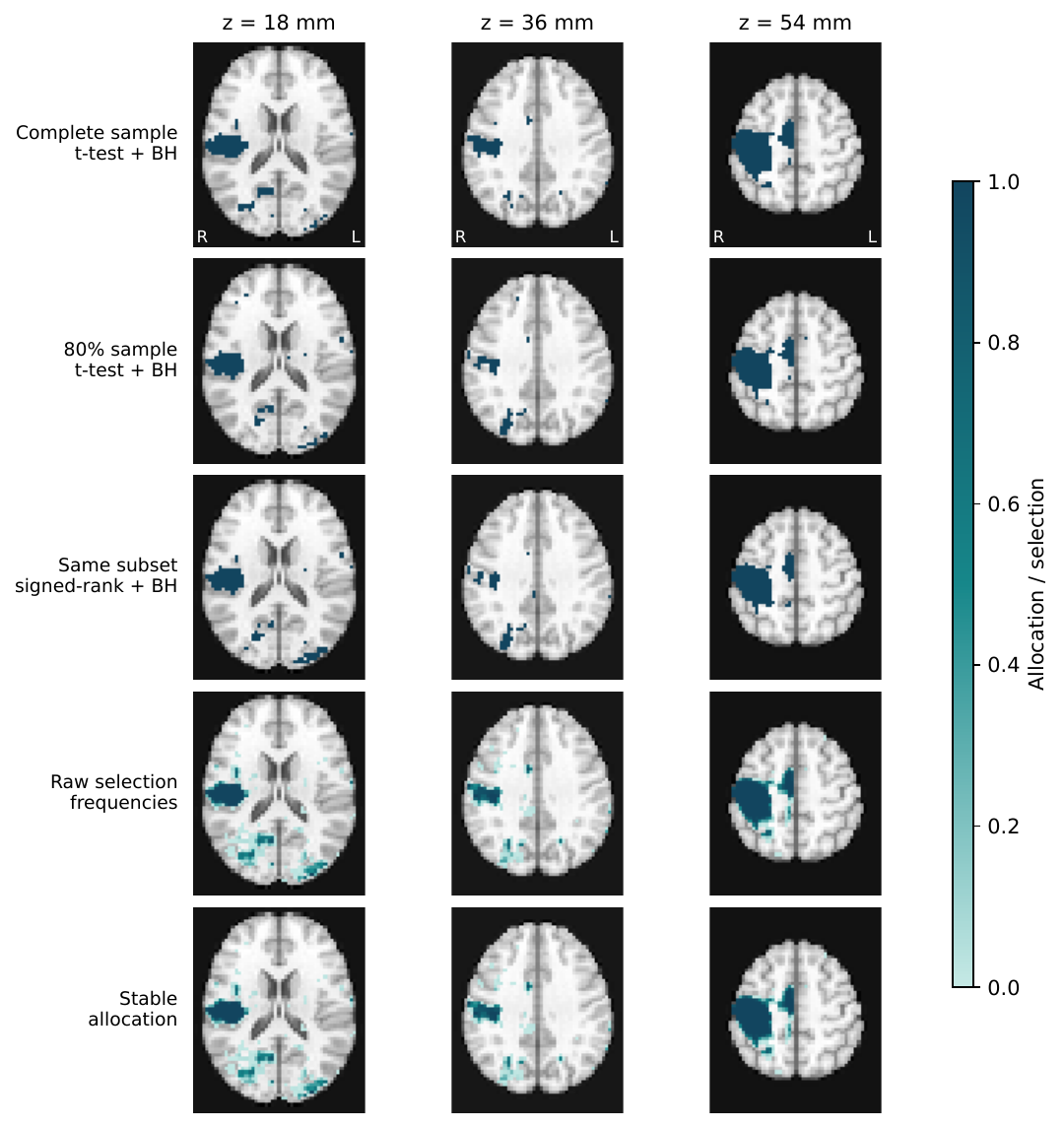}
\caption{BH sensitivity maps for the fixed left-versus-right contrast.
The first row uses the full 47-person discovery sample. The next two
rows use the same first 37-person subset with two group statistics.
The final rows show raw selection frequencies and lifted allocations
over all 120 analyses. Axial slices and color scales are fixed across
rows. Colored voxels carry positive selection or allocation; the
gray background is the anatomical template.}
\label{fig:imaging}
\end{figure}

Discovery counts range from 2,810 to 5,685. The 1,786 voxels selected
in every run receive one, 58,710 never-selected voxels receive zero,
and 7,473 receive fractional allocations. Total allocation is $4,067.9$
units, compared with 4,669 discoveries from complete-discovery t-test BH.
Each persistent voxel has normalized allocation $.000246$ before lifting
and a full unit afterward.
\Cref{fig:imaging} shows how the two statistics change binary boundaries
and how the wrapper summarizes those changes.

\begin{figure}[tb]
\centering\includegraphics[width=.62\textwidth]{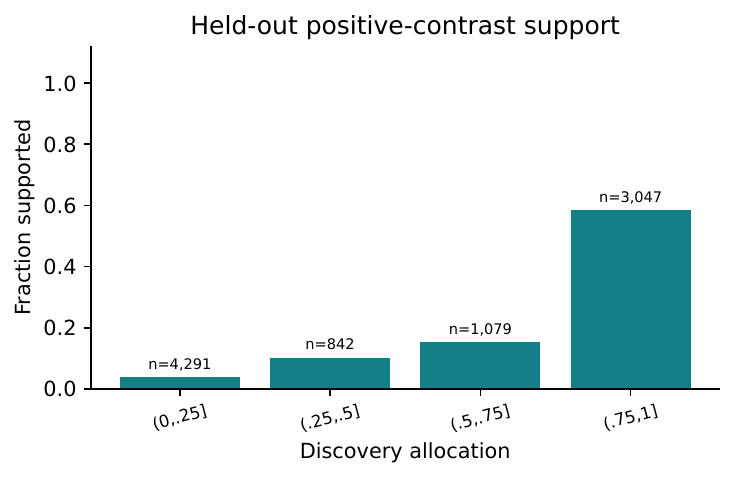}
\caption{Held-out positive-contrast support across imaging allocation
bins. Each positively allocated voxel is counted in one bin. No binomial
uncertainty bars are used because
voxels are spatially dependent.}
\label{fig:imaging-support}
\end{figure}

Held-out support means one-sided t-test BH rejection at $.10$ among
the other 47 participants. The report is fixed before those p-values
are examined. Support rises from $4.0\%$ in the lowest positive
allocation bin to $58.4\%$ in the highest (\Cref{fig:imaging-support}).
The supported share is $46.5\%$ for stable allocation and $43.0\%$ for
the reference BH set. This split assesses reproducibility within one
source cohort. It is not an external replication study, and no
independence across voxels is assumed for these descriptive fractions.

\begin{table}[tb]\centering\small
\begin{tabular}{llrrr}\toprule
Application&Report&Total&Supported&Share\\\midrule
Leukemia&Reference BH&716.0&217.0&30.3\%\\
&Raw frequencies&554.5&173.6&31.3\%\\
&Stable allocation&478.9&157.8&32.9\%\\\midrule
Imaging&Reference BH&4,669.0&2,006.0&43.0\%\\
&Raw frequencies&4,189.0&1,911.1&45.6\%\\
&Stable allocation&4,067.9&1,892.4&46.5\%\\\bottomrule
\end{tabular}
\caption{Allocation effort and held-out support. Total and supported
columns are allocation units; share is their ratio. The support rules
are fixed reference BH analyses in the separate assessment samples.
Raw frequencies are shown as a descriptive comparator and have no
generic stable FDR guarantee.}
\label{tab:applications}\end{table}
\FloatBarrier

\section{A BH example where raw frequencies fail}\label{sec:known-truth}

A synthetic example isolates the effect of differing discovery counts.
There are ten clear signals and a block of 90 correlated null hypotheses.
Each analysis applies ordinary BH at $.10$. The signal p-values are zero,
an idealization of strong evidence, and all nulls share one uniform
p-value within an analysis. Changes of statistic move this null block
together. Every run has valid p-values satisfying PRDS and selects
either the ten signals or all 100 hypotheses. The latter occurs with
probability $.10$, so each run's FDR is exactly $.09$.

The run statistics are correlated at $.8$ because they reuse a shared
Gaussian measurement; each run combines it with a different independent
Gaussian measurement from the same observed data set. The full model
and calibration appear in \Cref{sec:failure-protocol}.
Let $Q$ be the fraction of the $B$ analyses selecting all 100 hypotheses,
$F$ the raw frequency report, and $A$ the wrapper's allocation.
Both reports assign one to every signal. Each null receives $Q$ in
the raw report and $Q/(10-9Q)$ after normalization and lifting.
Their realized false shares are
\begin{equation}\label{eq:bh-failure-loss}
 \sFDP_N(F)=\frac{9Q}{1+9Q},\qquad
 \sFDP_N(A)=.9Q.
\end{equation}
For example, when 12 of 120 runs select the null block, raw frequencies
assign $.10$ to each null: almost half the total allocation is then
assigned to nulls. The wrapper assigns about $.011$ to each null,
giving a null share of $.09$.

\begin{figure}[tb]
\centering\includegraphics[width=\textwidth]{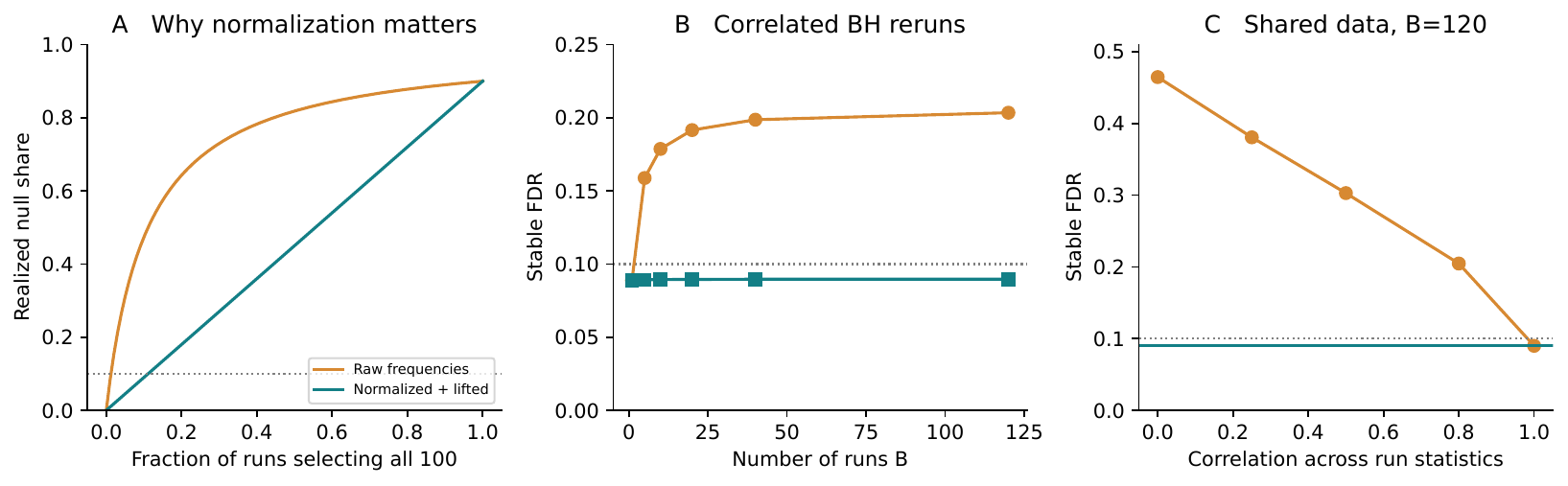}
\caption{Failure of raw frequencies despite valid BH reruns.
Orange denotes raw averaging, teal denotes normalization and lifting,
and dotted lines mark the nominal $.10$. Panel A compares the realized
shares in \Cref{eq:bh-failure-loss}. Panel B averages over 100,000
independent data sets with run-statistic correlation $.8$; error bars
are 1.96 Monte Carlo standard errors. Panel C evaluates the expected
risk by numerical integration at $B=120$, varying that correlation.
Every individual run has FDR $.09$ in every panel.}
\label{fig:wrapper-sim}\end{figure}

At $B=120$, the empirical stable FDR is $.2034$ for raw frequencies
(Monte Carlo standard error $.0010$), compared with $.0897$ for the
wrapper (standard error $.0006$). Numerical integration gives $.2047$
for the raw report, while the wrapper's exact risk is $.09$ for every
$B$. An analytic lower bound of $.1059$ also establishes that the raw
risk exceeds $.10$ (\Cref{sec:failure-protocol}). Repeating the analyses
estimates the raw frequency profile more precisely while preserving its
excess null allocation. In this example, all rejection sets are nested
and their union is their largest member.

\subsection{When are raw frequencies sufficient?}

Equal rejection counts across runs make raw averaging valid: its stable
FDR equals the average FDR of the runs, regardless of overlap between
their selected hypotheses. More generally, suppose every possible
batch is either entirely empty or consists of nonempty sets whose
largest count is at most $c$ times the smallest. If every run controls
FDR at $\alpha$, its raw frequency report $F$ satisfies
\begin{equation}\label{eq:raw-bound-main}
 \sFDR_P(F)\leq
 \min\left\{\frac{Bc}{B+c-1}\alpha,
             \frac{c\alpha}{1+(c-1)\alpha}\right\}\leq c\alpha.
\end{equation}
Thus a factor of two in counts gives at most $2\alpha$, without
any condition on the union. \Cref{sec:pooling-details} gives a short
proof. \Cref{sec:raw-extensions} gives further exact-control conditions
and establishes sharpness of the linear coefficient.

Count variation also helps explain why raw frequencies perform
reasonably well in the real-data examples. For a given nonempty
batch, put $\bar k=B^{-1}\sum_b|r_b|$ and
$\gamma=\max_b|r_b|/\bar k$. The raw false share is at most
$\gamma$ times the mean run-wise false share. An even more direct
comparison normalizes the raw and wrapped reports to total mass one
and computes half the sum of their absolute coordinate differences.
Call this quantity $\delta$. For any possible set of true nulls,
the realized false shares of those two reports differ by at most
$\delta$ (\Cref{sec:pooling-details}).

\begin{table}[tb]\centering\small
\begin{tabular}{lrr}\toprule
Application&Count factor $\gamma$&Maximal share gap $\delta$\\\midrule
Leukemia&1.715&.0502\\
Neuroimaging&1.357&.0170\\\bottomrule
\end{tabular}
\caption{Observed count imbalance and distance between raw and wrapped
reports. The final column bounds the absolute difference in allocation
share for every subset of hypotheses, including the true nulls.
Values are rounded.}
\label{tab:raw-closeness}\end{table}

In leukemia, this maximal difference is about 5.0 percentage points; in imaging
it is 1.7 points (\Cref{tab:raw-closeness}). The separate-data supported
shares differ by 1.6 and 0.9 points, respectively
(\Cref{tab:applications}). These diagnostics explain the similarity
of the observed reports. The FDR bound in \Cref{eq:raw-bound-main}
requires the count condition to hold across possible data sets;
the observed values alone describe the realized sensitivity analysis.

\FloatBarrier
\section{Stabilizing BH over chosen level from a single sort}\label{sec:stablebh}

We now describe how to stabilize the BH procedure to the choice of level at which the FDR is control.


For a vector $d\geq0$, write $s=\sum_i d_i$ and $v=\max_i d_i$.
The continuous lift $\mathcal L_\rho(d)$ returns $d$ for $s\leq1$
and $d/v$ for $s\geq1+\rho$. For $1<s<1+\rho$ it interpolates
between $d/s$ and $d/v$ using weight $t^2(3-2t)$ on $d/v$, where
$t=(s-1)/\rho$. Its normalized coordinates are $d_i/(1\vee s)$.
\Cref{sec:perturb} gives the general certificate construction and its
risk bound.

Let $p_{(1)}\leq\cdots\leq p_{(m)}$. The entry level of the $j$th
ordered hypothesis into BH is the BH-adjusted p-value
\[
 q_{(j)}=\min_{k\geq j}\frac{m p_{(k)}}{k}.
\]
Thus the selection at $\lambda$ is exactly $\{i:q_i\leq\lambda\}$.
Let $\Lambda$ be uniform on $[L,H]\subset(0,1]$, and write
$\ell_j=(L\vee q_{(j)})\wedge H$ for $j=1,\ldots,m$ and
$\ell_{m+1}=H$. On the open interval $(\ell_k,\ell_{k+1})$ there
are exactly $k$ selections whenever this interval has positive
length. The certificate in \Cref{eq:pert-cert}, which averages the level-scaled normalized BH indicators, is
\begin{equation}\label{eq:stable-path}
 d_{(j)}
 =\frac{\alpha}{H-L}
 \sum_{k=j}^{m}\frac{1}{k}\log\!\left(\frac{\ell_{k+1}}{\ell_k}\right).
\end{equation}
A usual BH sort, one reverse cumulative minimum, and one reverse
cumulative sum give all the coordinates. No grid of levels or
Monte Carlo draws is needed.

For $0<\delta<1$ use the band
$L=\alpha(1-\delta)$, $H=1\wedge\{\alpha(1+\delta)\}$.
Set $s_*=\alpha\log(H/L)/(H-L)$ and use $\rho=s_*-1$.
Since $\E[\Lambda]\leq\alpha$ and $\Lambda$ is nondegenerate,
$s_*=\alpha\E[1/\Lambda]>1$. Define Stable-BH to be
$\mathcal L_\rho(d)$. If at least one feature enters BH by $L$,
then $s=s_*$ and the lift is full. All hypotheses entering by $L$
receive one, and those with entry levels in $(L,H)$ receive
strictly fractional allocations. Features entering at or after $H$
receive zero. The band can also be tilted toward a preferred side
of $\alpha$ by changing the mixing density and integrating it in
\Cref{eq:stable-path}.

\begin{theorem}[Stable-BH]\label{thm:stablebh}
Under PRDS and superuniform null p-values, Stable-BH controls sFDR
at $\alpha |N|/m$. For every fixed $\delta\in(0,1)$, the allocation
is jointly continuous in $(p,\alpha)\in[0,1]^m\times(0,1)$.
Away from BH decision boundaries, Stable-BH converges to the
binary BH indicator as $\delta\downarrow0$.
\end{theorem}

\begin{proof}
For each fixed $\lambda$, the classical BH guarantee gives
$\E[\FDP_N(R_\lambda)]\leq\lambda |N|/m$ \citep{BY01}.
Apply \Cref{thm:perturb}. The sorted values, the clamped entry
levels and all terms in \Cref{eq:stable-path} are continuous.
Tied raw p-values have equal entry levels and equal allocations,
so mapping the sorted allocations back to the original coordinates
preserves continuity. The ramp width is positive and continuous
in $\alpha$. If the BH set is constant in a neighborhood of $\alpha$,
then sufficiently narrow bands have this same set at every level.
For a nonempty set all its members have the same positive
certificate and $s=s_*$, giving their binary allocation. An empty
set yields $d=0$.
\end{proof}

\begin{figure}[htbp]
 \centering\includegraphics[width=.79\textwidth]{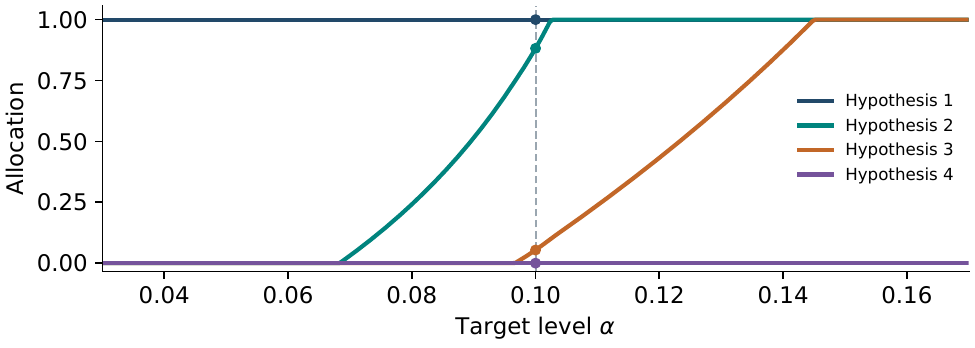}
 \caption{Stable-BH allocations as the target level varies for \Cref{ex:stablebh}. The markers show the allocations at $\alpha=.10$ with relative bandwidth $\delta=.20$.}
 \label{fig:bh-path}
\end{figure}

\begin{example}[Level sensitivity for BH]\label[example]{ex:stablebh}
At $\alpha=0.10$ with $p=(.001,.041,.087,.40)$, BH gives
$(1,1,0,0)$. The band $[.08,.12]$ and \Cref{eq:stable-path} give
Stable-BH allocations approximately $(1,.882,.054,0)$.
The entry levels of the second and third features are $.082$
and $.116$, respectively. \Cref{fig:bh-path} shows the
allocations as the target level varies.
\end{example}

\begin{figure}[htbp]
 \centering\includegraphics[width=.82\textwidth]{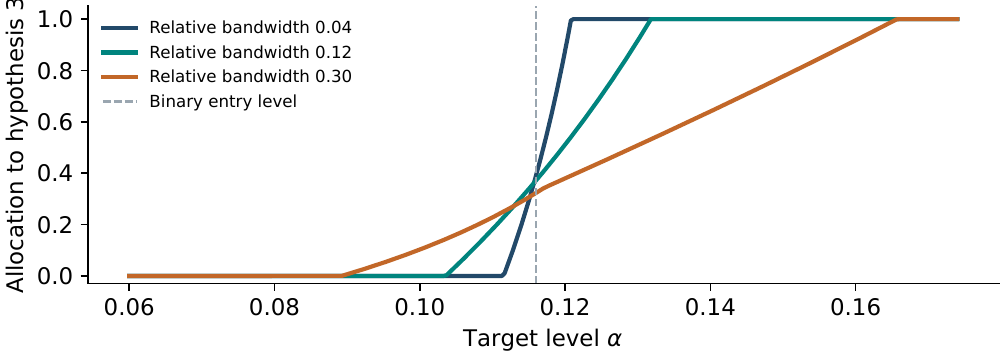}
 \caption{Allocation to hypothesis 3 of \Cref{ex:stablebh} under three choices of level bandwidth. The other inputs are held fixed. The vertical line marks its binary BH entry level.}
 \label{fig:bandwidth-discussion}
\end{figure}

\paragraph{The level bandwidth as a reporting choice.}
The relative bandwidth specifies which levels enter the sensitivity analysis. Wider bands can allocate to a borderline hypothesis at smaller targets and spread its transition over a longer range. \Cref{fig:bandwidth-discussion} holds the p-values fixed and varies this choice.

\section{Kernelized BH under PRDS}\label{sec:pbh}


\subsection{A kernel extension of dependency control}

A convenient way to specify a kernel is through an auxiliary distribution.
This is an analytic representation; the procedure requires no auxiliary
random draw.

\begin{definition}[Unit-area kernel]\label[definition]{def:kernel}
Let $\nu$ be a probability distribution on $[0,\infty)$ with mean one.
Its survival kernel is
\begin{equation}\label{eq:kernel}
 \kappa(u)=\nu([u,\infty)),\quad u\geq0.
\end{equation}
Then $0\leq\kappa\leq1$, $\kappa$ is nonincreasing, and
$\int_0^\infty\kappa(u)\dd u=1$.
\end{definition}

Writing $\delta_1$ for a point mass at one, three examples are
\[
 \begin{array}{c|c|c}
 \nu&\kappa(u)&\text{allocation profile}\\\hline
 \delta_1&\one\{u\leq1\}&\text{hard cutoff}\\
 \operatorname{Exp}(1)&e^{-u}&\text{exponential taper}\\
 \operatorname{Unif}[0,2]&(1-u/2)_+&\text{linear taper}.
 \end{array}
\]
The exponential kernel will be our default.

\begin{lemma}[Kernel PRDS lemma]\label[lemma]{lem:kernel}
Assume \Cref{eq:prds} and let $i\in N$. If
$T:[0,1]^m\to(0,\infty)$ is measurable, finite, and nonincreasing, then
\begin{equation}\label{eq:kernel-lemma}
 \E\left[\frac{\kappa(P_i/T(\bm P))}{T(\bm P)}\right]\leq1.
\end{equation}
\end{lemma}
\begin{proof}
The Blanchard--Roquain dependency-control lemma gives, for each $z>0$,
\begin{equation}\label{eq:BR}
 \E\left[\frac{\one\{P_i\leq zT(\bm P)\}}{T(\bm P)}\right]\leq z.
\end{equation}
Indeed, $\{T<t\}$ is increasing, so \Cref{eq:prds} supplies the required
conditional stochastic monotonicity. The contribution at $z=0$ is
zero because a null p-value is positive almost surely. Tonelli's theorem and
\Cref{eq:kernel} now yield
\[
 \E\left[\frac{\kappa(P_i/T)}{T}\right]
 =\int\E\left[\frac{\one\{P_i\leq zT\}}{T}\right]\nu(\dd z)
 \leq\int z\,\nu(\dd z)=1.
\]
\end{proof}

\begin{theorem}[Stable self-consistency under PRDS]\label{thm:psc}
Suppose \Cref{eq:prds} holds on $N$. Fix $\alpha\in(0,1)$ and let
$a:[0,1]^m\to[0,1]^m$ be measurable and set $A=a(\bm P)$.
Suppose it satisfies
\begin{equation}\label{eq:psc}
 A_i\leq\kappa\left(\frac{mP_i}{\alpha \mathcal D(A)}\right),
 \qquad i\in\ms.
\end{equation}
If $\mathcal D(A)$ is a nonincreasing function of $\bm P$, then
\begin{equation}\label{eq:psc-bound}
 \E\left[\frac{A_i}{\alpha \mathcal D(A)}\right]\leq\frac1m
 \quad(i\in N),\qquad
 \sFDR_P(A)\leq\alpha\frac{m_0}{m}.
\end{equation}
\end{theorem}
\begin{proof}
Apply \Cref{lem:kernel} with $T=\alpha \mathcal D(A)/m$. Since
$A_i/(\alpha \mathcal D(A))\leq\kappa(P_i/T)/(mT)$, the coordinate bound follows.
Summing it over $N$ proves the second assertion.
\end{proof}

\subsection{A continuous default: finite-round kernelized BH}

\begin{definition}[$K$-round kernelized BH]\label[definition]{def:finite}
For an input $p\in[0,1]^m$, fix a positive integer $K$ and a unit-area
kernel $\kappa$. Starting from
$r_0=1$, compute, for $k=0,\ldots,K-1$,
\begin{equation}\label{eq:iteration}
 a_i^{(k+1)}=\kappa\left(\frac{mp_i}{\alpha r_k}\right),\qquad
 r_{k+1}=1\vee\sum_i a_i^{(k+1)}.
\end{equation}
The output is $a^{(K)}$. We use $\kappa(u)=e^{-u}$ and $K=25$ as a
simple default.
\end{definition}

\begin{theorem}[Validity and continuity]\label{thm:finite}
Under PRDS, $K$-round kernelized BH obeys \Cref{eq:psc-bound} for every fixed
$K\geq1$. Its coordinates and its total allocation are nonincreasing in
the p-values and nondecreasing in $\alpha$. If $\kappa$ is continuous,
the allocation is jointly continuous in $(p,\alpha)\in[0,1]^m\times(0,1)$.
For the exponential kernel and $p_i>0$, its $i$th allocation lies in $(0,1)$.
The computational cost is $O(Km)$.
\end{theorem}
\begin{proof}
For fixed $(p,\alpha)$, the map
$F(r)=1\vee\sum_i\kappa(mp_i/(\alpha r))$ is nondecreasing on $[1,m]$.
Since $r_1\geq r_0$, induction gives $r_{k+1}\geq r_k$.
Consequently, with $a=a^{(K)}$ and $\mathcal D(a)=r_K$,
\[
 a_i=\kappa(mp_i/(\alpha r_{K-1}))
 \leq\kappa(mp_i/(\alpha r_K)).
\]
Induction also shows that every $r_k$ and every $a_i^{(k)}$ is
nonincreasing in $p$ and nondecreasing in $\alpha$. Thus
\Cref{thm:psc} applies. Continuity follows by composing continuous functions,
including $x\mapsto1\vee x$. The remaining assertions follow from the
exponential formula and the number of operations per round.
\end{proof}

For the exponential kernel, the common scale
$t_k=\alpha r_k/m$ has a simple interpretation:
$a_i^{(k+1)}=e^{-p_i/t_k}$. A p-value equal to $t_k$ receives allocation
$e^{-1}$, and increasing a p-value by $t_k$ multiplies its allocation by
$e^{-1}$. The scale increases as the aggregate evidence supports a larger
allocation. Increasing $K$ increases every allocation while retaining the
guarantee. Fixing $K$ also makes continuity immediate and avoids a
data-dependent numerical stopping rule.

\begin{figure}[htbp]
 \centering\includegraphics[width=\textwidth]{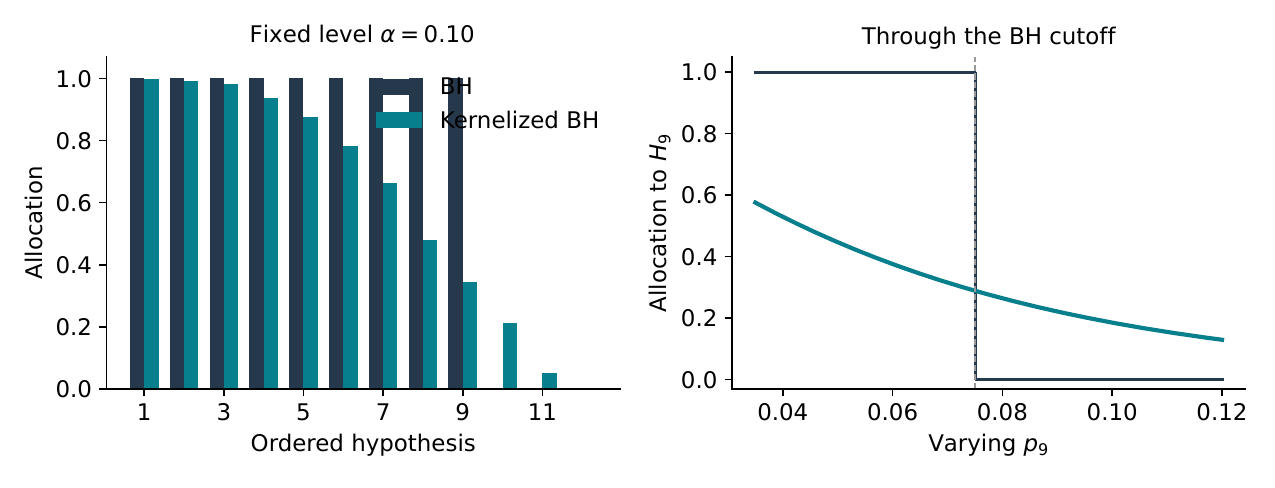}
 \caption{Binary and stable allocations for \Cref{eq:examplep} at
 $\alpha=0.10$. Kernelized BH uses the exponential kernel and $K=25$.
 The right panel changes only $p_9$. The dashed line marks the BH cutoff.}
 \label{fig:allocations}
\end{figure}

Consider the deterministic p-value vector
\begin{equation}\label{eq:examplep}
 (.0001,.0005,.001,.004,.008,.015,.025,.045,.065,.095,.18,.55).
\end{equation}
At level $\alpha=0.10$, BH selects the first nine hypotheses.
Twenty-five-round exponential kernelized BH assigns total mass approximately
$7.319$. Its allocation to hypothesis 9 is about $0.344$, and hypothesis
10 receives about $0.211$. \Cref{fig:allocations} compares the outputs and
varies $p_9$ while holding the other eleven p-values fixed. BH changes its
decision on hypothesis 9 at $p_9=0.075$; the kernel allocation changes
continuously.

\subsection{The maximal step-up version and ordinary BH}\label{sec:maximal}

There is also a direct maximal analog of the BH step-up rule. For a
continuous kernel, define
\begin{equation}\label{eq:maximal}
 F_{p,\alpha}(r)=\sum_i\kappa\left(\frac{mp_i}{\alpha r}\right),\qquad
 r_* =\max\{r\in[1,m]:r\leq1\vee F_{p,\alpha}(r)\},
 \quad a_i^*=\kappa\left(\frac{mp_i}{\alpha r_*}\right).
\end{equation}
The set is nonempty and compact. At its maximum,
$r_*=1\vee F_{p,\alpha}(r_*)=\mathcal D(a^*)$: a strict inequality in the opposite
direction would, by continuity, allow a larger $r$, unless $r_*=m$,
where $F(m)\leq m$. The feasible sets expand when any p-value decreases,
so $r_*$ is nonincreasing in $p$. Hence \Cref{thm:psc} proves validity.

For the hard kernel, the analogous maximum is
$r_*=1\vee R_{\rm BH}$, where
\begin{equation}\label{eq:BH}
 R_{\rm BH}=\max\bigl(\{0\}\cup
 \{j\in\ms:p_{(j)}\leq\alpha j/m\}\bigr).
\end{equation}
The resulting vector is exactly the binary BH selection. Thus the maximal
construction extends BH across kernels. Finite-round iteration from one can
select a smaller fixed point, and even the infinite-iteration limit can
differ from the maximum in \Cref{eq:maximal}.

\begin{example}[A discontinuity of the maximal rule]\label[example]{ex:maximal-jump}
Take $m=4$ and equal p-values $p_1=\cdots=p_4=p$. Put
$c=4p/\alpha$, so $F(r)=4e^{-c/r}$. For $r>0$,
\[
 \max_{r>0}\frac{F(r)}r=\frac{4}{ec},
 \qquad\text{attained at }r=c.
\]
If $p>\alpha/e$, then $c>4/e$ and $F(r)<r$ for every $r>0$.
Thus the maximal feasible scale in \Cref{eq:maximal} is $r_*=1$.
At $p=\alpha/e$, $c=4/e>1$ and $F(c)=c$, so $r_*=4/e$.
Consequently the allocation at the critical point is $e^{-1}$ in every
coordinate, while its limit from $p>\alpha/e$ is $e^{-4/e}$.
Both are fractional, but the maximal rule jumps. Each finite-round map
in \Cref{def:finite} is continuous at this point.
\end{example}

\subsection{Total allocation and the BH baseline}\label{sec:bh-total}

The exponential kernel replaces the binary indicator by a different
allocation profile. The finite-round updates find a self-consistent
allocation, and the maximal-scale version maximizes the scale along
that kernel profile. They do not maximize total allocation over a
feasible region containing the ordinary BH vector. For example,
with one hypothesis, $p=\alpha/2$ gives BH allocation one and
exponential kernel allocation $e^{-1/2}$. The larger ambient
decision space alone gives no total-allocation comparison.

There is a simple allocation that does retain every BH rejection.
For a p-value input $p$, let $r_{\rm BH}$ be its level-$\alpha$ BH set
and write $r=|r_{\rm BH}|$.
If $r=0$ the allocation is zero, and if $r=m$ it is one everywhere.
For $0<r<m$, set
\begin{equation}\label{eq:bh-fill}
 t_r=1\wedge\frac{\alpha r^2}{(m-r)(m-\alpha r)},\qquad
 a_i^+=
 \begin{cases}
  1,&i\in r_{\rm BH},\\
  t_r,&i\notin r_{\rm BH}.
 \end{cases}
\end{equation}
We call this \emph{BH with additional allocations}. The allocation
to unselected hypotheses is uniform because this certificate uses
only the BH set and its size.

\begin{theorem}[A maximum-total extension of BH]\label{thm:bh-fill}
Under superuniform null p-values and PRDS, the random output
$A^+=a^+(\bm P)$ of \Cref{eq:bh-fill} controls
sFDR at $\alpha$. It dominates the ordinary BH indicator coordinatewise,
and its total allocation is
\[
 s(a^+)=\min\!\left\{m,\frac{r}{1-\alpha r/m}\right\}\geq r.
\]
The inequality is strict whenever $0<r<m$. At the observed data,
it maximizes total allocation among all vectors that retain the
observed BH rejections at one and satisfy
\begin{equation}\label{eq:bh-anchored-region}
 L_S(a)\leq
 \frac{|r_{\rm BH}\cap S|}{1\vee r}+\alpha\left(1-\frac{|S|}{m}\right)
 \quad\text{for every }S\subseteq[m].
\end{equation}
The guarantee here is at $\alpha$, including when $m_0<m$.
\end{theorem}

\begin{proof}
Let $R$ denote the random BH set and $r=|R|$. Write $B=\one_R$. For every $S\subseteq N(P)$, the BH
dependency-control argument gives
\[
 \E_P[L_S(B)]\leq\alpha |S|/m.
\]
Indeed, for a null $i$ the BH indicator divided by $1\vee r$
is bounded by
$\one\{P_i\leq\alpha(1\vee r)/m\}/(1\vee r)$, and
\Cref{eq:BR} bounds its expectation by $\alpha/m$.
Thus any allocation satisfying \Cref{eq:bh-anchored-region}
has expected null loss at most
$\alpha m_0/m+\alpha(1-m_0/m)=\alpha$.

For $r>0$, a vector retaining the BH set has total $r+T$, where
$T$ is its additional allocation. The constraint for $S=R^c$
requires
\[
 \frac{T}{r+T}\leq\frac{\alpha r}{m},
 \qquad
 T\leq\frac{\alpha r^2}{m-\alpha r}.
\]
The unit caps also require $T\leq m-r$. The uniform choice in
\Cref{eq:bh-fill} attains the smaller bound. To check its remaining
constraints, write $x=|R\cap S|$, $y=|R^c\cap S|$.
The difference between the two sides of
\Cref{eq:bh-anchored-region} is affine in $(x,y)$.
It suffices to check the four vertices of
$[0,r]\times[0,m-r]$. At $S=R^c$ this is the displayed bound,
at $S=[m]$ both losses equal one, and the other two vertices
satisfy the inequality immediately. If $r=0$, the full-set
constraint forces zero allocation. This proves validity and
optimality over the stated region.
\end{proof}

\begin{figure}[htbp]
 \centering\includegraphics[width=\textwidth]{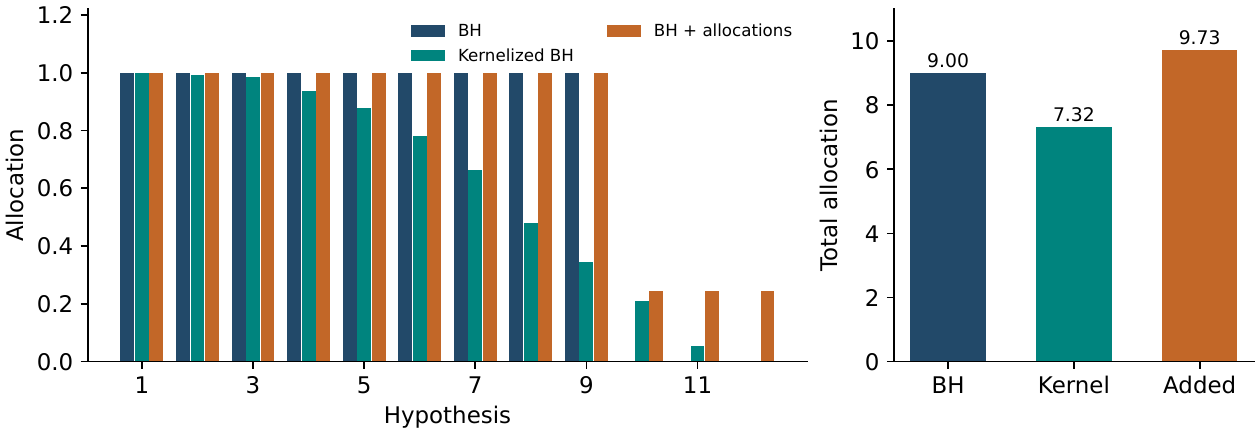}
 \caption{BH, exponential kernelized BH, and BH with additional
 allocations at the same target $\alpha=.10$. The right panel
 shows their total allocations.}
 \label{fig:bh-total}
\end{figure}

\begin{example}[Total allocation can be increased]\label[example]{ex:bh-fill}
For $m=12$, $\alpha=.10$, and the p-values in
\Cref{eq:examplep}, BH makes nine rejections. Kernelized BH
has total about $7.32$. The allocation in \Cref{eq:bh-fill}
retains the nine rejections and assigns $.243$ to each of the
other three, for total $9.73$. This objective counts allocation
to all hypotheses. \Cref{fig:bh-total}
shows the three allocation profiles.
\end{example}

The enlargement in \Cref{eq:bh-anchored-region} is an instance
of the anchoring construction developed in \Cref{sec:anchor}.
It changes the normalized shares, whereas the lift in
\Cref{def:lift} preserves the normalized shares above one.
A continuous allocation that dominates the BH total at the
same level is ruled out by the saturated global-null guarantee.

\begin{proposition}[Continuity and same-level dominance]\label[proposition]{prop:no-continuous-bh}
Fix $m\geq1$ and $0<\alpha<1$. No continuous map
$a:[0,1]^m\to[0,1]^m$ both controls sFDR at $\alpha$ for
independent uniform p-values under the global null and satisfies
$s(a(p))\geq |R_{\rm BH,\alpha}(p)|$ for every $p$.
\end{proposition}

\begin{proof}
Under the independent uniform global null, BH makes at least
one rejection with probability exactly $\alpha$ \citep{BH95}. On this event
total dominance forces stable FDP one. Validity therefore forces
$s(a)=0$ almost surely on the complement. The BH no-rejection
region is open, and continuity and the positive uniform density
give $s(a)=0$ throughout it. Consider a boundary point with
$p_1=\alpha/m$ and all other p-values sufficiently close to one.
It is approached from the no-rejection region, while BH rejects
at the boundary. Continuity requires total allocation zero there,
and dominance requires at least one.
\end{proof}

For continuity with a lower-level BH baseline, a one-sided level
band gives a direct alternative. Fix $0<c<d<1$, mix levels uniformly
on $[c\alpha,d\alpha]$, and use the generic certificate and a ramp
$0<\rho\leq1/d-1$. The allocation is continuous and controls
sFDR at $\alpha m_0/m$. Whenever BH at $c\alpha$ has a discovery,
all runs are nonempty and the certified mass is at least $1/d$,
so the lift is full. Every BH rejection at $c\alpha$ is selected
throughout the band and receives one. For example,
$[.8\alpha,.95\alpha]$ and $\rho=.05$ give a continuous
allocation dominating BH at $.8\alpha$.

\section{Data-dependent levels, including for ordinary BH}\label{sec:alpha}

The dependency-control argument acts on the combined scale
$T=\alpha\mathcal D(A)/m$. It therefore allows $\alpha$ itself to depend on the
p-values, provided this combined scale remains nonincreasing.

\begin{theorem}[Monotone data-dependent levels]\label{thm:alpha}
Suppose PRDS holds on $N$. Let $\widehat\alpha(\bm P)>0$ and
$a:[0,1]^m\to[0,1]^m$ be measurable, with $A=a(\bm P)$. Suppose
\begin{equation}\label{eq:adaptive-sc}
 A_i\leq\kappa\left(\frac{mP_i}{\widehat\alpha \mathcal D(A)}\right),
 \qquad
 \bm P\longmapsto\widehat\alpha(\bm P)\mathcal D(A)
 \ \text{is nonincreasing}.
\end{equation}
Then
\begin{equation}\label{eq:relative}
 \E\left[\frac{A_i}{\widehat\alpha \mathcal D(A)}\right]\leq\frac1m
 \quad(i\in N),\qquad
 \E\left[\frac{\sFDP_N(A)}{\widehat\alpha}\right]\leq\frac{m_0}{m}.
\end{equation}
If $\widehat\alpha\leq q$ almost surely for a deterministic $q$, then
$\sFDR_P(A)\leq q m_0/m$.
\end{theorem}
\begin{proof}
Use \Cref{lem:kernel} with $T=\widehat\alpha \mathcal D(A)/m$ and sum the
coordinate inequalities. The final assertion follows by multiplying
the relative loss by $\widehat\alpha\leq q$.
\end{proof}

\begin{corollary}[BH at a monotone random level]\label[corollary]{cor:BHalpha}
Let $\widehat\alpha:[0,1]^m\to(0,1)$ be nonincreasing, and apply ordinary
BH at level $\widehat\alpha(\bm P)$. Under PRDS,
\begin{equation}\label{eq:BHrandom}
 \E\left[\frac{\FDP_N(R_{\rm BH}(\widehat\alpha))}
 {\widehat\alpha}\right]\leq\frac{m_0}{m}.
\end{equation}
If $\widehat\alpha\leq q$, its ordinary FDR is at most $qm_0/m$.
The same assertions hold for $K$-round kernelized BH run at
$\widehat\alpha(\bm P)$.
\end{corollary}
\begin{proof}
BH is nonincreasing in $p$ and nondecreasing in its level. Thus the composed
rejection count is nonincreasing in $p$, and
$\widehat\alpha(1\vee R_{\rm BH})$ is nonincreasing. Apply
\Cref{thm:alpha} with the hard kernel. For kernelized BH,
\Cref{thm:finite} supplies the same monotonicities.
\end{proof}

\Cref{cor:BHalpha} is an immediate consequence of the existing PRDS lemma,
with the random level absorbed into its scale. It permits evidence-dependent
risk tolerance within the original BH procedure. The guarantee is an
unconditional expected ratio. For example, with one uniform null p-value,
take $\widehat\alpha(p)=q$ for $p\leq q$ and
$\widehat\alpha(p)=\varepsilon$ otherwise, where $0<\varepsilon<q<1$.
BH rejects with probability $q$, and \Cref{eq:BHrandom} holds with equality.
However,
$\E[\widehat\alpha]=q^2+\varepsilon(1-q)<q=\FDR$.
Thus $\E[\FDP/\widehat\alpha]\leq1$ should be interpreted as the
relative-risk guarantee it states.

\subsection{A simple evidence-to-level policy}\label{sec:policy}

An analyst can specify a risk ceiling $q$, a lower fraction
$\eta\in[0,1)$, and an allocation scale $c>0$. First compute a pilot
allocation $b$ using $K_0$-round exponential kernelized BH at the ceiling $q$.
Let $S_0=\sum_i b_i$ and choose
\begin{equation}\label{eq:policy}
 \widehat\alpha(p)=q\left\{\eta+(1-\eta)\frac{S_0(p)}{S_0(p)+c}\right\}.
\end{equation}
Finally, run $K$-round exponential kernelized BH at $\widehat\alpha(p)$.
Both $K_0$ and $K$ are fixed in advance.

The especially simple choice $\eta=0$ and $c=1$ gives
\[
 \widehat\alpha=q\frac{S_0}{1+S_0}.
\]
It needs only a risk ceiling, using the reference allocation unit as its mass
scale. One pilot unit gives half the ceiling; three units give three quarters.
The exponential pilot has $S_0>0$, so the chosen level is positive. A positive
$\eta$ provides a lower tolerance when this is useful in the application.

The pilot mass is nonincreasing in the p-values. Hence
$\widehat\alpha$ is nonincreasing and lies in $[q\eta,q)$.
It moves continuously from $q\eta$ toward $q$ as the aggregate evidence
increases. The parameter $c$ is the pilot allocation mass at which the policy
has moved halfway through that range. For example, $q=0.05$, $\eta=0.2$,
and $c=5$ give
\[
 \begin{array}{c|rrrr}
 S_0&0&1&5&20\\\hline
 \widehat\alpha&0.0100&0.0167&0.0300&0.0420.
 \end{array}
\]
By \Cref{cor:BHalpha}, the final procedure controls expected relative sFDR and ordinary sFDR at ceiling $q m_0/m$. Since the pilot, the policy,
and the final finite-round rule are continuous, the entire map from p-values
to allocations is continuous.

This policy replaces the choice of one conventional level by a specified
relationship between evidence and willingness to allocate. The ceiling
expresses the application's tolerance for null allocations, and $c$ expresses
an interpretable mass of promising leads. The data determine the level
within that policy. More generally, any positive continuous nonincreasing
level policy can be used in the same way.

\begin{center}
\fbox{\begin{minipage}{0.93\textwidth}
\textbf{A practical exponential kernelized BH procedure.}
\begin{enumerate}[leftmargin=*,itemsep=2pt]
\item Specify a risk ceiling $q$, a lower fraction $\eta$, a pilot mass
scale $c$, and fixed round counts $K_0,K$.
\item Compute the pilot $b^{\rm pilot}$ by $K_0$ rounds of
\Cref{eq:iteration} at level $q$, and set $S_0=\sum_i b_i^{\rm pilot}$.
\item Compute $\widehat\alpha$ using \Cref{eq:policy}, then compute a
baseline $b$ by $K$ rounds at level $\widehat\alpha$.
\item Report $b$. The anchored upgrade developed in \Cref{sec:anchor}
can subsequently enlarge this baseline.
\end{enumerate}
Under PRDS, its random output $B$ satisfies
$\E[\sFDP_N(B)/\widehat\alpha]\leq m_0/m$ and
$\sFDR_P(B)\leq q m_0/m$. The full allocation map is continuous.
\end{minipage}}
\end{center}

\begin{figure}[htbp]
 \centering\includegraphics[width=\textwidth]{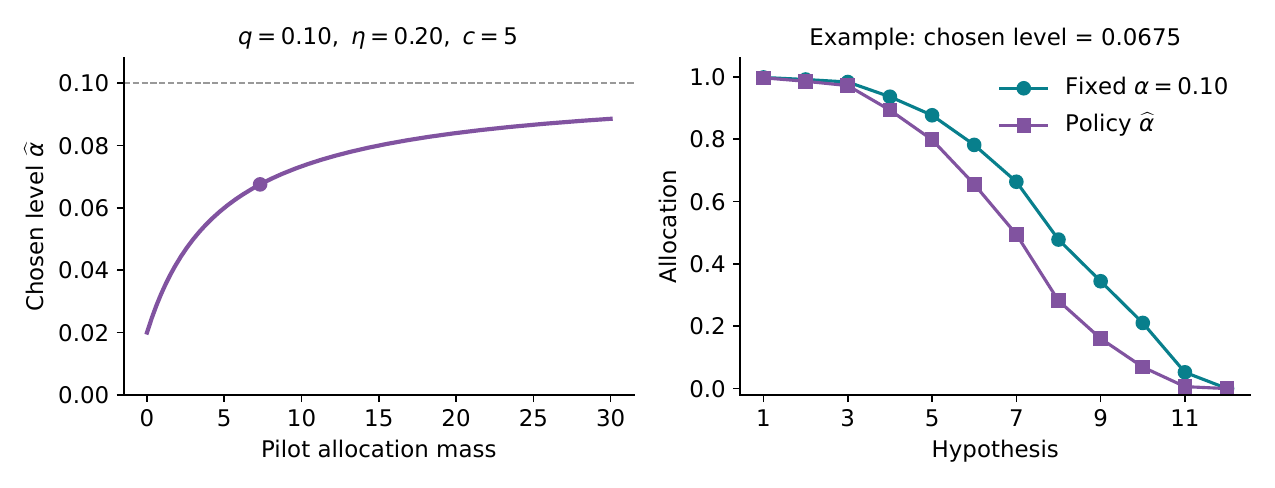}
 \caption{A continuous evidence-to-level policy and its resulting allocations.
 The point on the left corresponds to the pilot mass in the deterministic
 example.}
 \label{fig:adaptive}
\end{figure}

\Cref{fig:adaptive} shows the policy in \Cref{eq:policy}, with
$q=0.10$, $\eta=0.20$, $c=5$, and $K_0=K=25$. The chosen level is
approximately $0.06753$ for \Cref{eq:examplep}. The final allocation remains
continuous in the p-values and obeys both the relative guarantee and the
ordinary sFDR ceiling. No null labels are assigned to this deterministic
example; the plots illustrate allocation behavior.

\FloatBarrier
\section{Stable-eBH under arbitrary dependence}\label{sec:ebh}

We now describe a stable variant of the e-BH procedure~\citep{WR22}.

An e-value for $H_i$ is a nonnegative random variable $E_i$ satisfying
$\E_P[E_i]\leq1$ for every $P\in H_i$. No dependence assumption is imposed
on the vector $(E_1,\ldots,E_m)$. We assume finite observed e-values for
the algorithms; infinite values can be handled by limits.

\begin{definition}[Compound validity]\label[definition]{def:compound}
A nonnegative random vector $C=(C_1,\ldots,C_m)$ is compoundly valid if
\begin{equation}\label{eq:compound}
 \frac1m\sum_{i\in N(P)}\E_P[C_i]\leq1
 \quad\text{for every }P\in\cP.
\end{equation}
Individual coordinates need not have expectation at most one under their
own null hypotheses.
\end{definition}

\begin{definition}[Stable e-self-consistency]\label[definition]{def:esc}
A random allocation $A$ is stable e-self-consistent at level $\alpha>0$ if
\begin{equation}\label{eq:esc}
 A_i\leq\frac{\alpha E_i}{m}\mathcal D(A),\qquad i\in\ms.
\end{equation}
\end{definition}

\begin{theorem}[Kernel Stable-eBH principle]\label{thm:esc}
Every measurable allocation satisfying \Cref{eq:esc} has
\begin{equation}\label{eq:esc-risk}
 \sFDR_P(A)\leq\alpha\frac{m_0}{m}.
\end{equation}
More strongly, for a fixed e-value vector, every family of allocations
satisfying \Cref{eq:esc} obeys the pointwise bound
\begin{equation}\label{eq:esup}
 \sup_{\alpha>0}\ \sup_{\substack{a\in[0,1]^m:\\a_i\leq\alpha E_i\mathcal D(a)/m\ (i\in[m])}}
 \frac{L_N(a)}{\alpha}\leq\frac1m\sum_{i\in N}E_i.
\end{equation}
The expectation of the left-hand side is therefore at most $m_0/m$,
whenever measurable, or in outer expectation otherwise.
\end{theorem}
\begin{proof}
Divide \Cref{eq:esc} by $\alpha \mathcal D(A)$ and sum over $N$. The right-hand
side is independent of the allocation and the level. Taking expectations
proves the claims.
\end{proof}

\subsection{Stable-eBH from a single sort}\label{sec:stable-ebh-path}

For an input vector $e$, write $e_{[1]}\geq\cdots\geq e_{[m]}$ for sorted e-values and
$q_{[j]}=\min_{k\geq j}m/(ke_{[k]})$, using $1/0=\infty$.
The e-BH rejection set at $\lambda$ is exactly $\{i:q_i\leq\lambda\}$.
We use these entry levels in \Cref{eq:stable-path} and apply the
same lift and level band as in Stable-BH. This is \emph{Stable-eBH}.
The calculation takes one sort and two reverse scans. For a fixed
bandwidth the map is continuous in the finite e-values and the
target level. Clamping at $H$ handles zero e-values, and ties
give equal allocations. As the band shrinks, the output equals
binary e-BH eventually whenever its rejection set is locally
constant in the level.

\begin{figure}[htbp]
 \centering\includegraphics[width=.79\textwidth]{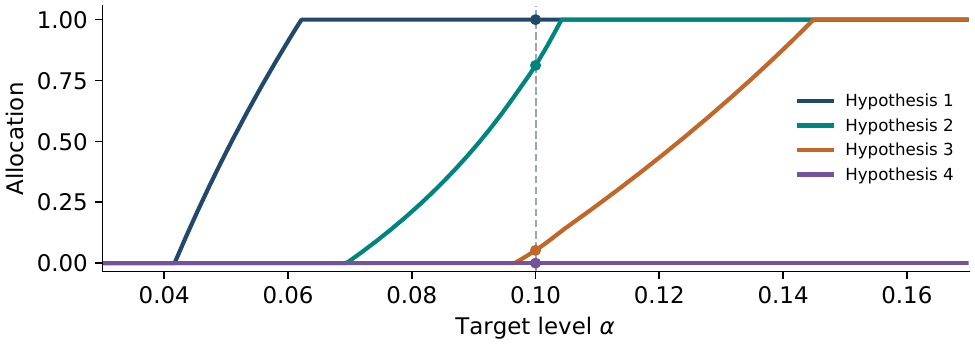}
 \caption{Stable-eBH level paths for e-values $(80,24,11.5,1)$. The markers show the allocations at $\alpha=.10$ with $\delta=.20$.}
 \label{fig:ebh-path}
\end{figure}

The fixed-level e-BH guarantee and \Cref{thm:perturb} give sFDR
at most $\alpha$ for compound e-values, with arbitrary dependence.
For ordinary e-values the same argument gives $\alpha |N|/m$.
For example, at $\alpha=.10$, the e-values $(80,24,11.5,1)$
have entry levels approximately $(.050,.083,.116,1)$.
Binary e-BH selects the first two, while Stable-eBH gives
approximately $(1,.812,.053,0)$ with the band $[.08,.12]$.
\Cref{fig:ebh-path} shows the e-value allocation paths.

\subsection{Exponential saturation}

Let $\psi(x)=1-e^{-x}$. Its two useful properties are
$0\leq\psi(x)\leq1$ and $\psi(x)\leq x$ for $x\geq0$.
Starting with $r_0=1$, kernel Stable-eBH computes
\begin{equation}\label{eq:ebhiteration}
 a_i^{(k+1)}=1-\exp\left(-\frac{\alpha e_i r_k}{m}\right),
 \qquad r_{k+1}=1\vee\sum_i a_i^{(k+1)}.
\end{equation}
For a fixed $K$, the output $a^{(K)}$ is continuous in $(e,\alpha)$.
The sequence $r_k$ is nondecreasing, and
$a_i^{(K)}\leq\alpha e_i r_{K-1}/m\leq\alpha e_i\mathcal D(a^{(K)})/m$.
At a random input $E$, its output $A^{(K)}=a^{(K)}(E)$ therefore
satisfies \Cref{thm:esc}. A finite positive e-value receives an
allocation strictly between zero and one. Larger evidence has diminishing
returns, with the allocation approaching one.

One may also use the limiting fixed point. For $b_i=\alpha e_i/m$,
$F(r)=\sum_i(1-e^{-b_i r})$ is increasing and concave.
If any $b_i>0$, $F(r)/r$ is strictly decreasing on $(0,\infty)$.
Consequently, $r=1\vee F(r)$ has exactly one solution in $[1,m]$:
it is $1$ if $F(1)\leq1$, and otherwise it is the unique solution of
$F(r)=r$ above one. The iterates converge to it. This solution, and hence
the limiting allocation, is continuous in $(e,\alpha)$: every convergent
subsequence of solutions along converging parameters solves the limiting
equation, whose solution is unique. The all-zero e-value vector has solution
$r=1$ and is covered by the same argument.

The exponential taper for p-values and exponential saturation for e-values
express the same allocation convention in opposite evidence directions.
A p-value receives less allocation as it grows; an e-value receives more.
The validity arguments use, respectively, a unit-area kernel and the
inequality $\psi(x)\leq x$.

\section{Stable knockoffs}\label{sec:fuzzy-knockoffs}

Let $W\in\R^m$ be a vector of knockoff statistics. Under each model law
$P$, assume that conditional on $|W|$ and $W_{N(P)^c}$, the nonzero null
signs are independent fair coins. This is the sign symmetry used by
knockoff+ and Model-X knockoffs \citep{BC15,CFJL18}. We use a small
symmetric perturbation to smooth the selection boundary, then set the
allocation scale using the compound e-value certificate. Strong
features can receive a full unit of allocation within a single
knockoff copy.

\subsection{Smooth the boundary, then set the scale}

For any score vector $w$, let $R_\beta(w)$ be the ordinary knockoff+
rejection set at pilot level $\beta$. Let $L_\beta(w)$ count negative
scores beyond the selected threshold, with $L_\beta(w)=0$ when there
are no discoveries. For bandwidth $h>0$, add an auxiliary perturbation
$hU$, where the coordinates of $U$ are independent uniform variables on
$[-1,1]$, independent of $W$. The score for feature $i$ is
\begin{equation}\label{eq:kn-evidence}
 c_i(w)=\E_U\!\left[
 \frac{\one\{i\in R_\beta(w+hU)\}}{1+L_\beta(w+hU)}
 \right].
\end{equation}
At each perturbation we apply ordinary knockoff+ and record its
selected indicators divided by one plus the number of negative scores
at the threshold. This is the normalized version of the stopped
compound e-value of \citet{RB24}. A feature well above the cutoff
appears in nearly all the selected sets. Features near the cutoff
appear in only a fraction of them. The factor $(1+L_\beta)^{-1}$
gives more weight to runs with fewer negatives at their cutoff.

We apply the lift of \Cref{def:lift} to $d_i=\alpha c_i$, with
$s=\sum_i d_i$:
\begin{equation}\label{eq:kn-lift}
 a=\mathcal L_\rho(\alpha c).
\end{equation}
The stopped-count weight $1/(1+L_\beta)$ is at least the generic
weight $1/(\beta |R_\beta|)$ when the run makes discoveries.
In the full-scale regime, each feature receives its weighted
inclusion frequency relative to the most stable feature.

We use $\beta=\alpha/(1+\rho)$ as the default pilot. If every
perturbation gives a nonempty rejection set, then its selected count
divided by one plus the negative count is at least $1/\beta$.
Averaging gives $s\geq\alpha/\beta=1+\rho$. In that case a feature
selected in every perturbation receives one. If such a feature exists,
then a feature selected in some but fewer than all perturbations
receives a strictly fractional allocation. More generally, writing
$\omega=(1+L_\beta(w+hU))^{-1}\one\{R_\beta(w+hU)\ne\varnothing\}$,
a feature with $\E[\omega\one\{i\notin R_\beta(w+hU)\}]\leq
\varepsilon\E[\omega]$ receives at least $1-\varepsilon$ in the
full-scale regime. This follows from $c_i\geq(1-\varepsilon)\E[\omega]$
and $\max_j c_j\leq\E[\omega]$.

\begin{example}[Smoothing at a knockoff boundary]\label[example]{ex:kn-boundary}
Consider the scores
\[
 (4,3.5,3,2.6,2.4,2.2,1.86,1.24,.55,-1.55,-.9,-.65,-.5,-.35).
\]
At level $\alpha=.20$, ordinary knockoff+ selects the first seven.
With $h=.28$, $\rho=.05$, and $\beta=\alpha/(1+\rho)$, the first
six positive scores outrank every negative score under every perturbation.
The scores $1.86$ and $1.24$ can exchange order with the largest
negative magnitude. Their stable allocations are approximately $.90$
and $.10$, while the first six each receive one. The selected
threshold always has zero negatives, so the two fractional allocations
are the probabilities of outranking the perturbed negative.
\end{example}

\begin{figure}[htbp]
 \centering\includegraphics[width=\textwidth]{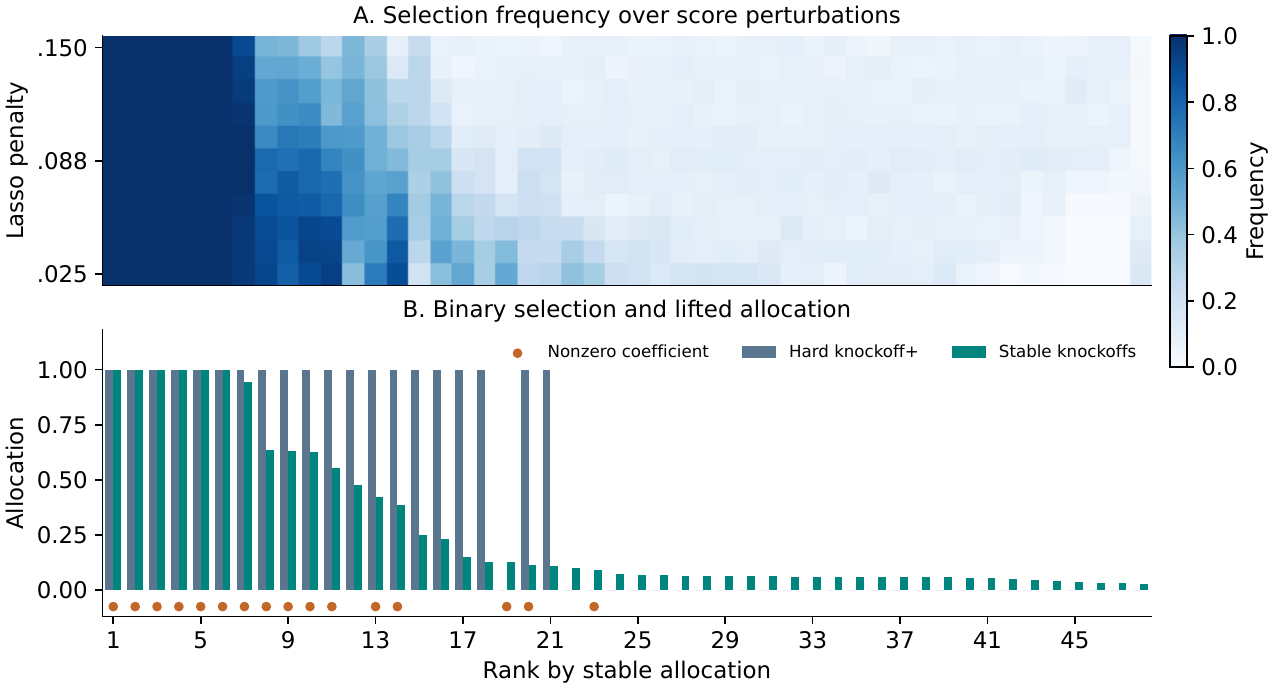}
 \caption{Stable knockoffs with a correlated design and heterogeneous
 signal strengths. Panel A shows selection frequencies across score
 perturbations at each lasso penalty. Panel B compares binary selection
 at the middle penalty with the lifted allocation. Orange dots mark
 true nonzero coefficients. Both panels use the same feature order.}
 \label{fig:kn-scan}
\end{figure}

\paragraph{A correlated regression with a lasso score.}
\Cref{fig:kn-scan} uses one model-X knockoff draw on a simulated
Gaussian regression with $360$ observations and $48$ predictors.
The predictors come in four independent Toeplitz blocks of size
12, with adjacent correlations $.25$, $.45$, $.60$, and $.72$.
Sixteen nonzero coefficients are spread across the four blocks.
Their magnitudes range from $.85$ to $.10$, with mixed signs.
The noise is independent standard Gaussian. For covariance $\Sigma$,
we use Gaussian knockoffs with cross-covariance $\Sigma-sI$,
where $s=\min\{1,1.8\lambda_{\min}(\Sigma)\}$ \citep{CFJL18}.

For each of eleven equally spaced lasso penalties between $.025$
and $.150$, we fit the lasso to the joint original and knockoff
design, and set $W_j=|\widehat b_j|-|\widehat b_{j+m}|$.
Swapping a pair swaps the fitted coefficients for the pair and
reverses the sign of $W_j$. We average the stopped certificates
over penalties and over $512$ symmetric perturbations per penalty,
using $\beta=.19$, $h=.11$, $\alpha=.20$, and $\rho=.05$.
Each penalty uses the same knockoff copy. Its stopped certificate
is valid by the sign argument below, and their average remains valid.
Hard knockoff+ uses the middle penalty and $\alpha=.20$.
The stable report assigns one to six features and then tapers
through a range of intermediate allocations. At ranks seven,
eleven, and fifteen, the allocations are $.94$, $.55$, and $.25$.

\subsection{Validity and the hard limit}

\begin{theorem}[Stable knockoff validity]\label{thm:fuzzy-knockoffs}
Under the knockoff sign symmetry, for each $0<\beta\leq\alpha<1$,
$h>0$, and $\rho>0$, let $a(w)$ be the construction in
\Cref{eq:kn-evidence,eq:kn-lift}. Then $A=a(W)$ has sFDR at most
$\alpha$, and the map $a:\mathbb R^m\to[0,1]^m$ is continuous.
The same bound holds if \Cref{eq:kn-evidence} is evaluated by a
fixed number of independent perturbations, with the risk expectation
then taken over the data, knockoffs, and auxiliary draws.
If $w$ has nonzero coordinates with distinct absolute values and
$\beta=\alpha/(1+\rho)$, the allocation converges to the binary
knockoff+ indicator at level $\beta$ as $h\downarrow0$.
\end{theorem}

\begin{proof}
The law of $W+hU$ is invariant under flipping any subset of the true
null coordinates. This follows by applying the same sign flips to
both $W$ and the independent symmetric $U$. Thus the perturbed null
signs are fair and independent conditional on the perturbed
magnitudes and non-null statistics. For every nonempty $S\subseteq
N(P)$, they are also fair and independent within $S$ conditional on
the perturbed magnitudes and scores outside $S$.

Write $C_i=c_i(W)$. For completeness, scan the perturbed magnitudes from small to large,
revealing the signs as their magnitudes are passed. The total sign
counts among surviving scores are available during the scan. Let
$V_S$ and $U_S$ denote the positive and negative counts within $S$.
The usual stopped-count argument gives a supermartingale
$V_S/(1+U_S)$. At state $(v,u)$ with $u>0$, removing the next
uniformly permuted sign gives conditional mean
\[
 \frac{v}{v+u}\frac{v-1}{1+u}+\frac{u}{v+u}\frac{v}{u}
 =\frac{v}{1+u}.
\]
When $u=0$ the ratio decreases. The knockoff+ threshold stops at the
first eligible state in the scan, with terminal ratio zero if no
threshold is eligible. Its denominator $1+L_\beta$ is at least
$1+U_S$ there. Initially $U_S$ is binomial with parameters
$|S|$ and $1/2$ since the perturbation has a density. The expected
initial ratio is $1-2^{-|S|}$ by the binomial identity
$\E[1/(1+U_S)]= (2-2^{-|S|})/(|S|+1)$. Therefore
\begin{equation}\label{eq:kn-budget}
 \E_P\left[\sum_{i\in S}C_i\right]\leq 1-2^{-|S|}.
\end{equation}
Averaging a fixed number of perturbations preserves this bound by
linearity. In particular, $mC_i$ are compound e-values.

By \Cref{eq:lift-identity},
\begin{equation}\label{eq:kn-lift-normalized}
 \frac{A_i}{\mathcal D(A)}=\frac{\alpha C_i}{1\vee\sum_j\alpha C_j}\leq\alpha C_i .
\end{equation}
Summing over the nulls and applying \Cref{eq:kn-budget} proves sFDR control.

The function $c_i(w)$ is the convolution of a bounded score with the
density of $hU$. Translations of that density are continuous in
$L^1$, so $c_i$ is continuous. In the lift, $g(s)=0$ for $s\leq1$
and the division by $v$ is used only when $s>1$, where
$v\geq s/m>0$. This gives continuity of the map $a$ for exact integration.
For the hard limit, sufficiently small bounded perturbations
preserve the signs and ordering of the absolute scores.
If $R_\beta(w)$ is nonempty, all its members have equal positive
scores $c_i(w)$ and the certified mass is at least $1+\rho$.
The lift therefore gives one on $R_\beta(w)$ and zero elsewhere.
If the set is empty, all scores are zero for sufficiently small $h$.
\end{proof}

One may also use $\beta=\alpha$ to match the original pilot
boundary. With that choice the mirror ratio can be exactly
$\alpha$, giving certified mass one in the hard limit. The ramp
then need not reach the full-scale regime. The slightly smaller
default pilot supplies the margin for the hard-limit statement
in \Cref{thm:fuzzy-knockoffs}. For a fixed evidence vector,
\Cref{eq:kn-lift-normalized} also gives expected relative stable FDP at most one for any positive data-dependent allocation level.
For every nonempty $S\subseteq[m]$, the random variable
$Z_S^{\rm kn}=\sum_{i\in S}C_i+2^{-|S|}$ has expectation at most
one whenever all hypotheses in $S$ are true. These are the
intersection certificates used in \Cref{sec:closure}.

The exact smoothing is deterministic given $W$. A practical version
takes $J$ perturbations, runs knockoff+ on each, and averages its
stopped evidence. It costs $O(Jm\log m)$, with no new knockoff
generation or feature fitting in these $J$ perturbations. The
Monte Carlo implementation may have small jumps as scores vary
with the auxiliary draws held fixed. Its finite-sample bound
includes the auxiliary randomness.

\subsection{The knockoff level path from a single sort}\label{sec:kn-level-path}

When the sensitivity analysis varies the level alone, we can
integrate the entire knockoff+ path after sorting the distinct
positive magnitudes from largest to smallest. At the $k$th cutoff
write $P_k,L_k$ for the cumulative positive and negative counts,
and $r_k=(1+L_k)/P_k$, with $r_k=\infty$ if $P_k=0$.
Knockoff+ takes the last eligible cutoff. A positive feature in
group $j$ thus enters at $q_j=\min_{k\geq j}r_k$.
Use the uniform level band $[L,H]$ of \Cref{sec:stablebh}.
Write $\ell_k=(L\vee q_k)\wedge H$ and $\ell_{K+1}=H$,
where $K$ is the number of magnitude groups. For a positive
feature $i$ in group $j$, the stopped-count certificate is
\[
 c_i=\frac{1}{H-L}\sum_{k=j}^K\frac{\ell_{k+1}-\ell_k}{1+L_k}.
\]
It is zero for nonpositive features. Positive-length intervals
have a constant cutoff, which gives the displayed formula by
integrating $1/(1+L_k)$ on those intervals. Ties in absolute value
enter as a group. The stopped-count bound used in
\Cref{thm:fuzzy-knockoffs} applies at each fixed level to the
original nonzero null signs. Averaging the bound and applying
$\mathcal L_\rho(\alpha c)$ gives sFDR control. This procedure
is continuous in the target level for fixed scores. Symmetric
score smoothing above also gives continuity in the scores.

For the scores in \Cref{ex:kn-boundary}, take target $.23$ and band
$[.184,.276]$. The seven largest positive scores are selected
throughout. The score $1.24$ enters at $.25$, after one negative
passes the cutoff. With $\rho=.05$ the seven stable features
receive one, and $1.24$ receives about $.165$.

\subsection{Pooling across knockoff draws}\label{sec:kn-pooling}

Across knockoff copies, normalized allocations can be combined by
\Cref{prop:normalized-pool}, extending the binary wrapper in
\Cref{alg:wrapper}. This applies to both ordinary knockoff+
indicators and the within-copy smooth allocations developed above.
Their common certificate supports an additional operation: averaging
the knockoff evidence before selecting the allocation scale.

Specifically, average the compound evidence across copies
and apply \Cref{eq:kn-lift} once to the average. The averaged
certificate remains valid and can support total allocation above
one. This extends the evidence aggregation of \citet{RB24} to
fractional decisions. Binary e-BH reaggregation introduces a final
selection threshold and an internal run-level choice; here the pooled
allocation profile carries its own error guarantee.
It also allows certified optimization with
the direct pool as a coordinatewise lower bound, since
$\bar Z_i\leq\alpha\bar C_i$ by \Cref{eq:kn-lift-normalized}.

The following regression example uses these certificates and lift.

\subsection{A regression allocation profile}

\begin{figure}[tb]
\centering
\includegraphics[width=\textwidth]{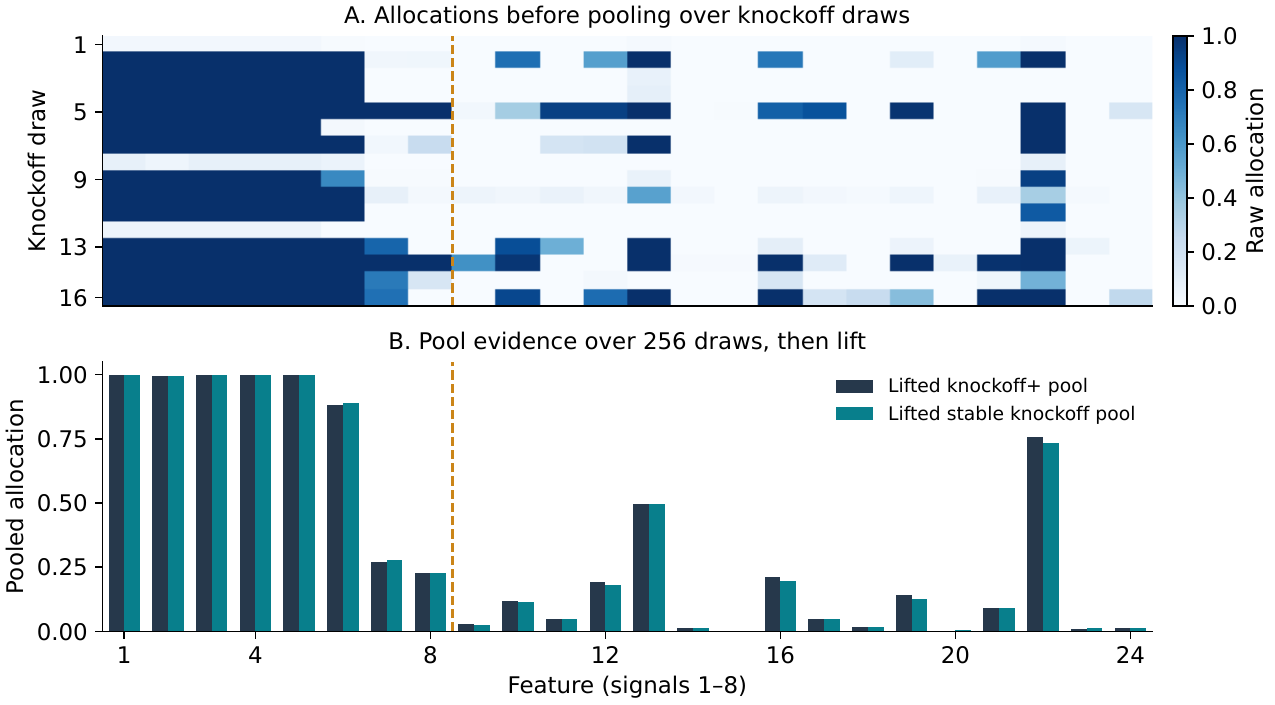}
\caption{Within-copy allocations and across-copy pooling on one
Gaussian regression data set. The top panel shows raw allocations
for the first $16$ copies. The bottom panel pools evidence across $256$ copies before
applying the lift, for both ordinary and score-smoothed pilots. The dashed line separates signals
from null features. The smoothed rule uses $128$ perturbations per
copy, at target level $\alpha=.20$, pilot level $\beta=.20/1.05$, score-perturbation bandwidth $h=.28$, and ramp width $\rho=.05$.}
\label{fig:kn-pool}
\end{figure}

Knockoffs produce different synthetic features on each rerun even
when all tuning choices and the observed data are fixed.
\Cref{fig:kn-pool} shows a controlled Gaussian regression example
with 220 observations and 24 independent standard normal features.
Eight signal coefficients decrease from $.75$ to $.32$; the other
coefficients are zero, and errors are independent standard normals.
We generate 256 Model-X knockoff copies. The figure uses both ordinary
knockoff evidence and score-smoothed evidence within a copy, followed
by evidence pooling and lifting. The statistic is the difference
of absolute marginal inner products with the normalized response.
\Cref{sec:fuzzy-knockoffs} gives these constructions and their assumptions.

The pooled report distinguishes a persistent core from features whose
allocations depend on a particular draw. A null feature also receives
substantial allocation on this data set. Stability under the randomization
is a property of the observed analysis; error control averages over the
sampling distribution as well. The known-null simulations in
\Cref{sec:fuzzy-knockoffs,sec:experiments} assess the latter guarantee,
including the cost in signal allocation from smoothing and subsampling.

\FloatBarrier

\subsection{Single-copy simulation}

We compare the score-smoothed rule and the single-sort level path with
ordinary knockoff+ at both $\alpha$ and the pilot $\beta$. Each independent data set contains
$150$ observations of $36$ independent standard normal features.
Under the global null all coefficients are zero. In the moderate
and strong regimes the first ten coefficients decrease linearly
from $0.40$ to $0.15$ and from $0.70$ to $0.24$, respectively.
We use the marginal-inner-product statistic of \Cref{fig:kn-pool} in this simulation.
We use $2400$ data sets per regime, one Model-X copy per data
set, and $128$ perturbations for smoothing. The target, pilot,
bandwidth, and ramp for score smoothing are those of \Cref{ex:kn-boundary}.
The level path uses $\delta=.20$ and $\rho=.05$.
The reported allocations are before pooling across knockoff copies.

\begin{table}[tb]
\centering\small
\begin{tabular}{@{}llrrrr@{}}
\toprule
Signals & Method & sFDR & True mass & Top four & Partial \\
\midrule
Global null & Knockoff+ at $\alpha$ & 0.032 & 0.00 & -- & 0.00 \\
Global null & Knockoff+ at $\beta$ & 0.015 & 0.00 & -- & 0.00 \\
Global null & Stable knockoffs & 0.020 & 0.00 & -- & 0.38 \\
Global null & Level-path knockoffs & 0.024 & 0.00 & -- & 0.17 \\
Moderate & Knockoff+ at $\alpha$ & 0.158 & 4.56 & 0.599 & 0.00 \\
Moderate & Knockoff+ at $\beta$ & 0.135 & 3.72 & 0.478 & 0.00 \\
Moderate & Stable knockoffs & 0.141 & 3.02 & 0.388 & 5.76 \\
Moderate & Level-path knockoffs & 0.146 & 3.54 & 0.458 & 2.60 \\
Strong & Knockoff+ at $\alpha$ & 0.165 & 6.63 & 0.854 & 0.00 \\
Strong & Knockoff+ at $\beta$ & 0.146 & 5.94 & 0.756 & 0.00 \\
Strong & Stable knockoffs & 0.148 & 5.36 & 0.685 & 6.23 \\
Strong & Level-path knockoffs & 0.153 & 5.77 & 0.735 & 2.70 \\
\bottomrule
\end{tabular}

\caption{Single-copy simulation. True mass is the mean total
allocation to non-nulls. Top four is the mean allocation per
feature among the four largest coefficients. Partial is the mean
number of allocations in $(0.01,0.99)$. The reproducible CSV
includes Monte Carlo standard errors for the risk estimates.}
\label{tab:kn-simulation}
\end{table}

In the strong-signal setting, the largest smoothed allocation is
at least $0.99$ in $70.2\%$ of data sets. In these data sets the
four strongest signals receive $0.947$ on average per feature.
The unconditional average is $0.685$, and the mean total
allocation to non-nulls is $5.36$, compared with $5.94$ for the
hard pilot and $6.63$ for ordinary knockoff+ at $\alpha$.
In the moderate setting, the largest smoothed allocation reaches
$0.99$ in $41.8\%$ of data sets. Thus the close-to-one behavior is
most apparent when a single copy provides enough evidence to
support a nearly full unit of allocation. The comparison with the
hard pilot shows the loss in average true allocation from the
smoothing and scaling in both signal regimes.
The level-path version has mean true allocation $5.77$ in the
strong-signal setting and about $2.70$ partial allocations per
data set, compared with $6.23$ partial allocations under score
smoothing. A level band probes how many ranks the cutoff can move
through with scores fixed. Score perturbations can also change
which features occupy those ranks.

\section{Stable e-closure and universal representations}\label{sec:closure}

For nonempty $S\subseteq\ms$, let $H_S=\bigcap_{i\in S}H_i$.
An \emph{intersection e-collection} is a family $(Z_S)$ of nonnegative
random variables such that
\begin{equation}\label{eq:intersection}
 \E_P[Z_S]\leq1\quad\text{for every }P\in H_S.
\end{equation}
Its members may depend on all the data. The construction below extends the
e-closure principle to the allocation space. It also follows from the
continuous-closure theorem of \citet{SMARA26} by using the present loss and
the box $[0,1]^m$.

\begin{definition}[Stable e-closure]\label[definition]{def:closure}
For an intersection e-collection $Z$ and $\alpha>0$, define
\begin{equation}\label{eq:closure}
 \cA_\alpha(Z)=\{a\in[0,1]^m:L_S(a)\leq\alpha Z_S
                  \text{ for every nonempty }S\subseteq\ms\}.
\end{equation}
\end{definition}

\begin{theorem}[Simultaneous stable e-closure]\label{thm:closure}
For every $P\in\cP$ with nonempty null set $N$,
\begin{equation}\label{eq:closure-uniform}
 \E_P\left[\sup_{\alpha>0}\ \sup_{a\in\cA_\alpha(Z)}
       \frac{L_N(a)}{\alpha}\right]\leq1,
\end{equation}
using outer expectation if needed. In particular, an arbitrary positive
data-dependent level $\widehat\alpha$ and any measurable selection
$\widehat A\in\cA_{\widehat\alpha}(Z)$ satisfy
$\E_P[L_N(\widehat A)/\widehat\alpha]\leq1$.
If $\widehat\alpha\leq q$, then $\sFDR_P(\widehat A)\leq q$.
\end{theorem}
\begin{proof}
For every feasible pair $(a,\alpha)$, the constraint indexed by the actual
null set gives $L_N(a)/\alpha\leq Z_N$. This holds simultaneously over
all such pairs. Under $P\in H_N$, \Cref{eq:intersection} gives
$\E_P[Z_N]\leq1$. For $N=\varnothing$ the loss is zero.
\end{proof}

The feasible region contains zero and is compact. Thus an analyst can
maximize a continuous utility of $a$ over it. Additional budget, fairness,
group, or baseline-dominance constraints can be imposed as long as the
selected allocation remains in the region. Under standard joint
measurability conditions, compactness supplies measurable maximizers.
The simultaneous statement allows the utility and the level to be chosen
after viewing the data.

For example, marginal e-values yield
$Z_S=|S|^{-1}\sum_{i\in S}E_i$. Every stable e-self-consistent allocation
belongs to this closure because
\[
 L_S(a)\leq\frac{\alpha}{m}\sum_{i\in S}E_i
          \leq\frac{\alpha}{|S|}\sum_{i\in S}E_i.
\]
Optimizing over this region subject to $a\geq b$ therefore improves any
baseline $b$ satisfying \Cref{eq:esc}.

\begin{figure}[htbp]
 \centering\includegraphics[width=.72\textwidth]{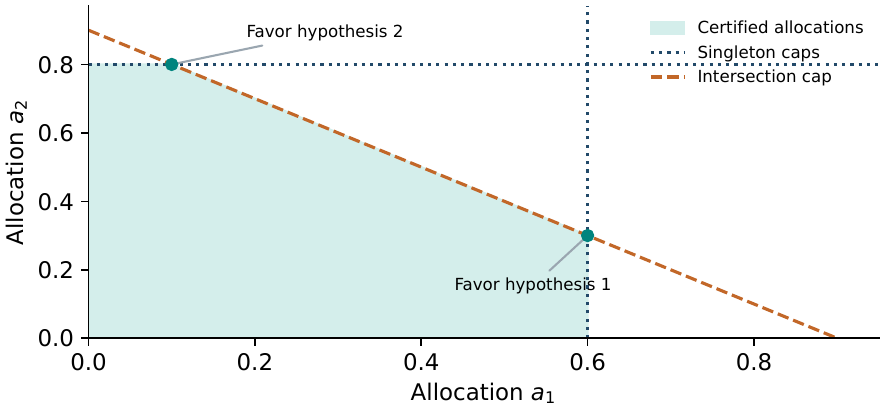}
 \caption{Stable e-closure for a realized two-hypothesis certificate. The marked vertices maximize utilities that give twice the weight to hypothesis 1 or 2.}
 \label{fig:closure-geometry}
\end{figure}

For $m=2$, consider a realized certificate with
$\alpha Z_{\{1\}}=.6$, $\alpha Z_{\{2\}}=.8$, and
$\alpha Z_{\{1,2\}}=.9$. The joint constraint implies
$a_1+a_2\leq.9$, so the denominator is one.
\Cref{fig:closure-geometry} shows the feasible region. Maximizing
$2a_1+a_2$ selects $(.6,.3)$, whereas maximizing $a_1+2a_2$
selects $(.1,.8)$. Both choices use the same certificate.

\subsection{Completeness for arbitrary inputs}

The input to an allocation procedure can be a vector of p-values, test
statistics, fitted models, or the full data. The following representation
depends only on the resulting allocation and its guarantee.

\Needspace{12\baselineskip}
\begin{theorem}[Universal intersection representation]\label{thm:complete}
Fix $\alpha>0$ and a measurable allocation $A=a(X)$. The following are
equivalent:
\begin{enumerate}[label=(\roman*)]
 \item $\E_P[L_{N(P)}(A)]\leq\alpha$ for every $P\in\cP$.
 \item There is an intersection e-collection $Z$ with
       $A\in\cA_\alpha(Z)$ almost surely under every $P$.
\end{enumerate}
Under (i), the canonical choice is
\begin{equation}\label{eq:selfref}
 Z_S(X)=L_S(A)/\alpha.
\end{equation}
More generally, let $\cB(X)\subseteq[0,1]^m$ be a measurable allocation
family. It satisfies, for every $P$,
\[
 \E_P\left[\sup_{b\in\cB(X)}L_{N(P)}(b)\right]\leq\alpha
\]
if and only if it is contained in a stable e-closure at level $\alpha$.
\end{theorem}
\begin{proof}
The implication (ii)$\Rightarrow$(i) follows from \Cref{thm:closure}.
For the converse, if $P\in H_S$, then $S\subseteq N(P)$. Since
$L_S(A)\leq L_{N(P)}(A)$, \Cref{eq:selfref} is an e-value for $H_S$.
It satisfies all closure constraints with equality. For the family version,
use $Z_S=\sup_{b\in\cB(X)}L_S(b)/\alpha$. Monotonicity in $S$ gives its
intersection validity, and the reverse direction again uses the constraint
at $N(P)$. Supremum measurability is assumed, or outer expectations are used.
\end{proof}

This proof uses only that $S\subseteq T$ implies $L_S(a)\leq L_T(a)$.
The same construction therefore applies to any nonnegative measurable loss
family with this property, on any measurable action space. Using a single
e-collection for several such losses also gives simultaneous control across
losses by the same constraint-at-the-truth argument. This is the general
continuous-loss perspective of SMARA and e-closure.

\subsection{Compound e-values}

\begin{theorem}[Universal compound representation]\label{thm:compound}
Let $C$ satisfy compound validity as in \Cref{def:compound}. Every
measurable allocation satisfying
\[
 A_i\leq\frac{\alpha C_i}{m}\mathcal D(A),\qquad i\in[m],
\]
controls sFDR at $\alpha$, by the argument of \Cref{thm:esc}.
Conversely, for any allocation $A$ and
fixed $\alpha>0$, define
\begin{equation}\label{eq:canonical-compound}
 C_i=\frac{m A_i}{\alpha \mathcal D(A)}.
\end{equation}
Then $A_i=\alpha C_i\mathcal D(A)/m$, and $C$ is compoundly valid if and only if
$A$ controls sFDR at level $\alpha$.
\end{theorem}
\begin{proof}
Dividing the displayed coordinate inequalities by $\alpha \mathcal D(A)$
and summing over $N$ gives
\[
 L_N(A)/\alpha\leq m^{-1}\sum_{i\in N}C_i.
\]
For \Cref{eq:canonical-compound}, this inequality is equality pointwise.
Taking expectations proves both directions.
\end{proof}

Thus every valid method admits an exact self-consistency certificate through
compound e-values. The theorem represents an allocation as a member of a
certified region; different utilities can select different members. The
binary counterpart and its connections to empirical Bayes are developed
by \citet{IWR24}.

\begin{remark}[Connection to the false evidence rate]\label[remark]{rem:FER}
For a fixed selection of all $m$ hypotheses, Koning's false evidence
proportion is $m^{-1}\sum_{i\in N}C_i$. Substituting
\Cref{eq:canonical-compound} gives $L_N(A)/\alpha$. Hence sFDR has
an exact fixed-selection FER representation. The bounded allocation and its
normalization are the additional choices that determine the procedures here.
\end{remark}

For a measurable allocation family $\cB(X)\subseteq[0,1]^m$,
closure gives the exact simultaneous representation in
\Cref{thm:complete}. A common compound vector certifying
all family members requires the coordinatewise envelopes to satisfy
\begin{equation}\label{eq:envelope}
 \E_P\left[\sum_{i\in N(P)}\sup_{a\in\cB(X)}
          \frac{a_i}{\alpha \mathcal D(a)}\right]\leq1.
\end{equation}
This is necessary by self-consistency and sufficient by taking each
coordinate to be $m$ times its displayed supremum. It can exceed the
expectation of the supremum of the sum. For instance, with two true nulls,
the deterministic family $\{(\alpha,0),(0,\alpha)\}$ has simultaneous
stable FDP $\alpha$ for $0<\alpha<1$, whereas the left-hand side of
\Cref{eq:envelope} is two.

\FloatBarrier
\section{Optimizing allocation objectives}\label{sec:optimization}

An investigator may want to maximize total allocation or to give
more weight to hypotheses that are inexpensive to follow up. We first
describe how the error constraint limits the available allocations.
We then optimize for a fixed certificate at the observed data and
over rules subject to the full expectation constraints.

\subsection{Allocations with the same normalized shares}

Define $\pi(a)=a/\mathcal D(a)$ and let
$\Delta_{\leq1}=\{z\in[0,\infty)^m:\sum_i z_i\leq1\}$.

We first characterize all allocations with a given normalized vector.

\begin{lemma}[Normalized allocations and their representatives]\label[lemma]{lem:normalized-geometry}
The image of $[0,1]^m$ under $\pi$ is exactly $\Delta_{\leq1}$.
For $z\in\Delta_{\leq1}$ with $\sum_i z_i<1$, the only representative
is $a=z$. If $\sum_i z_i=1$, its representatives are exactly
\begin{equation}\label{eq:normalized-fiber}
 a=t z,\qquad 1\leq t\leq\frac1{\max_i z_i}.
\end{equation}
For every $S\subseteq[m]$, $L_S(a)=\sum_{i\in S}\pi_i(a)$.
\end{lemma}
\begin{proof}
Every normalized allocation has mass at most one; taking $a=z$
realizes every point of the simplex. If $s(a)\geq1$, the normalized
mass is one. Thus a normalized mass below one forces $a=z$.
On the unit-mass face, $a$ must be proportional to $z$; its total
is $t\geq1$, and the unit coordinate caps give
$t\leq1/\max_i z_i$. The loss identity follows by definition.
\end{proof}

Consequently, the full risk constraints on a normalized random rule
are the linear inequalities
\begin{equation}\label{eq:opt-risk}
 Z=z(X)\in\Delta_{\leq1},\qquad
 \E_P\left[\sum_{i\in N(P)}Z_i\right]\leq\alpha
 \quad(P\in\cP).
\end{equation}
Their feasible set is convex in the normalized rule. The representatives
in \Cref{eq:normalized-fiber} describe the possible allocation scales
along a certified direction. Maximizing a concave utility of $Z$ under
a reference law is a convex problem; in a finite experiment with linear
utility it is a linear program. Objectives measured in the original
allocations also depend on the choice of scale.

\subsection{Maximizing total allocation over a certificate}

At a given data point, let $c\subseteq\Delta_{\leq1}$ be a nonempty
compact polytope of certified normalized allocations. For stable e-closure
write $e_S$ for the observed intersection certificates; the polytope
is specified by $z(S)\leq\alpha e_S$ for every nonempty $S$, where
$z(S)=\sum_{i\in S}z_i$. For compound e-self-consistency it is specified
by $z_i\leq\alpha e_i/m$. The maximum total allocation can be computed
by at most two linear programs.

\begin{proposition}[Exact certified total allocation]\label[proposition]{prop:opt-mass}
Let $u=\max_{z\in c}\sum_i z_i$. If $u<1$, a maximizer $z^\star$ of
this program gives an optimal allocation $a^\star=z^\star$.
If $u=1$, solve
\begin{equation}\label{eq:opt-balanced}
 \min_{z,t}\ t
 \quad\text{subject to}\quad z\in c,\qquad
 \sum_i z_i=1,\qquad z_i\leq t\quad(i\in[m]).
\end{equation}
The optimal total allocation is $1/t^\star$, attained by
$a^\star=z^\star/t^\star$.
\end{proposition}
\begin{proof}
By \Cref{lem:normalized-geometry}, normalized mass below one has
a unique representative, while the largest total on the unit-mass
face is $1/\max_i z_i$. Whenever $u=1$, this total is at least one and is
maximized by \Cref{eq:opt-balanced}.
\end{proof}

For coordinate caps $z_i\leq d_i$, with $d_i\in[0,1]$, the low-mass
solution is $a=d$ when $\sum_i d_i<1$. Otherwise $t^\star$ is the
smallest $t$ with $\sum_i(d_i\wedge t)=1$, and
$a_i^\star=(d_i\wedge t^\star)/t^\star$. Sorting the caps or scalar
root finding gives this solution. In particular, $d_i=\alpha c_i$
gives a total-allocation optimizer for the stable knockoff scores under
their compound certificate.

For weighted sums, concave utilities of $a$, or constraints
such as $a\geq b$ and $a\in[0,1]^m$, it is convenient to work in
the original allocations. An intersection constraint is
$\sum_{i\in S}a_i\leq\alpha e_S$ on $s(a)\leq1$ and
$\sum_{i\in S}a_i\leq\alpha e_S s(a)$ on $s(a)\geq1$.
Both branches are polyhedra. Linear objectives give two linear programs,
and concave objectives give two convex optimization problems. The
best feasible solution across the branches is optimal under the
certificate. \Cref{sec:rawanchor} gives a linear-size formulation for
the raw anchored region. This pointwise approach is related to the
polytope optimization for null allocation rates in \citet{SMARA26}.

If a baseline belongs to the certified region, we can impose $a\geq b$.
Maximizing a weighted sum with strictly positive weights then yields
a coordinatewise maximal allocation in that region. A data-dependent
choice of objective is permitted by the simultaneous guarantee in
\Cref{thm:closure}. The certificate specifies the feasible region on
which this optimality statement holds.

\begin{figure}[htbp]
 \centering\includegraphics[width=\textwidth]{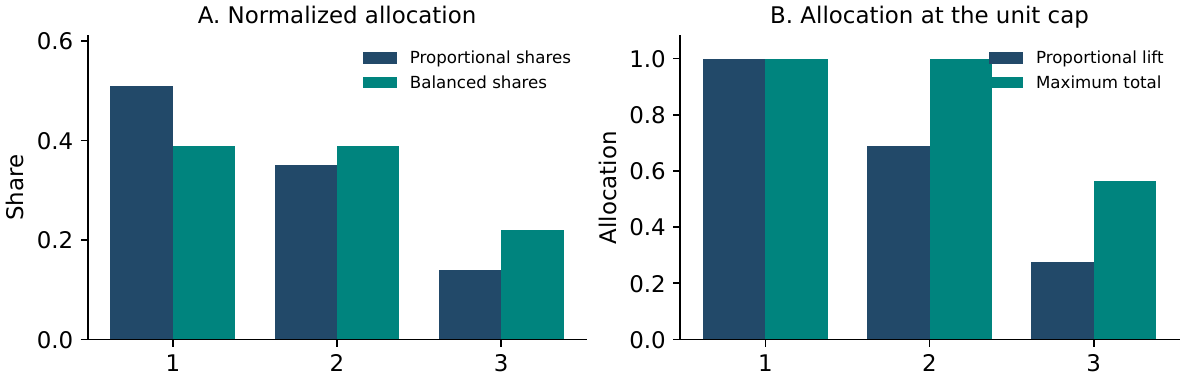}
 \caption{Proportional lifting and maximum total allocation under the same coordinate caps. Balancing the normalized shares allows two coordinates to reach one.}
 \label{fig:optimization-geometry}
\end{figure}

For coordinate caps $d=(.80,.55,.22)$, the proportional lift gives
$a=(1,.6875,.275)$, of total mass about $1.96$. The balanced solution
has $t^\star=.39$ and normalized shares $(.39,.39,.22)$. Its total
allocation is about $2.56$, with two coordinates at one.
\Cref{fig:optimization-geometry} compares the normalized and lifted
allocations.

\subsection{Optimizing under the full expectation constraints}

Consider a finite observation space and a finite model. In addition
to $z$ introduce the inverse scale $t=1/\mathcal D(a)$ at each observation.
The attainable pairs are precisely those satisfying
\begin{equation}\label{eq:opt-complement}
 0\leq z_i\leq t,\qquad \frac1m\leq t\leq1,\qquad
 \sum_i z_i\leq1,\qquad (1-t)\Bigl(1-\sum_i z_i\Bigr)=0.
\end{equation}
Indeed, $a=z/t$ is in the allocation box. If $t=1$ its mass is at
most one. If $\sum_i z_i=1$ its mass is $1/t$. In either case the
proposed pair is recovered from $a$. The full sFDR constraints
are \Cref{eq:opt-risk}, and baseline dominance is the linear constraint
$z_i(x)\geq t(x)b_i(x)$ at every $x$.

Each assignment of the mass regime to the observations gives a
compact polytope in $(z,t)$. On that polytope the expected total
allocation under a reference law $\mu$ is the sum of the low-mass
terms $\mu(x)\sum_i z_i(x)$ and the high-mass terms $\mu(x)/t(x)$.
This objective is convex on each branch polytope, so a maximum occurs
at a vertex. Enumerating regime assignments and their vertices thus
gives an exact optimizer of expected total allocation. Its cost can
grow exponentially with the size of the experiment.

A coordinatewise admissible majorant of any valid baseline can be
computed with one binary variable per observation and a linear objective.
Use a binary $s(x)$ to impose
$t(x)\geq1-(1-1/m)s(x)$ and $\sum_i z_i(x)\geq s(x)$ in place of
the product constraint. With $\mu$ of full support, maximize
\begin{equation}\label{eq:opt-admissible}
 \E_\mu\!\left[\sum_i Z_i-T\right],\qquad Z=z(X),\quad T=t(X),
\end{equation}
subject to the full risk constraints and baseline dominance. Writing $A=Z/T$ and $S=s(A)$ for the random allocation and its total, the
integrand equals $S-1$ for $S\leq1$ and $1-1/S$ for $S\geq1$,
so it strictly increases with $S$. An optimizer exists by compactness.
Any valid rule dominating it coordinatewise and increasing an allocation
at some observation would strictly increase \Cref{eq:opt-admissible}.
This proves admissibility under the full constraints for this experiment.
The objective in \Cref{eq:opt-admissible} prioritizes coordinatewise
improvements through a strictly increasing transform of total allocation.
Optimizing expected total allocation itself uses the vertex problem above.

\subsection{Continuity and coordinatewise improvement}

The exact total-allocation optimizer can jump at the boundary between
mass regimes. For example, with two normalized caps $z_1,z_2\leq d$,
the optimal total is $2d$ for $d<1/2$. At $d=1/2$ it is two. On
a compact data domain, a common Lipschitz bound defines a compact
class of allocation maps by Arzelà--Ascoli. Within this class, the
sFDR constraints and baseline dominance are closed under uniform
limits. Maximizing expected total allocation over this class with a
full-support reference law therefore gives an admissible majorant
within the prescribed class, whenever the baseline belongs to it.

Smoothness also permits explicit local improvements. Suppose a
continuous coordinatewise nonincreasing rule $b(p)$ is continuously
differentiable on an open ball where $s(b)>1$, $\max_i b_i<1$, and
$\partial b_i/\partial p_j<0$ for every $i,j$. For a nonzero nonnegative
smooth function $\varphi$ supported on a compact subset of the ball,
let $a_i=(1+\epsilon\varphi)b_i$. For sufficiently small $\epsilon>0$,
the strict margins on the support preserve the caps and all derivative
signs. Outside the support the allocation equals the baseline, so the
rule is continuous and coordinatewise nonincreasing. Multiplicative
scaling preserves every realized $L_S$ where $s(b)>1$. Thus the rule
dominates $b$ without changing its stable FDP.

For fixed $\alpha$, exponential kernelized BH with $m\geq2$ and $K\geq2$
has such a ball near a vector of small positive p-values. There the
first scale update exceeds one and has strictly negative derivatives
in all p-values. Subsequent scale updates retain these strict signs,
and every final allocation coordinate has strictly negative derivatives.
The local improvement gives inadmissibility within the continuous
monotone class under any model giving the improvement region positive
probability.

\FloatBarrier
\section{Discussion}\label{sec:discussion}

A sensitivity analysis can reveal both the findings that persist
across reasonable choices and the ones that depend on a particular
choice. The wrapper summarizes both kinds of findings on a common
allocation scale with stable FDR control.
Dividing each run's selections by its discovery count gives each run
the same weight in the error calculation, even when some runs report
many more findings than others. The lift lets scientists use those
same relative shares to prioritize follow-up.

In the leukemia and imaging examples, findings with higher allocations
were more often supported by a separate group of patients or
participants. Raw frequencies gave similar relative shares in these
two analyses, while the synthetic BH example shows how much the
error can change when false discoveries tend to occur in runs with
longer lists. Reporting the analysis choices, the discovery counts,
and any empty runs allows readers to judge what stability was
assessed \citep{YK20}. Held-out support measures reproducibility in
these examples. A prospective study could examine whether using the
allocations to choose follow-up experiments improves the number of
confirmed findings per unit of effort.

Methods with random steps have another immediate use for these
reports. A scientist can redraw knockoffs, change a random split, or
vary the learner in a conditional randomization test
\citep{CFJL18,Liu22,Tansey22}. Each rerun must retain its error guarantee
for the same hypotheses. For a single set of test results, the
continuous procedures developed here can also grade findings near a
selection boundary. The more total allocation one asks for, the
harder it can be to preserve continuity. For example, the obstruction
in \Cref{sec:bh-total} shows why a continuous allocation cannot always
match the total from ordinary BH at the same level. Efficient
optimization over structured families of hypotheses and validation of
allocations against the actual costs of follow-up are directions for
further work.

\paragraph{Acknowledgment.}
The author used OpenAI's GPT models to assist with drafting and editing
this manuscript.

\FloatBarrier

\appendix

\section*{Appendix Outline}

\Cref{app:wrapper} proves the wrapper and raw-frequency guarantees
and gives guidance on choosing the number of reruns.
\Cref{app:certificates} develops continuous lifting, quantifies its
sensitivity to reruns, and relates the certificates to compound
e-value aggregation. \Cref{sec:anchor} gives procedures that expand
the allocations available from a valid baseline.
\Cref{app:empirical} documents the data, calibrations, and simulations.
\Cref{sec:other-metrics} develops stable familywise error and stable
false coverage as examples of other reporting goals.

\section{The wrapper and its computation}\label[appendix]{app:wrapper}

\subsection{The wrapper guarantee and raw frequency reports}\label{sec:pooling-details}

\subsubsection{Why normalization and lifting preserve the guarantee}

Use the deterministic notation of \Cref{alg:wrapper}. The averaged
normalized vector $d$ has mass equal to the fraction of nonempty runs.
If that fraction is below one, the output is $a=d$, and its denominator
is one. If all runs are nonempty, $d$ has mass one and the output
is $a=d/v$, where $v=\max_id_i>0$. Its total mass is $1/v\geq1$.
Thus, in both cases,
\[
 \frac{a_i}{1\vee\sum_j a_j}=d_i.
\]
Summing over the true nulls and taking expectations gives exactly
\Cref{eq:wrapper-guarantee}. An always-selected index maximizes every
run's normalized contribution and hence the average; it receives one.
An index selected in no run has zero contribution and receives zero.

\subsubsection{Averaging raw frequencies}

For realized rejection sets let $k_b=|r_b|$,
$h_b=|r_b\cap N|/(1\vee k_b)$, and
$f_i=B^{-1}\sum_b\one\{i\in r_b\}$. Then
\begin{equation}\label{eq:raw-identity}
 L_N(f)=\frac{\sum_bk_bh_b}{B\vee\sum_bk_b}.
\end{equation}
Equal counts make this the ordinary average of the $h_b$, including
the all-empty case. For nonempty counts within a factor $c$, put
$M=\max_bk_b$ and $\bar h=B^{-1}\sum_bh_b$. Since
$\sum_bk_b\geq M+(B-1)M/c$,
\[
 L_N(f)\leq\frac{Bc}{B+c-1}\bar h.
\]
For the other bound, let $x_b=k_b/\min_\ell k_\ell\in[1,c]$
and $u=\sum_bx_bh_b\leq Bc\bar h$. Then
\[
 L_N(f)=\frac{u}{u+\sum_bx_b(1-h_b)}
 \leq\frac{u}{u+B(1-\bar h)}
 \leq\frac{c\bar h}{1+(c-1)\bar h}.
\]
This last function is increasing and concave. Taking expectations
and using Jensen's inequality and $\E[\bar H]\leq\alpha$, where
$\bar H$ is the random average run-wise FDP, proves
\Cref{eq:raw-bound-main}. The all-empty case contributes zero.
The count assumption is uniform over possible batches; comparable
positive counts alone do not cover a mixture of empty and nonempty runs.

\subsubsection{An observable distance between the reports}

Suppose all runs are nonempty. Let $a$ be the wrapper output,
$f$ the raw frequencies, and $\bar k=B^{-1}\sum_bk_b$.
The normalized reports $p=a/\sum_ia_i=d$ and $q=f/\sum_if_i$
both have mass one. Define
\begin{equation}\label{eq:report-distance}
 \delta=\frac12\sum_i|p_i-q_i|.
\end{equation}
Their coordinate differences sum to zero, so the positive and negative
parts each sum to $\delta$. For every subset $S$,
\begin{equation}\label{eq:report-distance-bound}
 |L_S(a)-L_S(f)|\leq\delta
 \leq\frac1{2B}\sum_b\left|\frac{k_b}{\bar k}-1\right|.
\end{equation}
For the last inequality, write $z_{b,i}=\one\{i\in r_b\}/k_b$.
Then $p=B^{-1}\sum_bz_b$ and
$q=\sum_b k_bz_b/(B\bar k)$. Each $z_b$ has mass one, so the
triangle inequality proves the claimed bound. The first inequality
holds simultaneously for all subsets, including the unknown true-null
set and the subset supported in the separate assessment data.

\subsection{Further guarantees for raw frequency reports}\label{sec:raw-extensions}

\Cref{sec:pooling-details} gives the raw-frequency identity and the
proofs of the comparable-count and report-distance bounds. Two further consequences
are useful for general allocation pipelines: exact control with fixed
relative counts or conditional calibration, and sharpness of the
finite-$B$ coefficient. The sets below may have arbitrary dependence.
Write $K_b=|R_b|$ and $F_i=B^{-1}\sum_b\one\{i\in R_b\}$.

\begin{proposition}[Other exact guarantees for raw frequencies]\label[proposition]{prop:raw-exact}
Suppose every run has FDR at most $\alpha$. Raw frequencies control
stable FDR at $\alpha$ under either of the following conditions:
\begin{enumerate}[label=(\roman*),leftmargin=2em]
\item The relative counts are deterministic: there are fixed
$w_1,\ldots,w_B\geq0$ with $\sum_bw_b=1$ such that
$|R_b|=w_b\sum_\ell|R_\ell|$ almost surely.
\item Writing $K_b=|R_b|$, the stronger conditional guarantee
\[
 \E_P[\FDP_{N(P)}(R_b)\mid K_1,\ldots,K_B]\leq\alpha
 \quad\text{almost surely for every }b
\]
holds under every $P$.
\end{enumerate}
The first condition includes fixed rejection counts, even if they differ
between runs, and equal counts with a common random scale.
\end{proposition}
\begin{proof}
Put $T=\sum_bK_b$ and $H_b=\FDP_{N(P)}(R_b)$.
For the raw frequencies $F_i=B^{-1}\sum_b\one\{i\in R_b\}$,
$L_{N(P)}(F)=\sum_bK_bH_b/(B\vee T)$. Under (i), this is
$T/(B\vee T)\sum_bw_bH_b\leq\sum_bw_bH_b$.
Under (ii), condition on the counts to obtain
$\E_P[L_{N(P)}(F)\mid K]\leq\alpha T/(B\vee T)\leq\alpha$.
Taking expectations proves both claims.
\end{proof}

\begin{example}[A sharp bound even for nested rejection sets]\label[example]{ex:raw-sharp}
Fix an integer $c\geq2$, a signal set $S$ of size $k\geq1$, and a
disjoint null set $N$ of size $(c-1)k$. Suppose
$0<\alpha\leq(c-1)/(Bc)$. An observed label $J$ satisfies
\[
 \Prob(J=b)=\frac{\alpha c}{c-1},\quad b=1,\ldots,B,
 \qquad \Prob(J=0)=1-\frac{B\alpha c}{c-1}.
\]
Procedure $b$ selects $S\cup N$ if $J=b$, and $S$ otherwise.
Each procedure has FDR exactly $\alpha$. Every batch consists of
nested, nonempty sets with count ratio at most $c$, and their union
is their largest set. When $J\ne0$, exactly one run selects the nulls,
so the raw report's stable FDP is $(c-1)/(B+c-1)$. Therefore
\[
 \sFDR(F)=\frac{Bc}{B+c-1}\alpha.
\]
This attains the linear finite-$B$ bound in \Cref{eq:raw-bound-main},
so its coefficient is sharp uniformly over $\alpha$. For $c=2$, $B=5$, and
$\alpha=.10$, the raw report has stable FDR $1/6$, although each
procedure has FDR $.10$ and the largest set is the union.
The wrapper in \Cref{alg:wrapper} has stable FDR exactly $\alpha$.
\end{example}

\subsection{Choosing the number of reruns}\label{sec:stability}

The number of reruns, $B$, sets the precision with which an analysis
summarizes its chosen perturbations. There is no minimum $B$ required
for stable FDR control: the guarantee holds for every fixed positive
integer. The perturbations determine which scientific sensitivity is
examined, and $B$ determines how precisely it is summarized.

The first step is to specify the choices and ranges the report should
represent. A small finite grid can be evaluated completely, with the
same number of repetitions per configuration. Both applications here
use balanced collections of 120 analyses. For a randomized pipeline,
the run count can be planned from the available computing budget and
the desired precision, with that count fixed before inspecting the
rerun results.

Ordinary selection frequencies give a useful scale for planning. With
independent random perturbations on the fixed observed data, their
Monte Carlo standard error is at most $1/(2\sqrt B)$ for each hypothesis:
five percentage points at $B=100$ and 2.5 at $B=400$. Stable allocations
also weight each run by its discovery count and apply a lift. Their
computation noise therefore depends on those counts and on the rescaling.
The normalized average has mean squared error proportional to $1/B$,
so four times as many runs halve its root mean squared error.
\Cref{sec:stability-proofs} describes the variance calculation and
resampling protocol. \Cref{sec:certificate-approx} gives concentration,
replacement, and continuity bounds.

A practical report can assess this noise by repeatedly resampling its
completed runs and recomputing the allocations. For a balanced grid,
resampling within each configuration preserves the intended balance.
For a random mixture of configurations, resampling whole runs also
reflects variation in how often each configuration appears. Differences
between the resulting reports indicate how precisely the fixed report
has been computed. These diagnostics describe the specified sensitivity
analysis, with the scientific data held fixed.

\begin{figure}[tb]
 \centering\includegraphics[width=.95\textwidth]{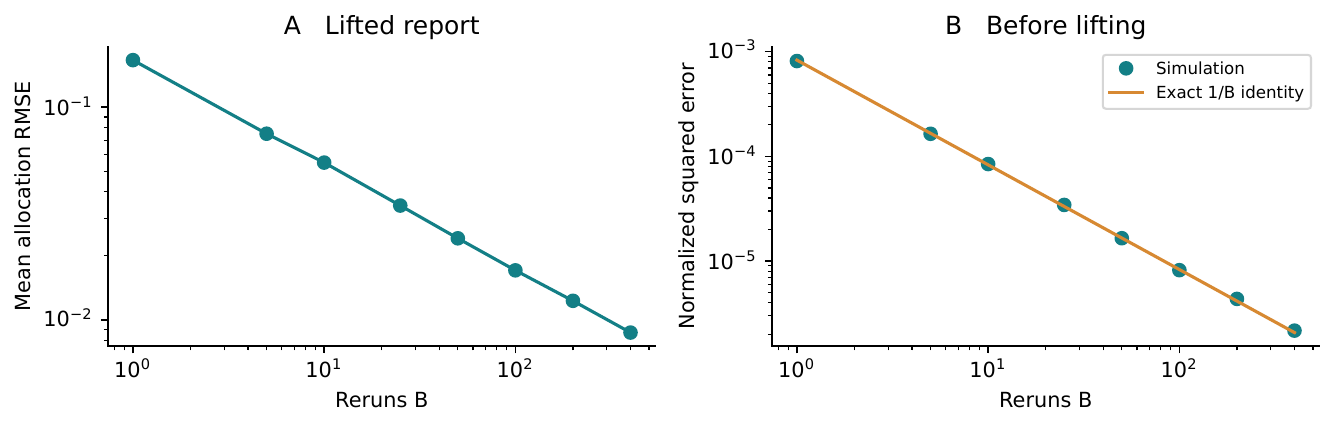}
 \caption{Computation stability for the leukemia sensitivity analysis.
 For each batch size, 200 independent batches are sampled with replacement
 from the fixed collection of 120 reruns. The reference report uses that
 collection's full empirical mean. Write $a^\star$ for that reference
 allocation and $A_B$ for a batch allocation. Allocation RMSE is
 $\{m^{-1}\sum_i(A_{B,i}-a_i^\star)^2\}^{1/2}$.
 Panel B shows the $1/B$ decrease in normalized mean squared
 computation error (\Cref{eq:mc-variance}). This resampling experiment
 treats the 120 completed runs as a random mixture, with each run equally
 likely; it holds the scientific data and that empirical law fixed.}
 \label{fig:computation-stability}
\end{figure}

The leukemia experiment in \Cref{fig:computation-stability} samples
independent batches from the empirical law on the 120 BH analyses.
The scientific observations are held fixed. Mean allocation RMSE is
$0.1666$ for one rerun, $0.0171$ for 100,
and $0.0087$ for 400.
These numbers give an application-specific guide to computation
precision. They describe approximation of the empirical sensitivity law;
they do not quantify uncertainty from collecting a new patient sample.
The patient-deletion and statistic comparisons are represented by the
allocation profile itself. The number of runs, the configuration balance,
discovery counts, and empty runs belong with the reported allocations.
The convergence diagnostic assesses a prespecified report. The fixed-run
error guarantee does not automatically extend to choosing when to stop
after inspecting successive outputs.

\FloatBarrier
\subsubsection{Precision calculation and resampling protocol}\label{sec:stability-proofs}

Conditional on the observed data $X=x$, let
$Z_i=\one\{i\in R\}/(1\vee|R|)$ for one random rerun, and
let $\mu=\E[Z\mid X=x]$. For $B$ conditionally independent reruns
with this same law, put $D_B=B^{-1}\sum_bZ^{(b)}$. Independence
and centering give
\begin{equation}\label{eq:mc-variance}
 \E[\|D_B-\mu\|_2^2\mid X=x]
 =\frac1B\E[\|Z-\mu\|_2^2\mid X=x].
\end{equation}
The cross terms in the squared norm have conditional expectation zero.
This proves the fourfold-run, half-root-error rule in
\Cref{sec:stability}. Independence is needed for this computation
statement; the error guarantee in \Cref{eq:wrapper-guarantee} permits
arbitrary dependence between runs.

For \Cref{fig:computation-stability}, the fixed collection of 120
leukemia analyses is treated as an empirical perturbation law, each
with probability $1/120$. For each
$B\in\{1,5,10,25,50,100,200,400\}$, 200 independent batches
are sampled with replacement from that collection. Each batch is
normalized and lifted, then compared with the full-collection report.
The result measures approximation of this empirical sensitivity law,
with the patient observations fixed. The archived computation table
records every batch. General certificate concentration, replacement,
and continuity bounds are in \Cref{sec:certificate-approx}.

\FloatBarrier

\section{Continuous lifting and allocation certificates}\label[appendix]{app:certificates}

\subsection{Risk certificates and continuous lifting}\label{sec:perturb}

The elementary lift in \Cref{alg:wrapper} switches scale at
total normalized mass one. A continuous lift can instead vary the
scale across a range of certificate masses. We first construct
certificates for mixtures of nominal levels.

Let $U$ index the perturbation, and let $R_U$ be a binary selection on the
same $m$ hypotheses at nominal level $\lambda_U>0$. Random seeds, retained
observations, choices of statistic, and nominal levels can all be part of
$U$. Define the random certificate $D=d(X)$ by
\begin{equation}\label{eq:pert-cert}
 D_i=d_i(X)=\alpha\,\E\left[
 \left.\frac{\one\{i\in R_U\}}{\lambda_U(1\vee |R_U|)}\,\right|\, X\right].
\end{equation}
We assume joint measurability and a finite certificate almost surely;
a level band bounded away from zero suffices. For a fixed number of runs,
the conditional expectation can be replaced by the arithmetic mean over
the runs.

The normalized vector $d/(1\vee\sum_i d_i)$ has mass at most one.
When $\sum_i d_i>1$, expanding it toward $d/\max_i d_i$ preserves
its normalized shares. The following lift interpolates between these
two scales. Its cubic weight has zero derivative at both ends of the
transition; validity uses only that this weight lies in $[0,1]$ and
vanishes below total mass one.

\begin{definition}[Continuous lift]\label[definition]{def:lift}
For $d\in[0,\infty)^m$, write $s=\sum_i d_i$ and $v=\max_i d_i$.
Fix a ramp width $\rho>0$, and put
\[
 t(s)=0\vee\left(\frac{s-1}{\rho}\wedge1\right),\qquad
 g(s)=t(s)^2\{3-2t(s)\}.
\]
The allocation $\mathcal L_\rho(d)$ is zero when $v=0$, and otherwise has
coordinates
\begin{equation}\label{eq:generic-lift}
 a_i=(1-g(s))\frac{d_i}{1\vee s}+g(s)\frac{d_i}{v}.
\end{equation}
\end{definition}

\begin{theorem}[Perturb, average, and lift]\label{thm:perturb}
For each $P$, suppose $\E_P[\FDP_{N(P)}(R_U)/\lambda_U]\leq 1$.
Then $A=\mathcal L_\rho(D)$ controls sFDR at $\alpha$.
Write $S=\sum_iD_i$. More precisely,
\begin{equation}\label{eq:lift-identity}
 \frac{A_i}{\mathcal D(A)}=\frac{D_i}{1\vee S},\qquad
 \sFDR_P(A)\leq\alpha\,\E_P[\FDP_{N(P)}(R_U)/\lambda_U].
\end{equation}
The same conclusion holds for arithmetic means of a fixed number of runs
if their mean expected relative FDP is at most one.
In particular, each run may be an ordinary FDR-controlling procedure
at its own prespecified level. There is no independence requirement
across runs. The lift is continuous in $(d,\rho)$ for $\rho>0$.
\end{theorem}

\begin{proof}
If $S\leq1$, $A=D$ and $\mathcal D(A)=1$. If $S>1$, both vectors in the
interpolation have total mass at least one, lie in the unit box, and
are proportional to $D$. Consequently $A/\mathcal D(A)=D/S$. The risk bound
follows from the tower property or linearity of expectation. Continuity
at $S=1$ follows from $g(1)=0$ and $\max_iD_i\ge S/m$ in the region $S>1$.
In a neighborhood of $d=0$ the deterministic lift equals $d$.
\end{proof}

\begin{corollary}[Persistent and absent findings]\label[corollary]{cor:wrapper-endpoints}
Consider $B$ runs at fixed levels $\lambda_b>0$. Suppose each run controls FDR at
$\lambda_b$ under every $P\in\cP$. Let
\[
 s_* = \frac{\alpha}{B}\sum_{b=1}^B\frac{1}{\lambda_b}>1,
 \qquad \rho=s_*-1,
 \qquad D_i=\frac{\alpha}{B}\sum_{b=1}^B
       \frac{\one\{i\in R_b\}}{\lambda_b(1\vee |R_b|)}.
\]
Then $A=\mathcal L_\rho(D)$ controls sFDR at $\alpha$.
If $i$ is selected in all $B$ runs, then $A_i=1$; if it is
selected in none, then $A_i=0$.
\end{corollary}
\begin{proof}
The risk bound follows from \Cref{thm:perturb}. If $i$ is always
selected, all runs are nonempty, so $S=\sum_j D_j=s_*=1+\rho$.
Moreover, $D_i\geq D_j$ for every $j$, term by term. Thus the lift
is full and $A_i=D_i/\max_j D_j=1$. If $i$ is never selected,
$D_i=0$, and the lift preserves this zero.
\end{proof}

At a common pilot $\beta<\alpha$, the calibration is simply
$\rho=\alpha/\beta-1$; $\beta=.9\alpha$ is one choice.
A level mixture with harmonic mean below $\alpha$ also satisfies
$s_*>1$, including a nondegenerate symmetric band about $\alpha$.
This choice makes the endpoint interpretation exact. At a common
pilot equal to $\alpha$, the certificate has total mass at most one,
and a continuous lift need not give persistent findings a full unit.

The guarantee holds for every fixed $B$, over the data and any
algorithmic randomization. Repeated draws approximate the conditional
average in \Cref{eq:pert-cert}, whose allocation also controls sFDR.
Fixing the number of runs and the perturbation scheme specifies which
sensitivity analysis is summarized.

\begin{figure}[htbp]
 \centering\includegraphics[width=.82\textwidth]{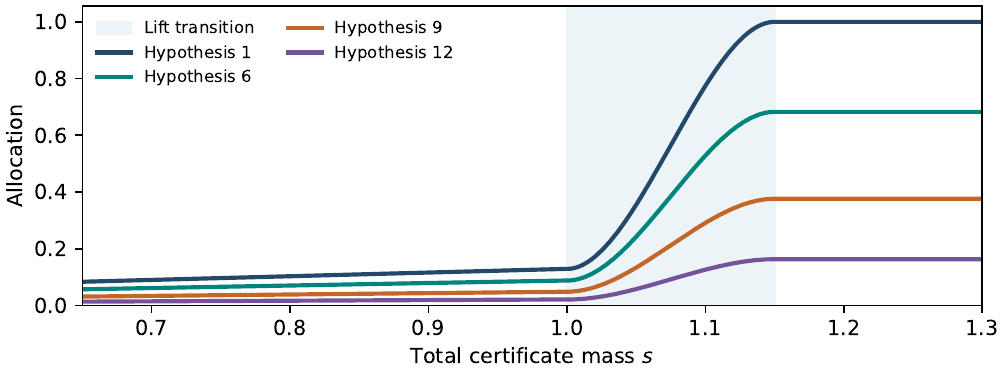}
 \caption{Continuous lifting of a fixed normalized certificate profile as its total mass varies. The shaded band is the transition from $s=1$ to $s=1+\rho$ for $\rho=.15$. The strongest features reach one at the end of the transition.}
 \label{fig:lift-ramp}
\end{figure}

The full lift is active when $S\geq1+\rho$. Write
$\omega=\one\{|R_U|>0\}/\{\lambda_U(1\vee |R_U|)\}$.
In this regime the allocation is the feature's weighted selection
frequency divided by the largest weighted selection frequency.
In particular, if
$\E[\omega\one\{i\notin R_U\}\mid X]\leq\varepsilon\E[\omega\mid X]$,
then $A_i\geq1-\varepsilon$. A feature selected in every nonempty
run has $A_i=1$. Below the full-scale regime the lift gives less
weight to the relative frequencies (\Cref{fig:lift-ramp}). At a common pilot $\beta$,
\[
 S=\frac{\alpha}{\beta}\Prob(R_U\ne\varnothing\mid X),
\]
so that the scale records how consistently the runs yield discoveries.
When using the empirical average, the same statements hold with the
empirical perturbation law and the guarantee is over data and randomization.

\begin{example}[Gaussian row weights and resampling]\label[example]{ex:gaussian-weights}
For i.i.d.\ rows, a common random subset drawn without replacement
gives a fresh sample of the retained size in distribution, conditional
on the indices. Each run recomputes the p-values and applies the
base procedure for that sample size. Separate masks across hypotheses
are possible if the new joint p-value law satisfies the dependence
condition of the base procedure. For instance, in a one-sided Gaussian
model with known variances and nonnegative cross-coordinate covariances,
the covariance between two subset sums is the original covariance
times their number of common rows. Conditional on the masks, the
Gaussian PRDS argument applies \citep{BY01}. With valid marginal
p-values and arbitrary dependence, BY can be used in place of BH.

With-replacement resampling requires calibration for the resampling
weights. For i.i.d.\ standard Gaussian observations $X_t$ and
independent bootstrap counts $C_t$ summing to $b$, the variance of
$\sum_t C_t X_t/\sqrt b$ conditional on the counts is
$\sum_t C_t^2/b$, which exceeds one whenever a row repeats. For this
model the exact denominator is $\sqrt{\sum_t C_t^2}$. Nonnegative
weights also preserve the sign of cross-coordinate covariances in
the Gaussian example, so weighted one-sided tests can be used with
BH. Other bootstrap statistics require their own null calibration
\citep{HW91}.

\end{example}

\begin{example}[Raw level averaging can exceed the target]\label[example]{ex:raw-bh}
Take nine independent uniform null p-values and one nonnull p-value
equal to $0.19\alpha$. Average the raw BH indicators at $0.9\alpha$
and $\alpha$ equally. Suppose eight null p-values exceed $\alpha$,
and denote the remaining one by $p$. For $p\leq0.09\alpha$, the
smaller level selects the null alone and the larger level selects
both the null and the nonnull. The averaged allocation has stable
FDP $2/3$. For $0.09\alpha<p\leq0.20\alpha$, only the larger level
selects the pair, and the stable FDP is $1/2$. The disjoint events
for the nine choices of the small null contribute
$1.035\alpha(1-\alpha)^8>\alpha$ at $\alpha=0.001$.
\end{example}

\subsection{Certificate approximation and continuity}\label{sec:certificate-approx}
For any nonnegative vectors $d,d'$ with positive maxima $v,v'$,
\begin{equation}\label{eq:normalization-lipschitz}
 \left\|\frac d v-\frac{d'}{v'}\right\|_\infty
 \leq \frac{2\|d-d'\|_\infty}{\max\{v,v'\}}.
\end{equation}
To verify this, suppose $v'\geq v$. Writing the difference as
$(d-d')/v'+(d/v)(v'-v)/v'$ bounds it by
$\|d-d'\|_\infty/v'+|v'-v|/v'$. The maximum map is
one-Lipschitz in the sup norm. Interchanging the vectors covers the
other ordering. 
\subsubsection{Approximating a normalized certificate}

The practical choice of run count and its resampling diagnostic appear
in \Cref{sec:stability,sec:stability-proofs}. The following bounds apply to any
conditionally i.i.d.\ sequence of normalized certificates.

\begin{proposition}[Certificate approximation]\label[proposition]{prop:certificate-approx}
Conditional on $X=x$, let $Z^{(1)},Z^{(2)},\ldots$ be i.i.d.\ vectors
with $\sum_iZ_i^{(b)}=1$ and $0\leq Z_i^{(b)}\leq1/r$ for a fixed
$r\geq1$. Write $\mu=\E[Z^{(1)}\mid X=x]$,
$\nu=\max_i\mu_i$, and $D_B=B^{-1}\sum_bZ^{(b)}$.
For $A_B=D_B/\max_iD_{B,i}$ and $a^\star=\mu/\nu$, every
$\eta\in(0,1)$ satisfies
\begin{equation}\label{eq:certificate-concentration}
 \Prob\left(\left.\|A_B-a^\star\|_\infty>
 \frac{2}{r\nu}\sqrt{\frac{\log(2m/\eta)}{2B}}\,\right|X=x\right)
 \leq\eta.
\end{equation}
Replacing one vector in a deterministic batch with another vector
obeying the same simplex and coordinate constraints gives
\begin{equation}\label{eq:certificate-replacement}
 \|a-a'\|_\infty\leq
 \min\left\{1,\frac{2}{Br\max\{v,v'\}}\right\},
 \quad v=\max_i d_i,\quad v'=\max_i d_i'.
\end{equation}
\end{proposition}
\begin{proof}
Conditional Hoeffding bounds and a union bound give
$\Prob(\|D_B-\mu\|_\infty>u\mid X=x)\leq2m e^{-2Br^2u^2}$.
Apply \Cref{eq:normalization-lipschitz} and
$\max\{\max_iD_{B,i},\nu\}\geq\nu$ to prove
\Cref{eq:certificate-concentration}. One replacement changes
every empirical coordinate by at most $1/(Br)$; the same normalization
inequality proves \Cref{eq:certificate-replacement}.
\end{proof}

\subsubsection{Changing the certificate law}

Let $Q,Q'$ be two laws on simplex-valued vectors with coordinates
in $[0,1/r]$. Write $d_Q=\E_Q[Z]$, $v_Q=\max_id_{Q,i}$,
and $a_Q=d_Q/v_Q$, with analogous notation for $Q'$.
With $\operatorname{TV}(Q,Q')=\sup_E|Q(E)-Q'(E)|$,
$\|d_Q-d_{Q'}\|_\infty\leq\operatorname{TV}(Q,Q')/r$.
Thus \Cref{eq:normalization-lipschitz} gives
\[
 \|a_Q-a_{Q'}\|_\infty\leq
 \min\left\{1,\frac{2\operatorname{TV}(Q,Q')}
 {r\max\{v_Q,v_{Q'}\}}\right\}.
\]
This is a deterministic comparison of normalized certificate laws.

\subsubsection{A global bound for the continuous lift}

Let $\mathcal L_\rho$ be \Cref{def:lift}, with fixed $\rho>0$.
For all $d,d'\in[0,\infty)^m$,
\begin{equation}\label{eq:continuous-lift-lipschitz}
 \|\mathcal L_\rho(d)-\mathcal L_\rho(d')\|_\infty
 \leq\left(2m+\frac{3}{2\rho}\right)\|d-d'\|_1.
\end{equation}
Indeed, write $s=\sum_i d_i$, $s'=\sum_i d'_i$,
$v=\max_i d_i$, and $v'=\max_i d'_i$.
Put $h(d)=d/(1\vee\sum_i d_i)$ and $u(d)=d/\max_i d_i$,
with $u(0)=0$. The map $h$ is two-Lipschitz from the $\ell_1$ norm
to the sup norm. The cubic ramp $g$ is $3/(2\rho)$-Lipschitz,
and $\|u(d)-h(d)\|_\infty\leq1$. In the decomposition
\[
 \mathcal L_\rho(d)-\mathcal L_\rho(d')
 =(1-g(s))\{h(d)-h(d')\}+g(s)\{u(d)-u(d')\}
 +(g(s)-g(s'))\{u(d')-h(d')\},
\]
the middle term is zero unless $s>1$, when
$\max\{v,v'\}\geq1/m$. The bound
\Cref{eq:normalization-lipschitz} therefore controls the first
two terms together by $2m\|d-d'\|_1$; when one vector is zero the
same bound follows directly. The last term contributes at most
$3\|d-d'\|_1/(2\rho)$. This proves
\Cref{eq:continuous-lift-lipschitz}, including across total mass one.
Together with coordinate concentration for bounded certificates, it
gives finite-batch bounds for continuous lifts without a nonempty-run
condition. Narrow ramps increase this worst-case Lipschitz constant.

\subsection{Extensions of the risk certificate}

\subsubsection{A post-hoc level certificate for a fixed allocation}

For a positive e-value input $e$, define
\[
 q_e(a)=\max_{i\in\ms}\frac{ma_i}{e_i\mathcal D(a)}.
\]
With the conventions that $0/0=0$ and a positive numerator divided by zero
is infinite, $a$ is e-self-consistent at every level
$\alpha\geq q_e(a)$. Hence \Cref{thm:esc} controls the expected
relative loss simultaneously over all allocations with their certified
levels. The anchored analog is obtained by taking the smallest $\alpha$
for which \Cref{eq:positivepart} holds; it can be found by scalar bisection.

\begin{figure}[htbp]
 \centering\includegraphics[width=.62\textwidth]{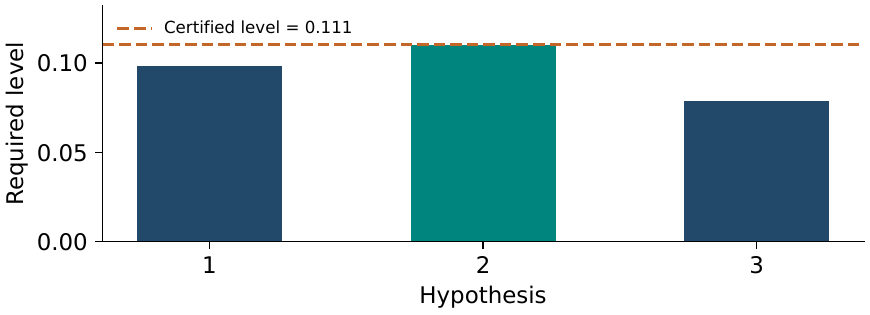}
 \caption{The certified level is the maximum of the three coordinatewise requirements for the given allocation.}
 \label{fig:certified-level}
\end{figure}

For $a=(1,.7,.2)$ and $e=(16,10,4)$ the three required levels are
approximately $(.099,.111,.079)$. The second coordinate sets the
smallest level at which the given allocation is certified
(\Cref{fig:certified-level}).

\subsubsection{Pooling intersection certificates}

The normalized pooling rule in \Cref{prop:normalized-pool} combines
allocations directly. One can also combine their intersection
certificates and optimize a new allocation. If
$Z_S^{(1)},\ldots,Z_S^{(B)}$ are intersection e-values, then
$Z_S=B^{-1}\sum_{b=1}^B Z_S^{(b)}$ is again an intersection
e-value. The resulting decision can be optimized for the desired utility.

\subsubsection{Weighted allocation units}

For prespecified positive weights $v_i$, one may define
$L_S^v(a)=\sum_{i\in S}v_i a_i/(1\vee\sum_jv_j a_j)$.
The general closure theorem and its completeness proof apply directly,
because this loss is monotone in $S$. The choice of units and caps should
reflect the application. Kernel self-consistency and anchoring can then be
developed for the corresponding weighted coordinate budget.

\subsection{Compound evidence and allocation certificates}\label{sec:related-details}

The normalization in \Cref{alg:wrapper} has an algebraic connection
to binary compound e-value aggregation.
\citet[Theorem~4.4 and Appendix~C.1]{IWR24} give the general compound
representation; \citet{RB24} use averaged knockoff evidence for binary
FDR control. For rejection sets $r_b$ from runs at level $\beta$,
their canonical evidence and its arithmetic mean are
\[
 c_i^{(b)}=\frac{m\one\{i\in r_b\}}{\beta(1\vee|r_b|)},
 \qquad \bar c_i=\frac m\beta d_i,
 \qquad d_i=\frac1B\sum_b\frac{\one\{i\in r_b\}}{1\vee|r_b|}.
\]
The compound validity of this construction follows from the FDR
guarantee of each run. Applying binary e-BH to $\bar c$ at $\alpha$
searches for a $k$ satisfying $d_{[k]}\geq\beta/(\alpha k)$.
At $\beta=\alpha$, a nonempty output requires total mass one
and equal positive coordinates. The allocation region defined by the
compound certificate can retain graded reports without that final
selection rule. \Cref{sec:closure,sec:optimization} describe the
resulting certification and optimization problems.

For power, \citet{RB24} recommend $\beta=\alpha/2$; their validity
theorem permits every $\beta\in(0,1)$. The continuous lift in
\Cref{def:lift} instead uses a level mixture and a ramp width to
control the scale continuously. These are explicit allocation choices,
with the bounds in \Cref{thm:perturb,cor:wrapper-endpoints}.

SMARA's continuous closure addresses strategic resource allocation,
participation, and null deterrence \citep{SMARA26}. Its loss uses the
null fraction of total resources, whereas the reference-unit floor
used here permits small allocations under the global null. Our
closure theorem specializes that general principle to this loss.
Koning's parallel E-measure work treats continuous evidence over
logically structured hypotheses \citep{Koning26}; \Cref{rem:FER}
gives the connection to its fixed-selection formulation.

\FloatBarrier

\section{Anchoring and explicit improvements}\label[appendix]{sec:anchor}

Anchoring enlarges a baseline's self-referential local e-values by accounting
for unused expectation budget \citep{Anchor26}. We first give a general
version for fractional allocations and then derive an explicit construction.

\begin{theorem}[Stable anchoring]\label{thm:anchor}
Let $B=b(X)\in[0,1]^m$ and $\alpha_0(X)>0$ be a baseline allocation and
its level. Suppose known constants $\eta_S\in[0,1]$ satisfy
\begin{equation}\label{eq:partial}
 \E_P\left[\frac{L_S(B)}{\alpha_0}\right]\leq\eta_S
 \quad\text{for every }P\in H_S.
\end{equation}
For any e-value $F_S$ for $H_S$, set
\begin{equation}\label{eq:anchored-general}
 Z_S^{\rm anc}=\frac{L_S(B)}{\alpha_0}+(1-\eta_S)F_S.
\end{equation}
Then $Z^{\rm anc}$ is an intersection e-collection and
$B\in\cA_{\alpha_0}(Z^{\rm anc})$ pointwise. Its closure contains the
self-referential closure generated by $L_S(B)/\alpha_0$.
Every measurable selection from the anchored closure has the guarantees
of \Cref{thm:closure}.
\end{theorem}
\begin{proof}
Under $H_S$, the expectation in \Cref{eq:anchored-general} is at most
$\eta_S+(1-\eta_S)=1$. The added term is nonnegative, so the baseline
and its entire self-referential closure satisfy the new constraints.
\end{proof}

For kernelized BH and kernel Stable-eBH, the coordinatewise bounds give
$\eta_S=|S|/m$. This spends the full certified intersection budget,
including the residual $1-|S|/m$. It enlarges the decision region while
retaining every baseline allocation.

\subsection{A continuous anchored improvement of kernelized BH}\label{sec:blend}

The simplest choice $F_S=1$ already gives an explicit improvement. It applies
directly to the p-value procedure, including at a monotone random level.

\begin{proposition}[Exponential headroom blend]\label[proposition]{prop:blend}
Suppose $B\in[0,1]^m$ and $\alpha_0>0$ satisfy
\Cref{eq:partial} with $\eta_S=|S|/m$. Write $T=\sum_i B_i$,
$M=\max_i B_i$, and $\Delta=mM-T\geq0$. Define
\begin{equation}\label{eq:blend}
 U=\begin{cases}
 \displaystyle\frac{\alpha_0 T(T-1)}{m\Delta+\alpha_0 T^2},& T>1,\\[6pt]
 0,&T\leq1,
 \end{cases}
 \qquad
 \delta=1-e^{-U},\qquad A_i=B_i+\delta(1-B_i).
\end{equation}
Then $A\geq B$ and $A$ belongs to the anchored closure generated by
$Z_S=L_S(B)/\alpha_0+1-|S|/m$. Thus
$\E[L_N(A)/\alpha_0]\leq1$. The corresponding map $(b,\alpha_0)\mapsto a$ is
continuous. If $T>1$, it strictly increases every coordinate with $B_i<1$.
\end{proposition}
\begin{proof}
For $T\leq1$, the output is the baseline. Suppose $T>1$ and write
$T'=T+\delta(m-T)\geq T$. The normalized share shift is
\[
 \frac{A_i}{T'}-\frac{B_i}{T}
 =\frac{\delta(T-mB_i)}{TT'}
 \geq-\frac{\delta\Delta}{T^2}\geq-\frac{\alpha_0}{m}.
\]
The last inequality uses $\delta\leq U$ and, for $\Delta>0$,
$U\Delta/T^2\leq\alpha_0(T-1)/(mT)\leq\alpha_0/m$;
it is immediate when $\Delta=0$. Both share vectors sum to one, so
for every $S$,
\[
 L_S(A)-L_S(B)
 =-\sum_{i\notin S}\left(\frac{A_i}{T'}-\frac{B_i}{T}\right)
 \leq\alpha_0(1-|S|/m).
\]
These are exactly the anchored constraints. Continuity holds also at
$T=1$, since $U\to0$ there. Strict increase follows from $U>0$ when
$T>1$ and $B_i<1$.
\end{proof}

\Cref{prop:blend} increases each coordinate by the same fraction of its
remaining headroom. It spends the proof budget by changing normalized shares
in a controlled way. It gives an $O(m)$ anchored kernelized BH procedure, with
the full level-$\alpha_0$ relative guarantee. It can also be applied to kernel Stable-eBH. The construction below uses the original e-values to obtain a more
evidence-specific redistribution.

\subsection{An anchored region from marginal e-values}\label{sec:rawanchor}

Starting from marginal e-values, an especially simple collection is
\begin{equation}\label{eq:rawanchor}
 Z_S^{\rm raw}=\frac{\sum_{i\in S}E_i+m-|S|}{m}.
\end{equation}
Under $H_S$, its expectation is at most one. Moreover, every baseline
satisfying \Cref{eq:esc} belongs to its stable e-closure, since
$L_S(b)/\alpha\leq m^{-1}\sum_{i\in S}E_i$.
Write $\cA_\alpha^{\rm raw}=\cA_\alpha(Z^{\rm raw})$.

\begin{proposition}[A linear-size description]\label[proposition]{prop:positivepart}
The anchored region is exactly
\begin{equation}\label{eq:positivepart}
 \cA_\alpha^{\rm raw}=
 \left\{a\in[0,1]^m:
   \sum_{i=1}^m\pos{\frac{ma_i}{\alpha \mathcal D(a)}-E_i+1}\leq m\right\}.
\end{equation}
Equivalently, with
$G_i=(E_i-1)_+$ and $C=\sum_i(E_i\wedge1)$, its condition is
\begin{equation}\label{eq:donationregion}
 \sum_i\pos{\frac{ma_i}{\alpha \mathcal D(a)}-G_i}\leq C.
\end{equation}
\end{proposition}
\begin{proof}
The subset constraint can be rewritten as
$\sum_{i\in S}\{ma_i/(\alpha \mathcal D(a))-E_i+1\}\leq m$.
The largest subset sum is the sum of positive parts, which proves
\Cref{eq:positivepart}. For $x\geq0$,
$(x-E_i+1)_+=(x-G_i)_++1-(E_i\wedge1)$.
Summing this identity proves \Cref{eq:donationregion}.
\end{proof}

This description makes linear optimization efficient. For a linear utility,
including the constraint $a\geq b$, one solves two linear programs and
keeps the better feasible solution. On $s(a)\leq1$, introduce $v_i\geq0$
and impose
\begin{equation}\label{eq:LPlow}
 v_i\geq ma_i/\alpha-E_i+1,\qquad \sum_i v_i\leq m.
\end{equation}
On $s(a)\geq1$, impose
\begin{equation}\label{eq:LPhigh}
 v_i\geq ma_i/\alpha-(E_i-1)s(a),\qquad
 \sum_i v_i\leq ms(a).
\end{equation}
Both programs have $O(m)$ variables and constraints, besides any application
constraints. The baseline is feasible in at least one branch. With positive
utility weights, an optimizer weakly increases every baseline coordinate by
construction and maximizes the chosen linear objective among such upgrades.

\subsection{A continuous anchored allocation without optimization}

An explicit construction retains continuity and fractional allocations.
It first reserves enough of the anchored budget to keep the baseline, then
redistributes the remainder.

\begin{theorem}[Smooth anchored kernel Stable-eBH]\label{thm:smoothanchor}
Let $E$ be a vector of marginal e-values, and let $B\in[0,1]^m$ satisfy
stable e-self-consistency at a possibly data-dependent level $\alpha>0$.
Define
\begin{align}
 D_B&=1\vee\sum_i B_i,& Q_i&=\frac{mB_i}{\alpha D_B},&
 G_i&=(E_i-1)_+,\label{eq:reserves}\\
 D_i^0&=(Q_i-G_i)_+,& C&=\sum_i(E_i\wedge1),&
 R&=C-\sum_i D_i^0.\nonumber
\end{align}
Choose any measurable weights $W_i\geq0$ with $\sum_i W_i=1$, and set
\begin{equation}\label{eq:donated}
 \widetilde E_i=G_i+D_i^0+RW_i.
\end{equation}
Starting with $R_0=D_B$, for a fixed $K\geq1$ compute
\begin{equation}\label{eq:headroom}
 A_i^{(k+1)}
 =B_i+(1-B_i)\left[1-\exp\left\{-\left(
       \frac{\alpha\widetilde E_i R_k}{m}-B_i\right)\right\}\right],
 \qquad
 R_{k+1}=1\vee\sum_i A_i^{(k+1)}.
\end{equation}
Then $\alpha\widetilde E_i R_k/m-B_i\geq0$ for every update,
$A^{(K)}\geq B$,
and $A^{(K)}\in\cA_\alpha^{\rm raw}$. In particular,
\begin{equation}\label{eq:anchor-risk}
 \E\left[\frac{L_N(A^{(K)})}{\alpha}\right]\leq1.
\end{equation}
If $\alpha\leq q$, its sFDR is at most $q$.
For fixed $K$ and uniform weights $W_i=1/m$, continuous input maps
$b(e)$ and $\alpha(e)$ give a continuous output map $a^{(K)}(e)$. Whenever $B_i<1$
and $\widetilde E_i>Q_i$, its first update strictly increases coordinate $i$.
\end{theorem}
\begin{proof}
Self-consistency gives $Q_i\leq E_i$, hence
$D_i^0\leq E_i\wedge1$, $R\geq0$, and
$\widetilde E_i\geq Q_i$. Let $D_i=D_i^0+RW_i$, so $D_i\geq0$ and
$\sum_i D_i=C$. For every null set $N$, pointwise,
\begin{equation}\label{eq:donationbound}
 \sum_{i\in N}\widetilde E_i
 \leq\sum_{i\in N}G_i+C
 =\sum_{i\in N}E_i+\sum_{i\notin N}(E_i\wedge1)
 \leq\sum_{i\in N}E_i+m-m_0.
\end{equation}
This gives compound validity, even though the donation weights, the baseline,
and the level may depend on all the data.

For a fixed scale $r\geq D_B$, put $V_i=\alpha\widetilde E_i r/m\geq B_i$.
Since $1-e^{-u}\leq u$ for $u\geq0$,
\[
 B_i\leq B_i+(1-B_i)(1-e^{-(V_i-B_i)})
 \leq B_i+(V_i-B_i)=V_i.
\]
The update is nondecreasing in $r$ and maps into $[B_i,1]$.
Thus $R_k$ is nondecreasing, and its final output satisfies
$mA_i^{(K)}/(\alpha \mathcal D(A^{(K)}))\leq\widetilde E_i=G_i+D_i$.
Consequently the positive parts in \Cref{eq:donationregion} sum to at most
$\sum_i D_i=C$, proving closure membership. Alternatively,
\Cref{eq:donationbound} directly gives
$L_N(A^{(K)})/\alpha\leq m^{-1}(\sum_{i\in N}E_i+m-m_0)$.
Taking expectations proves \Cref{eq:anchor-risk}. Continuity follows from
the displayed formulas; positive parts and the maximum with one are
continuous. Strict improvement follows when the exponent is positive and
$1-B_i>0$.
\end{proof}

The pointwise bounds also hold simultaneously over all levels, all
self-consistent baselines, and all donation policies based on the same
original e-value vector. This is the allocation version of spending the
intersection-wise slack in the original e-BH proof.

\subsection{Anchoring kernelized BH through p-value-derived e-values}

Let $B=b(\bm P)$ be a $K$-round kernelized BH allocation, at a fixed level $\alpha$ or
at a monotone policy $\widehat\alpha(\bm P)$. Write its level as $\alpha_0$.
\Cref{thm:alpha} gives the coordinatewise certificates
\begin{equation}\label{eq:p-lift}
 E_i^{\rm pilot}=\frac{m B_i}{\alpha_0\mathcal D(B)},\qquad
 \E_P[E_i^{\rm pilot}]\leq1\quad\text{whenever }P\in H_i.
\end{equation}
Here the p-value assumptions are imposed under every model law, with PRDS
on its true-null coordinates. These are marginal e-values, derived from the
full p-value vector. They satisfy
$B_i=\alpha_0 E_i^{\rm pilot}\mathcal D(B)/m$ exactly.

The raw anchored closure based on $E^{\rm pilot}$ is exactly the closure
used in \Cref{prop:blend}, since
$m^{-1}\sum_{i\in S}E_i^{\rm pilot}=L_S(B)/\alpha_0$.
Thus \Cref{prop:blend} gives a direct continuous upgrade, and the two linear
programs give an optimal linear-utility upgrade. The normalized share
constraints in the proof explain how an allocation can increase every
coordinate while redistributing the proof budget. Its validity does not
require the upgraded allocation to remain monotone in the p-values; the
intersection e-value certificate carries the guarantee.

The pilot e-values can also be retained for later decisions. Once they have
been constructed using a specified valid policy, stable e-closure permits
arbitrary subsequent positive level choices, with its simultaneous relative
guarantee. This separates the construction of valid evidence from the later
allocation objective.

\subsection{A numerical example}

\begin{figure}[htbp]
 \centering\includegraphics[width=\textwidth]{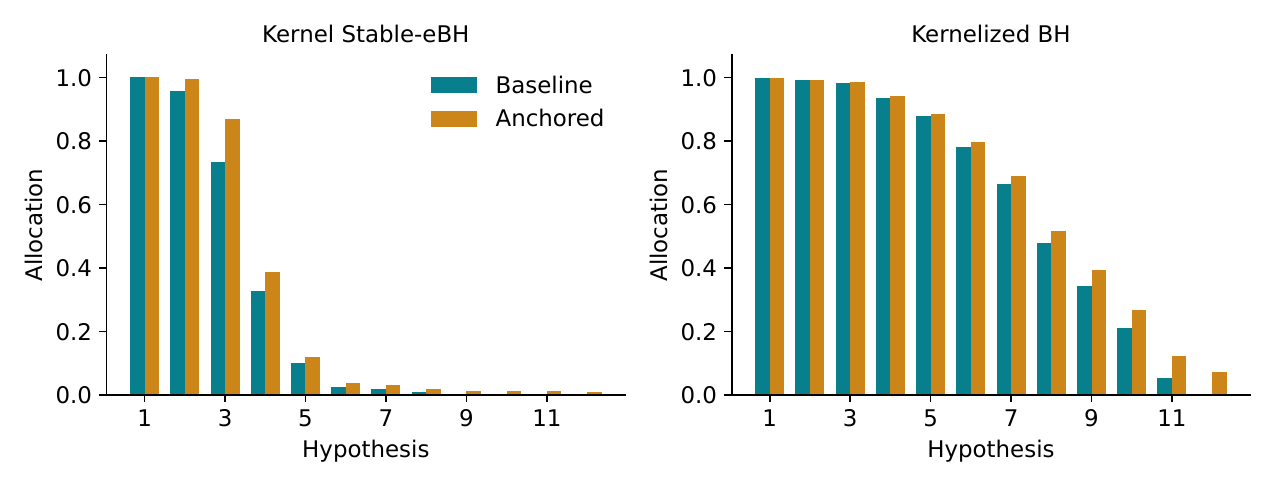}
 \caption{Explicit continuous anchoring of kernel Stable-eBH and kernelized BH.
 Both baselines use $\alpha=0.10$ and 25 rounds. The left panel uses 25
 donation updates; the right uses the explicit headroom blend.}
 \label{fig:anchoring}
\end{figure}

For a direct e-value example, we use
\[
 e=(500,120,50,15,4,1,.7,.3,.1,.03,.01,.001)
\]
at $\alpha=0.10$. \Cref{fig:anchoring} applies $K=25$ rounds of kernel Stable-eBH and then $25$ rounds of the smooth anchored upgrade, using uniform
donation weights. The second panel uses the exponential headroom blend of
\Cref{prop:blend} on the p-value allocation. In both cases every coordinate
weakly increases. These examples compare deterministic allocations; the
guarantees follow from the theorems.

\FloatBarrier

\section{Data, empirical protocols, and simulations}\label[appendix]{app:empirical}

\subsection{Data and empirical protocols}\label{sec:protocols}


\subsubsection{Leukemia calibration}
The matrices are the original discovery and test cohorts from the CRAN
\texttt{SIS} package, version 1.5. Provenance records source URLs,
SHA-256 digests, package description, and license. Probe identifiers
are source column numbers, not gene annotations. Floors and logarithms
are fixed label-blind transformations. Under equality of the original
expression distributions, the transformed observations and independent
tie keys are exchangeable between groups.

For the mean statistic, each Monte Carlo permutation independently
chooses a uniform group of the observed ALL size from the pooled sample.
The two-sided p-value is $(1+K_{\geq})/(K+1)$, where $K_{\geq}$ counts
permuted absolute mean differences at least as large as observed.
The observed labeling and independent random labelings are exchangeable
under the null; the plus-one rank p-value is therefore superuniform.
For the rank sum, lexicographic sorting of measurement and independent
uniform key gives distinct ranks. Dynamic programming evaluates the
exact rank-sum law for the two retained group sizes. These statements
give marginal validity; they do not impose a joint BH dependence law.

Patient subsets are sampled without replacement, independently of their
expression values, within each diagnostic group. The retained group
sizes are $(21,8)$, $(24,9)$, and $(27,11)$. Every run uses the same
7,129 probes. The test-cohort support rule uses floor $100$, the mean
statistic, 9,999 permutations, and BH at $.10$, with direction matched
to the complete-discovery mean log difference. Ties in permutation
statistics are included conservatively.

\subsubsection{The correlated BH counterexample}\label{sec:failure-protocol}

The observations are independent standard normals
$G,G_1,\ldots,G_B$. Analysis $b$ uses
\[
 Z_b=\sqrt\rho\,G+\sqrt{1-\rho}\,G_b,\qquad
 p_i^{(b)}=
 \begin{cases}0,&i\leq10,\\1-\Phi(Z_b),&i>10.\end{cases}
\]
Here $\Phi$ is the standard normal distribution function, and
the last 90 indices are true nulls. Every null p-value is uniform
within a run, and the block of identical null p-values is PRDS.
Every analysis uses the same observed vector, with a different
prespecified statistic. BH selects the null block exactly when
$Z_b\geq z_\alpha:=\Phi^{-1}(1-\alpha)$, so its FDR is $.9\alpha$.
We use $\alpha=.10$, $\rho=.8$, $B\leq120$, and 100,000 data sets.
The pointwise identity for the wrapper is checked on every data set.

Conditional on $G=g$, the number $K=BQ$ of high-count runs is
binomial with parameters $B$ and
\[
 q(g)=\Phi\!\left(\frac{\sqrt\rho\,g-z_\alpha}{\sqrt{1-\rho}}\right).
\]
Consequently, the exact raw risk is the one-dimensional integral
\[
 \int_{-\infty}^{\infty}\sum_{k=0}^B
 \frac{9k}{B+9k}\binom Bk q(g)^k\{1-q(g)\}^{B-k}\,
 \phi(g)\,\mathrm dg,
\]
where $\phi$ is the standard normal density. Gaussian quadrature
at 200 and 400 nodes agrees within $10^{-8}$ for the primary setting.
For an analytic lower bound, put
$(g_1,g_2,g_3,g_4)=(.5,1,1.5,\infty)$ and
$(k_1,k_2,k_3)=(1,12,48)$. Since $q(g)$ and $9k/(B+9k)$
are increasing, the integral is at least
\[
 \sum_{j=1}^3\{\Phi(g_{j+1})-\Phi(g_j)\}
 \frac{9k_j}{B+9k_j}
 \Prob\{\operatorname{Bin}(B,q(g_j))\geq k_j\}
 \approx .105917>.10.
\]
The counterexample script reproduces the simulation, quadrature, lower
bound, and figure. It also
checks the raw-frequency identity with empty runs and verifies the
finite-$B$ sharpness construction in \Cref{ex:raw-sharp} by exact enumeration.

\subsubsection{Neuroimaging calibration and provenance}
The archive includes the 94 contrast vectors restricted to the fixed
mask, original download URLs and SHA-256 digests, the original subject
split, the mask and anatomical template, and the source data license.
The source contrast is
\texttt{LeftAuditory\&VisualClickVsRightAuditory\&VisualClick}.
The grid is $53\times63\times46$, with affine diagonal $(-3,3,3)$ mm
and translation $(78,-111,-51)$ mm. The external MNI152 mask was
resampled before testing; it does not use observed contrast values.
The display slices have fixed axial coordinates 18, 36, and 54 mm.

At a voxel, the t-test uses the finite observations and its Student
reference law with $n_i-1$ degrees of freedom. Fewer than three usable
observations or zero sample variance gives $p=1$. Under the sufficient
Gaussian location model, conditioning on usable-observation indicators
leaves i.i.d.\ Gaussian measurements for that target population.
Informative measurement availability would require a model for a
broader population target.

The signed-rank statistic sums ranks of positive nonzero observations.
Independent uniform keys break ties of absolute magnitude. For $n_i$
usable nonzero observations, the reference distribution is
$\sum_{j=1}^{n_i}jB_j$, with independent $B_j\sim\mathrm{Bernoulli}(1/2)$.
The exact upper tail is evaluated by dynamic programming. Under the
zero-centered symmetric model, signs are independent fair coins
conditional on absolute values; under a nonpositive Gaussian location,
their conditional positive probabilities are at most one half, giving
a conservative upper tail. Zero observations are omitted.
All $67,969$ voxel hypotheses remain in each BH family.
The held-out p-values use the same t-test on the other 47 participants.
The split assesses participant-level reproducibility in a single study.

\subsection{Illustrations and numerical verification}\label{sec:experiments}

\subsubsection{Level and data sensitivity}\label{sec:path-sim}

We compare the path methods with their binary counterparts at
$\alpha=.10$ with $\delta=.20$. The simulated data have $m=40$
coordinates, the first eight of which are signals. For BH,
the test statistics are jointly Gaussian with unit variances,
pairwise correlation $.5$, and signal means $3.2$. For e-BH, the
Gaussian common factor has loadings of alternating sign
$\pm\sqrt{.5}$, and signal means are $4.5$. The e-values are
$\exp(2.5Z_i-2.5^2/2)$, which have null expectation one.
Both path comparisons use 16,000 independent data sets.

In a separate row-deletion experiment, each data set contains
80 i.i.d.\ Gaussian rows with pairwise cross-coordinate correlation
$.5$, unit marginal variances, and mean $3.2/\sqrt{80}$ in the
first eight coordinates. We use 40 prespecified masks of 64 rows
each, chosen without replacement independently of the data.
On each mask we form the one-sided p-values and run BH at
$\beta=.09$. We pool and apply \Cref{eq:generic-lift} with
$\rho=.05$. We compare with full-data BH at $.10$ over 2,400
independent data sets.
We also vary the statistic by applying 40 vectors of row weights
drawn independently of the data. Each weight is $.5$ or $1.5$ with
equal probability, and each vector is normalized to have Euclidean
norm one. Conditional on the weights the resulting $z$-statistics
have unit variances and nonnegative correlations. We use the same
pilot and lift as for row deletion. Both perturbations preserve the
null mean, and they probe sensitivity to different choices about
the contribution of each observation.

\begin{table}[tb]
 \centering\footnotesize
 \begin{tabular}{@{}lllrrrr@{}}
\toprule
Setting & Procedure & Risk & Estimate & Signal mass & Partial & Peak \\
\midrule
Gaussian PRDS & BH & sFDR & 0.068 & 6.73 & 0.00 & 0.97 \\
Gaussian PRDS & Stable-BH & sFDR & 0.067 & 6.68 & 0.85 & 0.96 \\
Mixed dependence & eBH & sFDR & 0.009 & 7.60 & 0.00 & 1.00 \\
Mixed dependence & Stable-eBH & sFDR & 0.009 & 7.59 & 0.19 & 1.00 \\
Row deletion & Full-data BH & sFDR & 0.062 & 6.73 & 0.00 & 0.97 \\
Row deletion & Subsampled Stable-BH & sFDR & 0.057 & 5.07 & 9.17 & 0.77 \\
Row reweighting & Weighted Stable-BH & sFDR & 0.057 & 5.12 & 9.15 & 0.77 \\
\bottomrule
\end{tabular}

 \caption{Empirical risks and allocation summaries at target $.10$.
 Signal mass is the mean total allocation to the eight nonzero means.
 Partial is the mean number of allocations in $(.01,.99)$.
 Peak is the fraction of data sets with a largest allocation of at
 least $.99$. The accompanying CSV includes risk standard errors.}
 \label{tab:stable-sim}
\end{table}

\begin{figure}[htbp]
 \centering\includegraphics[width=\textwidth]{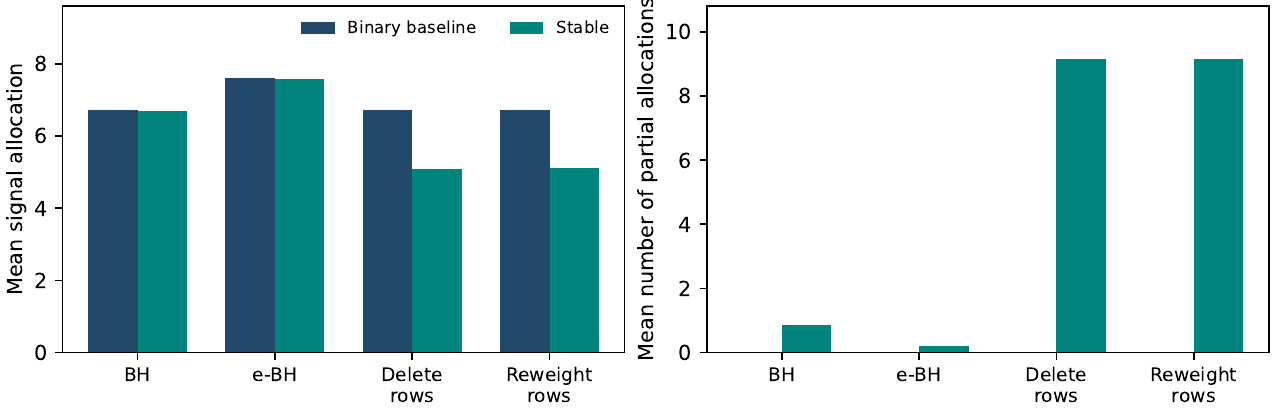}
 \caption{Allocation comparisons for the simulations in \Cref{tab:stable-sim}. Partial allocations lie in $(.01,.99)$.}
 \label{fig:simulation-allocation}
\end{figure}

In \Cref{tab:stable-sim,fig:simulation-allocation}, smoothing the level yields nearly the same
mean signal allocation as the corresponding binary procedure in each
path experiment. Row deletion and reweighting produce roughly nine partial allocations
per data set, and reduce mean signal allocation from $6.73$ to about
$5.1$. These perturbations measure stability to changes in the data
entering the test statistic, which can move even strong candidates
away from the full-unit regime.

\subsubsection{A PRDS simulation}

For a finite-sample check, let $m=60$, with the first ten hypotheses non-null.
We generate
\[
 Z_i=\mu_i+\sqrt\rho\,G+\sqrt{1-\rho}\,G_i,
 \qquad \mu_i=3\one\{i\leq10\},
\]
where $G,G_1,\ldots,G_m$ are independent standard normals and
$\rho\in\{0,0.5\}$. The one-sided p-values $P_i=\Phi(-Z_i)$ satisfy
PRDS. The e-values $E_i=e^{2Z_i-2}$ have null expectation one. We use
20,000 independent repetitions, $q=0.10$, and 25 rounds for each
allocation stage. For the adaptive rows, first compute kernel
Stable-BH at level $q$, let $S_0$ be its total allocation, and set
\[
 \widehat\alpha=q\left\{\eta+(1-\eta)\frac{S_0}{S_0+c}\right\},
 \qquad \eta=.20,\quad c=5.
\]
The data-dependent level guarantee is developed in \Cref{sec:alpha},
and the anchored upgrades are developed in \Cref{sec:anchor}.

\Cref{tab:sim} reports empirical sFDR, the empirical mean of stable FDP divided by the procedure's level, and expected non-null allocation.
For fixed-level methods the level is $0.10$; for adaptive methods it is the
displayed $\widehat\alpha$. The relative-risk bound is $50/60$ for BH,
kernelized BH, and unanchored kernel Stable-eBH, and one for the anchored procedures and BH with additional allocations.
Non-null mass describes allocation to useful hypotheses; it is measured in
the reference allocation units of \Cref{sec:metric}.

\begin{table}[tb]
 \centering\small
 \begin{tabular}{@{}rlrrr@{}}
\toprule
$\rho$ & Procedure & sFDR & Relative risk & Non-null mass\\
\midrule
0.0 & BH & 0.0835 & 0.835 & 7.93 \\
0.0 & Kernelized BH & 0.0800 & 0.800 & 6.98 \\
0.0 & BH + allocations & 0.0962 & 0.962 & 7.94 \\
0.0 & Adaptive kernelized BH & 0.0547 & 0.783 & 6.37 \\
0.0 & Anchored adaptive kernelized BH & 0.0605 & 0.867 & 6.38 \\
0.0 & Kernel Stable-eBH & 0.0743 & 0.743 & 4.34 \\
0.0 & Anchored kernel Stable-eBH & 0.0869 & 0.869 & 4.80 \\
0.5 & BH & 0.0692 & 0.692 & 7.59 \\
0.5 & Kernelized BH & 0.0687 & 0.687 & 6.61 \\
0.5 & BH + allocations & 0.0816 & 0.816 & 7.59 \\
0.5 & Adaptive kernelized BH & 0.0543 & 0.668 & 5.99 \\
0.5 & Anchored adaptive kernelized BH & 0.0603 & 0.750 & 5.99 \\
0.5 & Kernel Stable-eBH & 0.0657 & 0.657 & 4.22 \\
0.5 & Anchored kernel Stable-eBH & 0.0782 & 0.782 & 4.58 \\
\bottomrule
\end{tabular}

 \caption{Monte Carlo means over 20,000 repetitions. The reproducible CSV
 also records standard errors, total allocation, and relative-risk standard
 errors. All methods use the same observations within a dependence setting.}
 \label{tab:sim}
\end{table}

The accompanying script checks self-consistency, monotonicity of the
adaptive kernelized BH map, baseline dominance after anchoring, and the exact
positive-part closure constraints. It also solves the two linear programs
for the direct e-value example. The simulation checks the predicted behavior
in a concrete model; all validity statements above are finite-sample results
under their stated assumptions.

\FloatBarrier

\section{Stable familywise error and false coverage}\label[appendix]{sec:other-metrics}

Many error criteria admit a stable version once we specify how the loss
for a binary decision applies to a graded report. We illustrate the
idea with stable familywise error (stable FWER) and stable false
coverage (stable FCR). The two examples are
deliberately selective. Developing stable versions of other error
criteria, and choosing losses that fit the intended use of the report,
are future work.

\subsection{Stable familywise error and Stable-Holm}\label{sec:stableholm}

For familywise protection, the natural loss is the largest
allocation to a true null:
\[
 \operatorname{sFWER}_P(A)=\E_P\!\left[\max_{i\in N(P)}A_i\right],
\]
with the maximum over an empty set equal to zero. This agrees with
classical FWER on binary allocations. The largest allocation has
a threshold interpretation: with an independent uniform threshold
on $[0,1]$, it is the conditional probability that at least one
null allocation exceeds the threshold.

Suppose $\E_P[\one\{R_U\cap N(P)\ne\varnothing\}/\lambda_U]\leq1$.
A generic rule is
\begin{equation}\label{eq:sfwer-pool}
 A_i=1\wedge\alpha\,\E[\one\{i\in R_U\}/\lambda_U\mid X].
\end{equation}
For any true null $i$, the indicator in the conditional expectation
is at most $\one\{R_U\cap N(P)\ne\varnothing\}$. Taking the maximum
over true nulls and then expectations proves sFWER control at $\alpha$.
At a constant pilot $\beta\leq\alpha$, a selection frequency of
at least $\beta/\alpha$ gives a full unit. Ordinary selection
frequencies are also valid whenever $\E[\one\{R_U\cap N(P)\ne\varnothing\}]
\leq\alpha$, by the same argument without the scale factor.

The nested Holm rejection path \citep{Holm79} gives a single-sort
version. With ascending p-values, the entry levels are
\[
 h_{(j)}=\max_{k\leq j}(m-k+1)p_{(k)}.
\]
Use the centered level band of \Cref{sec:stablebh} and set
\[
 a_i=\Prob(\Lambda\ge h_i)
     =\frac{\{H-(L\vee h_i)\}_+}{H-L}.
\]
We call this \emph{Stable-Holm}; write $A=a(\bm P)$ for its random output.
The nested path gives the exact identity
\[
 \max_{i\in N}A_i
 =\Prob_\Lambda(R_\Lambda\cap N\ne\varnothing\mid X).
\]
Consequently sFWER is at most $\E\Lambda\leq\alpha$ under arbitrary
dependence. The allocation is continuous in $(p,\alpha)$, equals
one for hypotheses rejected throughout the band, and is computed
by a sort and prefix maxima. For $p=(.010,.030,.053,.6)$ and
$\alpha=.10$, the Holm entry levels are $(.04,.09,.106,.6)$.
The binary selection is $(1,1,0,0)$ and Stable-Holm with
$[.08,.12]$ gives $(1,.75,.35,0)$.

\begin{figure}[htbp]
 \centering\includegraphics[width=.79\textwidth]{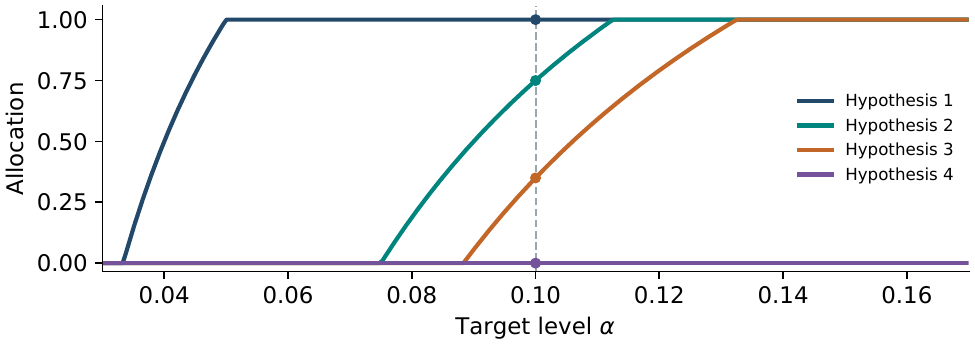}
 \caption{Stable-Holm allocations for p-values $(.010,.030,.053,.60)$. The markers show the allocations at $\alpha=.10$ with $\delta=.20$.}
 \label{fig:holm-path}
\end{figure}

\subsection{Stable false coverage for confidence-set reports}\label{sec:stablefcr}

Let $\theta_i=\theta_i(P)$ be the target parameter for index $i$.
A report consists of allocations $A$ and confidence sets $C_i$ for
the indices with $A_i>0$. Its stable false coverage rate is
\[
 \operatorname{sFCR}_P(A,C)
 =\E_P\!\left[\frac{\sum_i A_i\one\{\theta_i\notin C_i\}}{\mathcal D(A)}\right].
\]
On binary allocations this is the false coverage-statement rate
of \citet{BY05}. Different runs can select different parameters
and give different intervals for the same parameter.

\begin{theorem}[Pooling and lifting confidence reports]\label{thm:stablefcr}
Let run $U$ report $(R_U,\{C_{i,U}:i\in R_U\})$ at nominal
false coverage level $\lambda_U>0$. Suppose
\[
 \E_P\left[\frac{\sum_{i\in R_U}\one\{\theta_i\notin C_{i,U}\}}
 {\lambda_U(1\vee |R_U|)}\right]\leq1
\]
for each $P$. Form $D$ by \Cref{eq:pert-cert} and $A$ by
\Cref{eq:generic-lift}. If the reported measurable $C_i$ contains
$C_{i,U}$ almost surely over the perturbations selecting $i$, then
$\operatorname{sFCR}_P(A,C)\leq\alpha$. In particular, for finitely
many interval reports one can use the smallest interval containing
the reported intervals for each selected parameter.
\end{theorem}

\begin{proof}
For each $i$, noncoverage by the envelope implies noncoverage in
every contributing run, up to a perturbation null set. Thus
\[
 \frac{A_i\one\{\theta_i\notin C_i\}}{\mathcal D(A)}
 \leq D_i\one\{\theta_i\notin C_i\}
 \leq\alpha\,\E\!\left[
 \left.\frac{\one\{i\in R_U,\ \theta_i\notin C_{i,U}\}}
 {\lambda_U(1\vee |R_U|)}\,\right|X\right].
\]
Sum and take expectations. The finite mixture case uses the same
inequality with the empirical perturbation law.
\end{proof}

\subsection{Coverage profiles and alternatives to interval unions}
\label{sec:coverage-profiles}

A confidence report can retain the distribution of the reported
intervals. For the perturbed reports in \Cref{thm:stablefcr}, define
the run weight
\[
 \omega_i(U)=\frac{\one\{i\in R_U\}}{\lambda_U(1\vee|R_U|)},
 \qquad B_i=\E[\omega_i(U)\mid X],\qquad D_i=\alpha B_i .
\]
For $B_i>0$, the \emph{coverage profile} is
\begin{equation}\label{eq:coverage-profile}
 V_i(t)=\frac{\E[\omega_i(U)\one\{t\in C_{i,U}\}\mid X]}{B_i},
 \qquad t\in\mathbb R .
\end{equation}
When $B_i=0$ we set $V_i\equiv1$ and $A_i=0$.
Thus $V_i(t)\in[0,1]$ is the weighted fraction of the
feature's contributing intervals that contain $t$.
The report $(A_i,V_i)$ preserves the intervals' agreement
without selecting one envelope.

For this report define the profile false coverage rate
\begin{equation}\label{eq:profile-fcr}
 \operatorname{sFCR}^{\rm prof}_P(A,V)
 =\E_P\!\left[\frac{\sum_i A_i\{1-V_i(\theta_i)\}}{\mathcal D(A)}\right].
\end{equation}
If each $V_i$ is an indicator of a confidence set, this is
the stable false coverage rate already defined above.
For fractional profiles, a parameter contained in most
contributing intervals incurs a correspondingly smaller loss.

\begin{theorem}[Profiles and thresholded confidence sets]\label{thm:profile-fcr}
Under the assumptions of \Cref{thm:stablefcr},
$A=\mathcal L_\rho(D)$ satisfies
$\operatorname{sFCR}^{\rm prof}_P(A,V)\leq\alpha$.
For any measurable choices $0\leq\tau_i<1$, including choices
made after viewing the profiles, define
\[
 C_i^{\tau_i}=\{t:V_i(t)>\tau_i\},\qquad
 \widetilde D_i=(1-\tau_i)D_i .
\]
Then $\widetilde A=\mathcal L_\rho(\widetilde D)$ satisfies
$\operatorname{sFCR}_P(\widetilde A,C^\tau)\leq\alpha$.
The same bound holds if each $C_i^\tau$ is enlarged to its interval
hull, and if coordinates are masked before lifting.
\end{theorem}

\begin{proof}
The profile loss obeys
\[
 D_i\{1-V_i(\theta_i)\}
 =\alpha\,\E\left[
 \left.\frac{\one\{i\in R_U,\ \theta_i\notin C_{i,U}\}}
 {\lambda_U(1\vee |R_U|)}\,\right|X\right].
\]
Since $A_i/\mathcal D(A)\leq D_i$, summing and taking expectations
proves the profile bound. If $\theta_i\notin C_i^{\tau_i}$,
then $V_i(\theta_i)\leq\tau_i$, giving
\[
 (1-\tau_i)D_i\one\{\theta_i\notin C_i^{\tau_i}\}
 \leq D_i\{1-V_i(\theta_i)\}.
\]
The lift bounds the normalized thresholded allocation by
$(1-\tau_i)D_i$. Enlarging a set reduces its noncoverage loss.
Masking reduces the certificate before the same argument.
All inequalities hold pointwise for the chosen thresholds.
\end{proof}

At threshold $\tau=0$, positive membership means that at least
one contributing interval contains the value, recovering their
union. A positive threshold removes parts supported by only a
small fraction of the weighted reports. The resulting set can
be disconnected, and its interval hull provides a single interval.
The certificate factor $1-\tau$ records the price of using the
smaller set. Majority coverage, $\tau=1/2$, uses half the certificate.
These operations are related to weighted majority-vote merging
of uncertainty sets \citep{GR24}. Here the weights also account
for the selection count and nominal level of each run.

\begin{figure}[htbp]
 \centering\includegraphics[width=.9\textwidth]{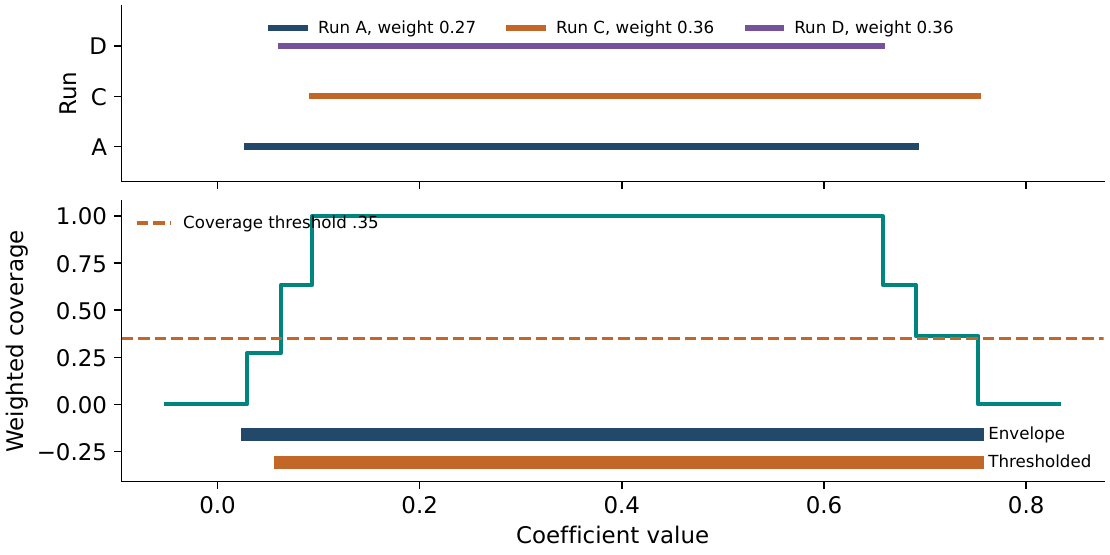}
 \caption{The coverage profile of feature 3 in the regression
 example of \Cref{sec:fcr-regression}. The top panel shows the
 contributing intervals and the bottom panel their weighted
 coverage fraction. Thresholding at $.35$ removes the portions
 supported only by the lowest-weight run. The horizontal bars compare the interval hull and the thresholded set.}
 \label{fig:coverage-profiles}
\end{figure}

\Cref{fig:coverage-profiles} uses the reports from the regression
example below. For this pilot level, multiplying by $1-\tau$
can move the certificate below the full-lift regime. To preserve
full-scale allocations for persistent features, the constituent
FCR pilots can instead use
$\beta=(1-\tau)\alpha/(1+\rho)$ when the same threshold is used
for all features. If every run makes a report, then
$\sum_i(1-\tau)d_i=1+\rho$. This offers a choice between a
coverage profile with its original allocation and a smaller
ordinary confidence set with a recalibrated allocation.

\paragraph{A single-sort Gaussian construction.}
Let $Z_i\sim N(\theta_i,1)$ independently. At level $\lambda$,
select by BH on the one-sided p-values $1-\Phi(Z_i)$, and for
each selected parameter report the two-sided interval
\begin{equation}\label{eq:normal-fcr}
 C_{i,\lambda}=
 \left[Z_i\ \pm\ \Phi^{-1}\!\left(1-\frac{\lambda |R_\lambda|}{2m}\right)\right].
\end{equation}
This is the FCR adjustment for simple selection \citep{BY05}.
To verify it directly, let $K_i$ be the BH rejection count after
replacing $P_i$ by zero. Whenever $i\in R_\lambda$, $K_i=|R_\lambda|$:
for all larger ranks, $P_i$ was already below their thresholds,
so replacing it by zero cannot create a larger feasible rank.
Conditional on $Z_{-i}$, the noncoverage probability for the
interval with count $K_i$ is $\lambda K_i/m$. The expected
contribution of $i$ to the FCR is therefore at most $\lambda/m$.

Use Stable-BH allocations and let $b_i=L\vee q_i$ for $a_i>0$.
The intervals are nested as $\lambda$ increases along the
selection path, since $\lambda |R_\lambda|$ is nondecreasing.
Their envelope is just $C_{i,b_i}$. Its count is the number
of entry levels at most $b_i$ and can be computed in a scan
through the sorted entry levels, processing ties together.
The procedure controls sFCR at $\alpha$. This symmetric
construction can include zero under one-sided selection. The
asymmetric construction below pairs the interval's sign
threshold with the selection rule.

\begin{figure}[htbp]
 \centering\includegraphics[width=\textwidth]{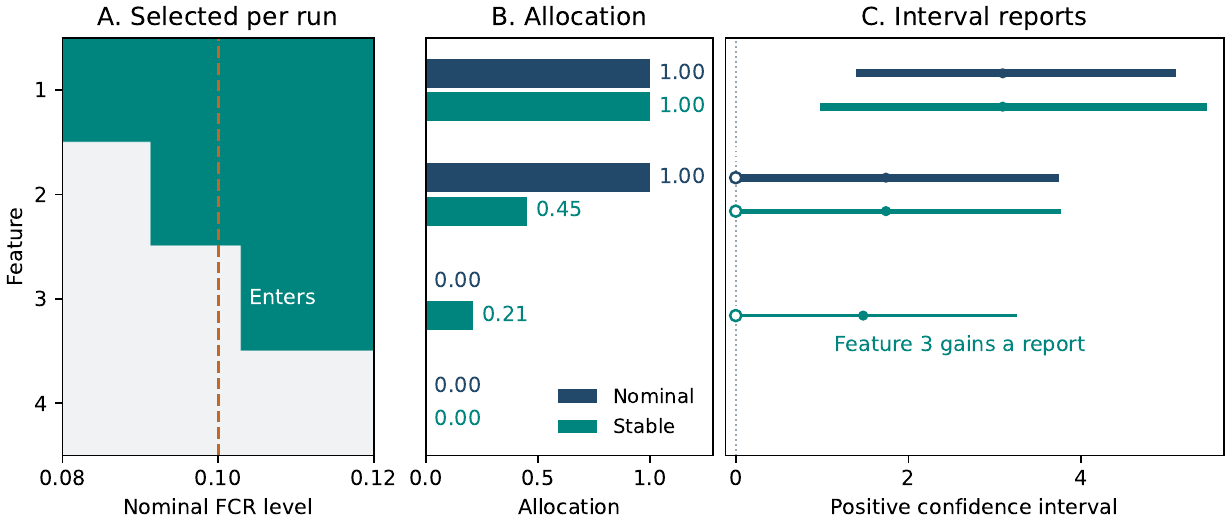}
 \caption{A stable confidence report changes which effects receive
 allocation. One-sided p-values are $(.001,.041,.070,.40)$ and
 $\psi=.9$. Left: positive MQC selections across 21 midpoint levels
 in $[.08,.12]$, with the nominal level $.10$ marked.
 Middle: binary reports at $.10$ and stable allocations.
 Right: hard MQC intervals and stable interval envelopes; points
 mark estimates and hollow endpoints exclude zero. Feature 3 has
 no nominal report but receives a positive interval and fractional
 allocation in the stable report.}
 \label{fig:stable-coverage}
\end{figure}

\paragraph{Sign-determining asymmetric intervals.}
For \Cref{fig:stable-coverage} the direction of interest is specified
as positive. We replace the symmetric family by the positive
modified quasi-conventional (MQC) intervals of \citet{WY20}.
For miscoverage $\gamma$, write
\[
 c=\Phi^{-1}(1-\gamma/2),\quad
 \bar c=\Phi^{-1}(1-\psi\gamma),\quad
 \widetilde c=\Phi^{-1}(1-(1-\psi)\gamma),\quad h=\bar c+c,
\]
and define $g_\gamma(t)=t+\Phi^{-1}(1-\gamma+\Phi(-\bar c-t))$.
We use $\psi=.9$ and $0<\gamma<.25$, for which the first
case of their Equation (4.2) applies. For $y\geq\bar c$, its
lower endpoint is
\[
 \ell_\gamma(y)=
 \begin{cases}
  0,&\bar c\leq y<\widetilde c,\\
  g_\gamma^{-1}(y),&\widetilde c\leq y\leq g_\gamma(h),\\
  h,&g_\gamma(h)<y<h+c,\\
  y-c,&h+c\leq y.
 \end{cases}
\]
Here $g_\gamma^{-1}(y)$ is the largest solution in $[0,h]$,
which can be computed by bisection. The reported interval is
$M_\gamma(y)=(\ell_\gamma(y),y+c]$. In particular, a lower
endpoint zero is open. For $y<\bar c$ we set
$M_\gamma(y)=\mathbb R$ and make no positive-direction report.
This enlargement of the unreported MQC intervals retains
their marginal coverage of at least $1-\gamma$.
The selected interval is asymmetric near the sign boundary.

At nominal FCR level $\lambda$, apply BH to $P_i=1-\Phi(Z_i)$
at level $\psi\lambda$, let $K_\lambda$ be the rejection count,
and report $M_{\lambda K_\lambda/m}(Z_i)$ on that set.
Every reported interval excludes zero, since
$P_i\leq\psi\lambda K_\lambda/m$ is exactly the positive-sign
threshold $Z_i\geq\bar c$.

\begin{proposition}[Positive MQC reports and stable aggregation]
\label[proposition]{prop:mqc-fcr}
For independent $Z_i\sim N(\theta_i,1)$ and $0<\lambda<.25$,
the positive MQC procedure above has FCR at most $\lambda$.
An arithmetic mean of these reports followed by the lift in
\Cref{thm:stablefcr} controls sFCR at the target level.
Every positively allocated interval envelope excludes zero.
\end{proposition}

\begin{proof}
Replace $P_i$ by zero and let $K_i$ be the BH count at
$\psi\lambda$. Whenever $i$ is selected, $K_i=K_\lambda$.
Conditional on $Z_{-i}$, the marginal interval
$M_{\lambda K_i/m}(Z_i)$ has noncoverage probability at most
$\lambda K_i/m$. Its contribution to the FCR is at most this
probability divided by $K_i$. Summing gives the bound.
The envelope is the interval hull of the actual contributing
MQC intervals. All of them lie in $(0,\infty)$, and this hull
preserves the open endpoint whenever its infimum is zero.
\end{proof}

The entry level for feature $i$ is $q_i/\psi$, where $q_i$ is
its one-sided BH-adjusted p-value. The MQC endpoints can vary
nonmonotonically with the level. For this illustration, we use
21 equally weighted midpoints of the band $[.08,.12]$, compute
the certificates and intervals at those levels, and then lift
the average certificate. The ramp width is
$\rho=21^{-1}\sum_j\alpha/\lambda_j-1$, so it is full when
every run makes a report. This finite mixture has the exact sFCR
guarantee of \Cref{thm:stablefcr}.

For a worked example, take one-sided p-values
$(.001,.041,.070,.40)$ and target $\alpha=.10$. The nominal
MQC run reports only features 1 and 2. Their entry levels are
$.004/.9$ and $.082/.9$; feature 3 enters at
$(.28/3)/.9\approx.1037$, just above the nominal level.
It is reported in 9 of the 21 perturbed runs and receives
positive allocation after aggregation. The stable allocations
are approximately $(1,.452,.211,0)$.
Thus the stable report adds a positive-direction interval for
feature 3 and reduces the attention assigned to feature 2,
which is absent at the lower levels. Every allocated interval
excludes zero. A hollow endpoint in \Cref{fig:stable-coverage}
denotes the open boundary of an interval of the form $(0,u]$.

\subsection{Regression pipelines that disagree on zero exclusion}
\label{sec:fcr-regression}

Suppose a downstream decision requires every positively allocated interval
to exclude zero. We illustrate the rule with four split-and-screen
regression pipelines applied to the same observations. The pipelines
vary the held-out fold and the lasso penalty. Each screens on its training
rows and fits the \emph{full} prespecified model on the held-out rows.
Thus every interval for feature $i$ targets the same coefficient $\theta_i$,
even when the screening sets differ. This use of independent data for
screening and inference is related to screen-and-clean methods
\citep{WR09}.

For pipeline $b$, let $S_b$ be the lasso screen. On the held-out
fold form marginal Student intervals at miscoverage $\beta/|S_b|$
for $i\in S_b$, and report those excluding zero:
$R_b=\{i\in S_b:0\notin C_{i,b}\}$. An empty screen gives an
empty report. Conditional on the training sample and held-out design,
the Gaussian linear model gives exact marginal coverage for the
full-model coefficient intervals. Bonferroni bounds the probability
of any miss in $S_b$ by $\beta$. The same bound applies after
zero exclusion, so each pipeline has FCR at most $\beta$.

We form $D$ from the four reports by \Cref{eq:pert-cert}, with equal
weights and level $\beta$, and take the interval hull $C_i$ across
runs that report feature $i$. If qualifying intervals disagree in
sign, their hull can contain zero. We therefore mask the certificate
and then lift:
\[
 \widetilde D_i=D_i\one\{0\notin C_i\},\qquad
 A=\mathcal L_\rho(\widetilde D),
\]
where an unreported coordinate has $\widetilde D_i=0$.
By \Cref{eq:lift-identity}, $A_i/\mathcal D(A)\leq\widetilde D_i\leq D_i$,
so the proof of \Cref{thm:stablefcr} applies to the filtered and
lifted output. Every allocated interval excludes zero. Applying the
mask before the lift is important because the lift recomputes the
scale after filtering.

\begin{figure}[!htbp]
 \centering\includegraphics[width=.94\textwidth]{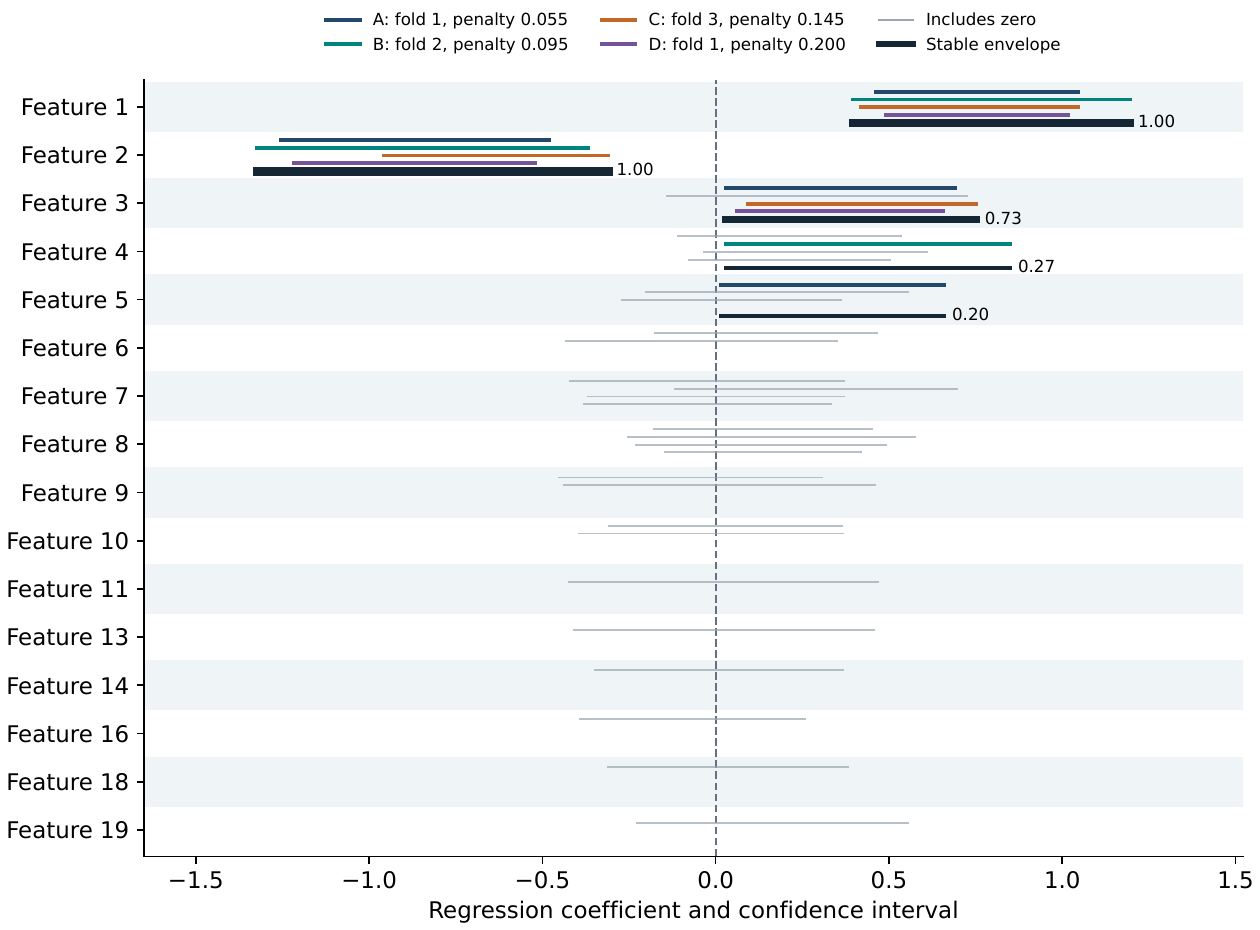}
 \caption{Zero exclusion under four split-lasso pipelines for one
 Gaussian regression data set. Gray intervals contain zero, colored
 intervals pass the filter, and blank entries indicate that a
 pipeline screened out the feature. Black segments are the envelopes
 of the passing intervals. Black line thickness and labels show the
 final stable allocations. The plot includes every feature screened
 by any pipeline.}
 \label{fig:fcr-regression}
\end{figure}

The example has $300$ observations and $20$ predictors. The design has
four independent blocks of five predictors with within-block correlation
$0.45^{|i-j|}$, and the noise variance is one. The first eight coefficients
are $(.68,-.52,.35,.27,.19,.15,.11,.08)$, and the others are zero.
Each held-out fold has $100$ observations. Pipelines A through D use
folds $(1,2,3,1)$ and lasso penalties $(.055,.095,.145,.20)$, with
objective $\|Y-Xb\|_2^2/(2n_{\rm train})+\lambda\|b\|_1$. The pilot
level is $\beta=.08$, the stable target is $\alpha=.10$, and the
ramp width is $.05$.

In the displayed data set, A reports features $\{1,2,3,5\}$,
B reports $\{1,2,4\}$, and C and D report $\{1,2,3\}$. The
stable report allocates $(1,1,.73,.27,.20)$ to features 1 through 5.
Features 1 and 2 pass in every pipeline. Feature 3 passes in three
runs, and features 4 and 5 each pass in one. Feature 4 receives
slightly more because its qualifying run divides its certificate
among three discoveries, compared with four for feature 5.
\Cref{tab:fcr-regression} reports a repeated-sampling check.

\subsection{Numerical verification of the extensions}
\label{sec:extension-sim}

The Holm comparison uses the BH observations of \Cref{sec:path-sim}.
For false coverage, independent $N(\theta_i,1)$ estimators have
$m=40$ and eight signal means $3.2$, with target $.10$.
Positive MQC uses $\psi=.9$, selection at $\psi\lambda$,
and 21 midpoint levels in $[.08,.12]$. The MQC comparison uses
2,400 data sets; Holm and symmetric intervals use 16,000.

\begin{table}[htbp]
 \centering\footnotesize
 \begin{tabular}{@{}lllrrrr@{}}
\toprule
Setting & Procedure & Risk & Estimate & Signal mass & Partial & Peak \\
\midrule
Gaussian PRDS & Holm & sFWER & 0.056 & 5.34 & 0.00 & 0.95 \\
Gaussian PRDS & Stable-Holm & sFWER & 0.056 & 5.33 & 0.43 & 0.95 \\
Gaussian intervals & Selected intervals & sFCR & 0.050 & 6.96 & 0.00 & 1.00 \\
Gaussian intervals & Stable envelopes & sFCR & 0.038 & 6.93 & 0.68 & 1.00 \\
Positive MQC & MQC report & sFCR & 0.092 & 6.88 & 0.00 & 1.00 \\
Positive MQC & Stable MQC & sFCR & 0.087 & 6.85 & 0.62 & 1.00 \\
\bottomrule
\end{tabular}

 \caption{Empirical familywise and false coverage risks at target $.10$.
 Signal mass, Partial, and Peak have the meanings of
 \Cref{tab:stable-sim}; the CSV includes standard errors.}
 \label{tab:stable-extension-sim}
\end{table}

For symmetric intervals, stable aggregation reduces estimated
sFCR from $.050$ to $.038$; allocation-weighted width rises
from $4.67$ to $4.86$. For positive MQC, the nominal and stable
risks are $.092$ and $.087$.

The regression comparison repeats \Cref{sec:fcr-regression} on
600 data sets. The stable envelope has empirical sFCR $.0074$
(SE $.0011$), and the pilot FCRs range from $.013$ to $.019$.
Mean mass is the number of binary reports or the total stable
allocation. The coverage profile retains the original allocation
scale, with profile loss $.0158$. Thresholding at $.35$ multiplies
the certificate by $.65$ and gives mean allocation $.81$.
The pilot adjustment following \Cref{thm:profile-fcr} restores
full-scale allocations for persistent reports.

\begin{table}[!htbp]
 \centering\footnotesize
 \begin{tabular}{@{}lrrr@{}}
\toprule
Report & Error & Monte Carlo SE & Mean mass \\
\midrule
A & 0.0132 & 0.0027 & 2.65 \\
B & 0.0129 & 0.0026 & 2.85 \\
C & 0.0174 & 0.0032 & 2.73 \\
D & 0.0194 & 0.0039 & 2.25 \\
Stable envelope & 0.0074 & 0.0011 & 2.40 \\
Coverage profile & 0.0158 & 0.0017 & 2.40 \\
Threshold $.35$ & 0.0073 & 0.0011 & 0.81 \\
\bottomrule
\end{tabular}

 \caption{False coverage after zero exclusion in the split-lasso
 regression experiment. Error is FCR for A--D, profile sFCR for the coverage-profile
 report, and ordinary sFCR for the envelope and thresholded reports.}
 \label{tab:fcr-regression}
\end{table}

\end{document}